\documentclass[aps]{revtex4}
\usepackage{amsfonts}
\usepackage{amsmath}
\usepackage{amssymb,epsf}
\usepackage{color}

\begin{document}

\title{Holographical aspects of dyonic black holes: \\
Massive gravity generalization }
\author{S. H. Hendi$^{1,2}$\footnote{
email address: hendi@shirazu.ac.ir}, N. Riazi$^{3}$\footnote{
email address: n$_{-}$riazi@sbu.ac.ir} and S.
Panahiyan$^{3}$\footnote{ email address:
sh.panahiyan@gmail.com}} \affiliation{$^1$ Physics Department and
Biruni Observatory, College of Sciences, Shiraz
University, Shiraz 71454, Iran\\
$^2$ Research Institute for Astronomy and Astrophysics of Maragha
(RIAAM), P.O. Box 55134-441, Maragha, Iran\\
$^3$ Physics Department, Shahid Beheshti University, Tehran 19839,
Iran}

\begin{abstract}
The content of this paper includes studying holographical and
thermodynamical aspects of dyonic black holes in the presence of
massive gravity. For the first part of paper, thermodynamical
properties of the bulk which includes black holes are studied and
the main focus is on critical behavior. It will be shown that the
existence of massive gravitons introduces remnant for temperature
after evaporation of black holes, van der Waals phase transition
for non-spherical black holes and etc. The consistency of
different thermodynamical approaches toward critical behavior of
the black holes is presented and the physical properties near the
region of thermal instability are given. Next part of the paper
studies holographical aspects of the boundary theory.
Magnetization and susceptibility of the boundary are extracted and
the conditions for having diamagnetic and paramagnetic behaviors
are investigated. It will be shown that generalization to massive
gravity results into the existence of diamagnetic/paramagnetic
phases in phase structure of the hyperbolic and horizon flat of
boundary conformal field theory.
\end{abstract}

\maketitle

\section{Introduction}

Considering different exact solutions for the massive objects,
special
attention is paid to the stringy black holes (see \cite%
{StringyBH1,StringyBH2,StringyBH3} for more details). These black
holes enjoy the duality of electric/magnetic charges and possibly
mass/dual mass \cite{DualMass}. In this regard, it was shown that
two constants of a Taub-NUT system
\cite{NUT1,NUT2,NUT3,NUT4,NUT5,NUT6} can be interpreted as a
gravitating dyon with both ordinary mass and its dual. In other
words, the role of Nut charge is the mass duality, such as the
duality between electric and magnetic charges in the $U(1)$
Maxwell theory \cite{Duality1,Duality2}. Dyonic black hole
solutions and their interesting properties have been
investigated in literature \cite%
{Dyonic1,Dyonic2,Dyonic3,Dyonic4,Dyonic5,Dyonic6,Dyonic7,Dyonic8,Dyonic9,
Dyonic10,Dyonic11,Dyonic12,Dyonic13,Dyonic14,Dyonic15,Dyonic16,Dyonic17,
Dyonic18,Dyonic19,Dyonic20,Dyonic21}. In this paper, we take into
account a dyonic system in the context of massive gravity and
obtain its exact black hole solutions.

It is well known that the usual general relativity can be
understood as a
unique theory of a massless spin-$2$ particle \cite%
{massless1,massless2,massless3,massless4}. This theory is
diffeomorphism invariant which implies a conserved stress-energy
tensor. But in the massive case, there is no restriction to have a
conserved source while its divergence can vanish for vanishing
mass. The idea of massive spin-$2$ particle is a historical
thought and the first academic works were done by Fierz and Pauli
\cite{Fierz}. Since this theory suffers the vDVZ (van
Dam-Veltman-Zakharov) discontinuity \cite{vDVZ1,vDVZ2,vDVZ3}, its
extension to nonlinear regimes was considered \cite{Vainshtein}.
Nevertheless these nonlinear extensions of massive gravity lead
to the presence of the Boulware--Deser ghost \cite{Boulware}. In
order to solve this problem, various candidates have been
suggested, among which we highlight the so-called DGP
(Dvali-Gabadadze-Porrati) model \cite{DGP1,DGP2,DGP3}, new massive
gravity \cite{NEW}, bi-gravity \cite{bi} and dRGT (de Rham,
Gabadadze and Tolley) theory \cite{dRGT1,dRGT2,dRGT3}. In Ref. \cite%
{SchmidtMay} the existence of ghost in dGRT massive gravity with St\"{u}
ckelberg fields was discussed and it was shown conclusively that
this theory is ghost free. There are more attentions to dGRT
massive gravity models and
its extensions (dGRT-like models) \cite%
{dRGT1,dRGT2,dRGT3,Hassan1,Hassan2,review,Arraut1,Arraut2} which
employ various models of reference metric for constructing the
massive terms in the context of black holes. One of the
interesting reference metrics was considered by Vegh in which the
role of massive graviton can be regarded as lattice \cite{Vegh}.
Such massive theory has been applied in various aspects
of gravitation and astrophysics \cite%
{HendiMass1,HendiMass2,HendiMass3,HendiMass4,HendiMass5,neut}. In
addition, from the cosmological point of view, massive gravity has
been considered
before \cite%
{bi,massivecosmologyA1,massivecosmologyA2,massivecosmologyB1,massivecosmologyB2}%
. It was also shown that the cosmological constant problem can be
solved in
the massive gravity context \cite%
{cosm1,cosm2,acc1,acc2,massivecosmologyA2,MassiveLambda1,MassiveLambda2}.
Regarding the points mentioned above, the extension of massive
gravity can open new theoretical avenues to have a deeper insight
in various aspects of gravitating systems.

On the other hand, analyzing the effects of massive gravity on
thermodynamics and phase transition of black holes is an
interesting
subject. Following the pioneering works of the Hawking and Bekenstein \cite%
{Hawking1,Hawking2,Bekenstein}, one may regard the black objects
as thermodynamical systems. In this regard, calculating conserved
and thermodynamic quantities of the black holes, examining the
validity of no-hair conjecture, and also the first law of
thermodynamics are very important. It is worthwhile to mention
that investigation of the Hawking phase transition and thermal
stability is an essential tool for considering a black hole as a
real and viable thermodynamical system.

Recent progress in the black hole thermodynamics \cite%
{Therm1,Therm2,Therm3,Therm4} and its relation to the AdS/CFT
correspondence \cite{AdS1,AdS2,AdS3,AdS4,AdS5,AdS6,AdS7} imply the
thermodynamical variability nature of the cosmological constant.
The thermodynamical quantity related to cosmological constant was
proposed to be pressure which is consistent with dimensional
analysis \cite{kubuiznak}. This opened up the possibility of
introduction of a van der Waals like behavior for the black holes.
In this regard, the reentrant of the phase transition
\cite{reentrant1,reentrant2,reentrant3,reentrant4}, existence of
the triple point \cite{triple1,triple2,triple3} and analogous heat
engines
\cite{heat1,heat2,heat3,heat4,heat5,heat6,heat7,heat8,heat9} were
investigated before. The van der Waals like behavior of the black
holes based on different gravitational models and matter fields
have been
investigated in literature \cite%
{van1,van2,van3,van4,van5,van6,van7,van8,van9,van10,van11,van12,van13,van14,
van15,van16,van17,van18,van19,van20,van21,van22,van23,van24,van25,van26,van27, van28,van29,van30,van31,van32,van33,van34,van35}%
. It was shown that depending on the gravity under consideration
and employed matter fields, the critical behavior of the system
may be modified and some conditions regarding the
existence/absence of van der Waals like behavior will appear.
Regarding the cosmological constant as a dynamical pressure also
modifies the role of the mass of black holes into enthalpy of the
system and first law of black hole thermodynamics. In this case,
the phase space is told to be extended. For a beautiful review
regarding the applications and implications of the extended phase
space, we refer the reader to Ref. \cite{Mann}.

Our main motivation in this paper is exploring the effects of the
graviton's mass on thermodynamical structure of the dyonic black
holes and holographical properties of the corresponding conformal
field theory.

Now, we are ready to embark on our central task, namely, that of
extending thermodynamic properties of dyonic black holes to the
massive theory of gravitation. The outline of the present paper is
as follows; First, the basic field equations governing Einstein
massive dyonic gravity are given and black hole solutions are
extracted. Then geometrical properties of the black holes are
investigated and thermodynamical quantities are obtained. Next,
some details regarding the enthalpy, temperature, heat capacity
and free energy are given. In Sec. \ref{vdW}, the van der Waals
like behavior, critical properties and conditions for
presence/absence critical behavior of these black holes are
investigated. Next, a comparative study regarding different phase
diagrams is done and the connection between van der Waals like
phase transitions and thermal stability of the solutions is
discussed. Section \ref{CFTsec}, is devoted to study holographical
aspects of conformal field theory. The magnetization properties
are extracted and conditions regarding paramegnatism/feromagnetism
are extracted. The paper is concluded with some closing remarks.

\section{Basic Equations and Dyonic Black hole solutions}

As we mentioned in the introduction, there are several approaches
toward constructing a massive theory of the gravity. One of the
most well-known theories was proposed by de Rham, Gabadadze and
Tolley which is based on a reference metric. Employing a modified
reference metric leads to construction of an interesting theory of
massive gravity which is essentially dRGT-like. In general, the
massive part of gravitational Lagrangian is given by
\begin{equation}
L_{massive}=m^{2}\sum_{i}^{4}c_{i}\mathcal{U}_{i}(g,\psi),
\label{Lmassive}
\end{equation}%
in which $c_{i}$'s are arbitrary constants. Since we are interested in $4$%
-dimensional solutions, $\mathcal{U}_{i}$ are symmetric
polynomials of the
eigenvalues of the $4\times 4$ matrix, $\mathcal{K}_{\nu }^{\mu }=\sqrt{%
g^{\mu \alpha }\psi_{\alpha \nu }}$ which can be written as
follows:
\begin{eqnarray}
\mathcal{U}_{1} &=&\left[ \mathcal{K}\right] ,  \notag \\
\mathcal{U}_{2} &=&\left[ \mathcal{K}\right] ^{2}-\left[ \mathcal{K}^{2}%
\right] ,  \notag \\
\mathcal{U}_{3} &=&\left[ \mathcal{K}\right] ^{3}-3\left[ \mathcal{K}\right] %
\left[ \mathcal{K}^{2}\right] +2\left[ \mathcal{K}^{3}\right] ,  \notag \\
\mathcal{U}_{4} &=&\left[ \mathcal{K}\right] ^{4}-6\left[ \mathcal{K}^{2}%
\right] \left[ \mathcal{K}\right] ^{2}+8\left[
\mathcal{K}^{3}\right] \left[
\mathcal{K}\right] +3\left[ \mathcal{K}^{2}\right] ^{2}-6\left[ \mathcal{K}%
^{4}\right] .  \label{ui}
\end{eqnarray}

Here, $\psi_{\alpha \nu}$ and $g^{\mu \alpha }$ are, respectively,
the reference metric and line element of spacetime. It is
worthwhile to mention
that depending on the choices of reference metric, the functional form of $%
\mathcal{U}_{i}$'s may be modified.

The phrase of dyonic comes from specific consideration of gauge
field theory. Here, in order to obtain topological dyonic
solutions, we consider electromagnetic gauge potential one-form
with the following explicit form
\begin{equation}
A=\left( \frac{q_{E}}{r_{+}}-\frac{q_{E}}{r}\right)
dt+q_{M}\Upsilon d\varphi ,  \label{gaugeP}
\end{equation}%
where $q_{E}$ and $q_{M}$ are two constants which are related to
the electric and magnetic charges, respectively. Depending on
choices of topology, $\Upsilon$ can be considered as
\begin{equation}
\Upsilon =\left\{
\begin{array}{cc}
\cos \theta & k=1 \\
\theta & k=0 \\
\cosh \theta & k=-1%
\end{array}%
\right. ,
\end{equation}%
where $k=1$, $0$ and $-1$ represent spherical, flat and hyperbolic
horizons of dyonic black holes, respectively. Based on Eq.
(\ref{gaugeP}), one finds $A_{t}(r=r_{h}) = 0$ which is a
necessary condition for regularity at the horizon, since $g_{tt}
\rightarrow 0$ when $r \rightarrow r_{h}$.

The $4$-dimensional action governing Einstein massive dyonic black
holes in the presence of the Maxwell electromagnetic field is
given by
\begin{equation}
\mathcal{I}=-\frac{1}{16\pi G_{4}}\int d^{4}x\sqrt{-g}\left[ \mathcal{R}%
-2\Lambda -F^{\mu \nu }F_{\mu \nu }+m^{2}\sum_{i}^{4}c_{i}\mathcal{U}%
_{i}(g,\psi)\right] ,  \label{Action}
\end{equation}%
where $\mathcal{R}$ is the scalar curvature, $\Lambda
=-\frac{3}{l^{2}}$ is the negative cosmological constant and
$G_{4}$ is the Newton constant in $4$ dimension. Also, $F_{\mu \nu
}=\partial _{\mu }A_{\nu }-\partial _{\nu }A_{\mu }$ is the
electromagnetic field tensor and $A_{\mu }$ is the gauge
potential.

Using the variational principle with the action of
Einstein--Maxwell massive gravity (\ref{Action}), it is a matter
of calculation to show that field equations are given by
\begin{eqnarray}
G_{\mu \nu }+\Lambda g_{\mu \nu }-\left[ 2F_{\mu \lambda }F_{\nu
}^{\lambda }-\frac{1}{2}g_{\mu \nu }F^{\sigma \rho }F_{\sigma \rho
}\right] +m^{2}\chi
_{\mu \nu } &=&0,  \label{Field equation} \\
&&  \notag \\
\partial _{\mu }\left( \sqrt{-g}F^{\mu \nu }\right) &=&0,
\label{Maxwell equation}
\end{eqnarray}%
where in the above equation, $G_{\mu \nu }$ is the Einstein tensor
and $\chi _{\mu \nu }$ is the contribution of massive terms in the
field equation, which is
\begin{eqnarray}
\chi _{\mu \nu } &=&-\frac{c_{1}}{2}\left( \mathcal{U}_{1}g_{\mu \nu }-%
\mathcal{K}_{\mu \nu }\right) -\frac{c_{2}}{2}\left(
\mathcal{U}_{2}g_{\mu \nu }-2\mathcal{U}_{1}\mathcal{K}_{\mu \nu
}+2\mathcal{K}_{\mu \nu
}^{2}\right) -\frac{c_{3}}{2}(\mathcal{U}_{3}g_{\mu \nu }-3\mathcal{U}_{2}%
\mathcal{K}_{\mu \nu }+  \notag \\
&&6\mathcal{U}_{1}\mathcal{K}_{\mu \nu }^{2}-6\mathcal{K}_{\mu \nu }^{3})-%
\frac{c_{4}}{2}(\mathcal{U}_{4}g_{\mu \nu
}-4\mathcal{U}_{3}\mathcal{K}_{\mu
\nu }+12\mathcal{U}_{2}\mathcal{K}_{\mu \nu }^{2}-24\mathcal{U}_{1}\mathcal{K%
}_{\mu \nu }^{3}+24\mathcal{K}_{\mu \nu }^{4}).
\end{eqnarray}

Here, our main motivation is studying thermodynamical and
holographical aspects of topological massive dyonic black holes.
Therefore, we consider the metric of $4$-dimensional spacetime as
\begin{equation}
ds^{2}=-f(r)dt^{2}+\frac{dr^{2}}{f(r)}+r^{2}d\Omega ^{2},
\label{Metric}
\end{equation}
in which
\begin{equation}
d\Omega ^{2}=\left\{
\begin{array}{cc}
d\theta ^{2}+\sin ^{2}\theta d\varphi ^{2} & k=1 \\
d\theta ^{2}+d\varphi ^{2} & k=0 \\
d\theta ^{2}+\sinh ^{2}\theta d\varphi ^{2} & k=-1%
\end{array}%
\right. .
\end{equation}

As for the reference metric, we employ the following ansatz which
was first introduced by Vegh \cite{Vegh} and then was generalized
by Cai, et al. in Ref. \cite{Cai2015}
\begin{equation}
\psi_{\mu \nu }=\frac{c^2}{r^2}g_{\mu \nu }|_{t=cte, r=cte} ,
\label{f11}
\end{equation}%
where $c$ is a positive constant. By using reference (\ref{f11})
and
spacetime (\ref{Metric}) metrics with Eq. (\ref{ui}), one can obtain $%
\mathcal{U}_{i}$ in the following forms
\begin{equation}
\mathcal{U}_{1}=\frac{2c}{r},\ \
\mathcal{U}_{2}=\frac{2c^{2}}{r^{2}},\ \ \mathcal{U}_{3}=0,\ \
\mathcal{U}_{4}=0.
\end{equation}

Using the obtained $\mathcal{U}_{i}$'s with metric (\ref{Metric}) and Eqs. (%
\ref{Field equation}) and (\ref{Maxwell equation}), one can
extract the metric function, $\psi(r)$, as
\begin{equation}
f(r)=k+\frac{r^{2}}{l^{2}}-\frac{2m_{0}}{r}+\frac{q_{E}^{2}+q_{M}^{2}}{r^{2}}%
+m^{2}\left( \frac{cc_{1}}{2}r+c^{2}c_{2}\right) ,  \label{ff(r)}
\end{equation}%
where $m_{0}$ is an integration constant which is related to the
total mass of massive dyonic black holes. In the absence of
massive parameter ($m=0$), obtained solutions (\ref{ff(r)}) reduce
to the usual topological dyonic black holes \cite{dyonicmassless}
\begin{equation}
f(r)=k+\frac{r^{2}}{l^{2}}-\frac{2m_{0}}{r}+\frac{q_{E}^{2}+q_{M}^{2}}{r^{2}}%
.
\end{equation}

The existence of singularity and horizon indicates that our
solutions can be intrinsically black holes. The presence of
singularity could be investigated by studying curvature scalars
for which we choose the Ricci and Kretschmann scalars. It is a
matter of calculation to show that for these black holes the
mentioned two scalars are
\begin{eqnarray*}
R &=&2\,{\frac{k-1}{{r}^{2}}}-{\frac{c\left( 3\,rc_{1}+2\,\,cc_{2}\right) {m}%
^{2}}{{r}^{2}}}-\frac{12}{l^{2}}\,, \\
R_{\alpha \beta \gamma \delta }R^{\alpha \beta \gamma \delta } &=&\frac{24}{%
l^{4}}+\frac{8(k-1)+4c(3rc_{1}+2cc_{2})m^{2}}{r^{2}l^{2}}+\frac{4(k-1)^{2}}{%
r^{4}}+\frac{8(k-1)\left( q_{E}^{2}+q_{M}^{2}\right)
}{r^{6}}+\frac{56\left(
q_{E}^{2}+q_{M}^{2}\right) ^{2}}{r^{8}} \\
&&-16\left[ (k-1)+\frac{6\left( q_{E}^{2}+q_{M}^{2}\right)
}{r^{2}}\right]
\frac{m_{0}}{r^{5}}+\frac{48m_{0}^{2}}{r^{6}}+2c^{2}\left[ \left(
rc_{1}+cc_{2}\right) ^{2}+c^{2}c_{2}^{2}\right] \frac{m^{4}}{r^{4}} \\
&&-\left[ \frac{4c(rc_{1}-2cc_{2})\left( q_{E}^{2}+q_{M}^{2}\right) }{r^{2}}%
+ \frac{16c^{2}c_{2}m_{0}}{r}-4c(k-1)(rc_{1}+2cc_{2})\right] \frac{m^{2}}{%
r^{4}},
\end{eqnarray*}%
which confirm the existence of essential singularity at the origin
since
\begin{eqnarray}
\lim_{r\longrightarrow 0}R &\longrightarrow &\infty , \\
\lim_{r\longrightarrow 0}R_{\alpha \beta \gamma \delta }R^{\alpha
\beta \gamma \delta } &\longrightarrow &\infty .
\end{eqnarray}

The Ricci and Kretschmann scalars are $\frac{-12}{l^{2}}$\ and $\frac{24}{%
l^{4}}$ for $r\longrightarrow \infty $. Considering $l^{2}=\frac{\pm 3}{%
\Lambda }$, the asymptotical behavior of these solutions is (a)dS for $%
\Lambda <0$ ($\Lambda >0$).

\begin{figure}[tbp]
$%
\begin{array}{cc}
\epsfxsize=7cm \epsffile{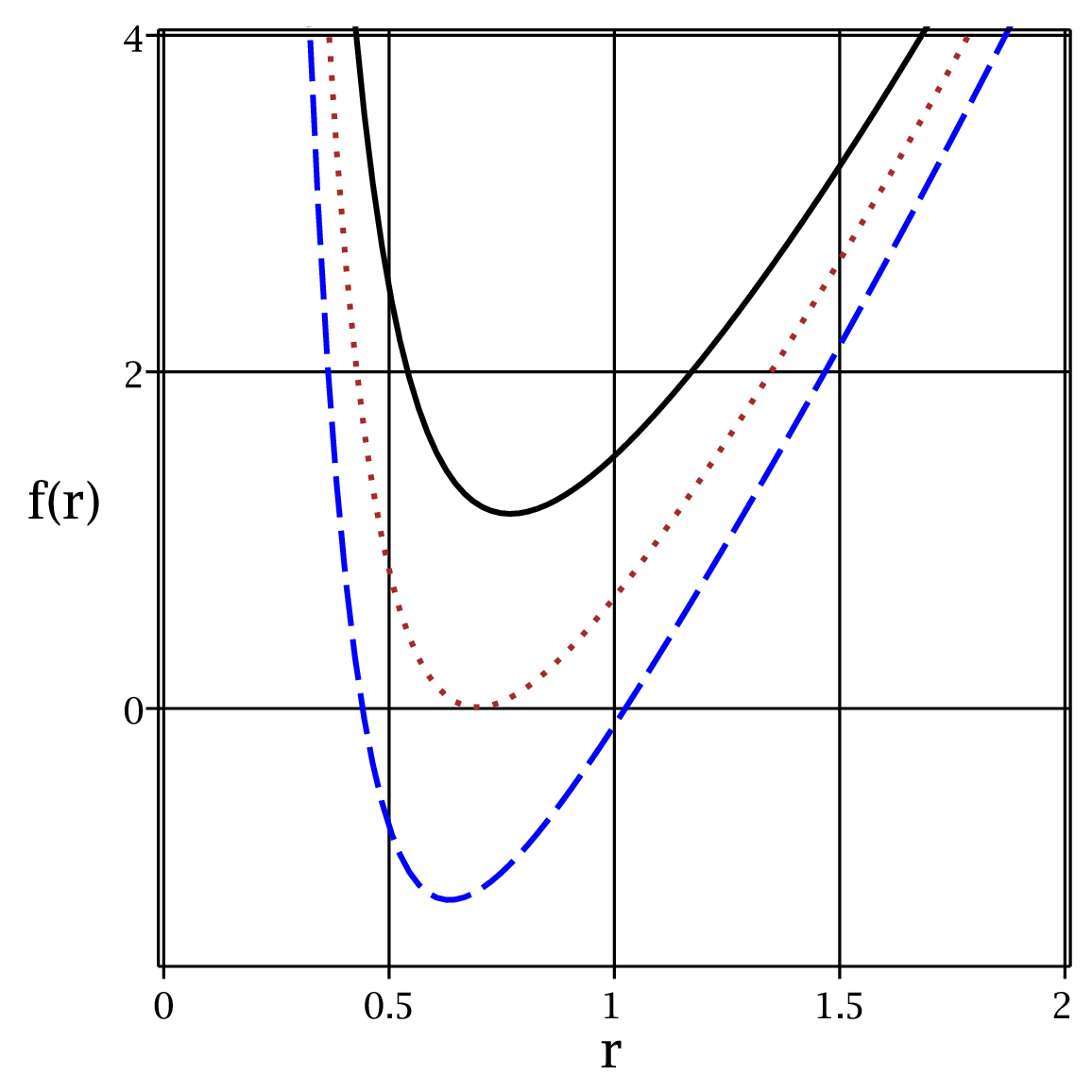} & \epsfxsize=7cm %
\epsffile{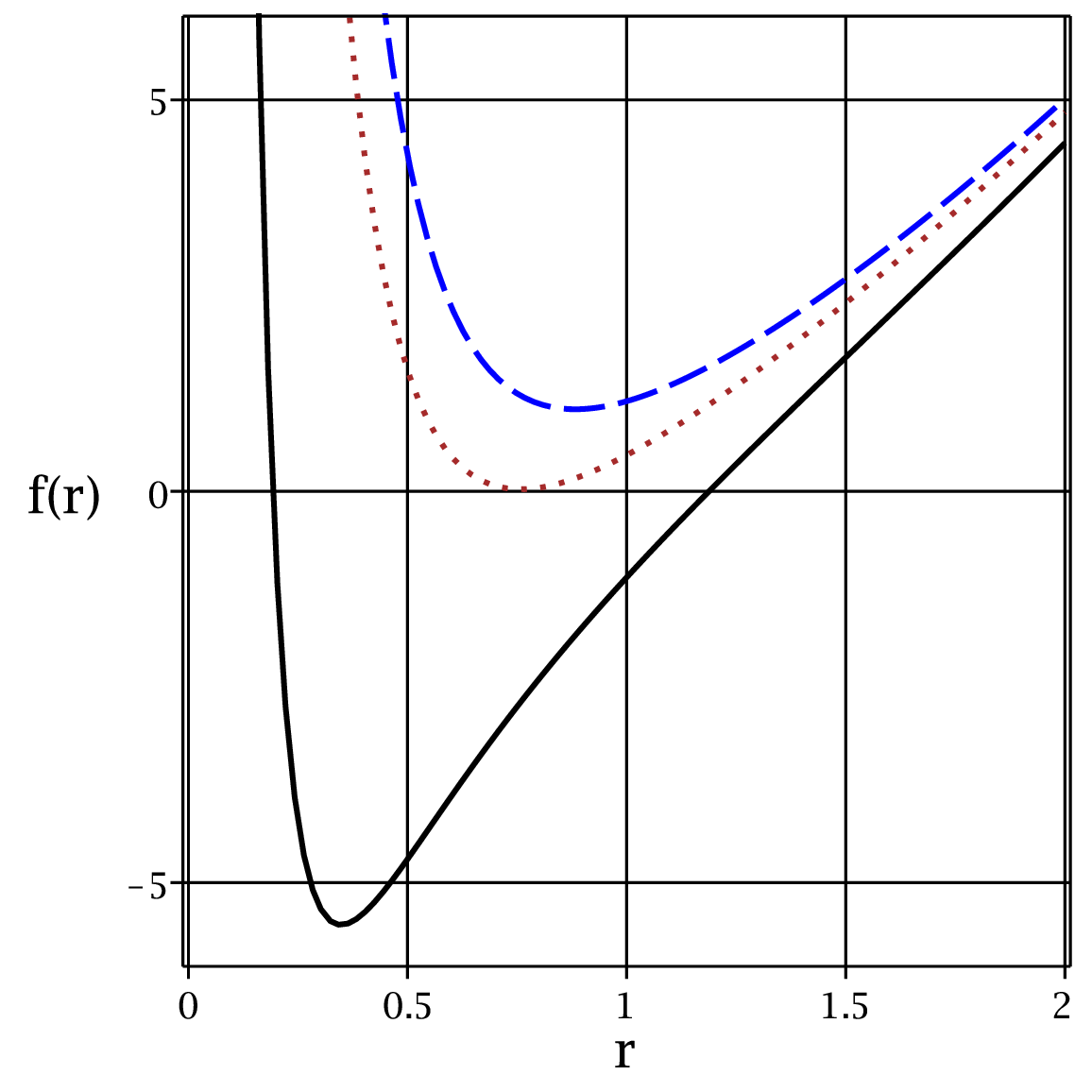}%
\end{array}
$%
\caption{$\protect\psi(r)$ versus $r$ for $b=1$, $Q _{E}=1$, $m=1$, $%
c=c_{1}=c_{2}=1$ and $k=1$; \newline
\textbf{Left diagram:} for $q_{M}=1$, $m_{0}=2$ (continuous line), $%
m_{0}=2.42 $ (dotted line) and $m_{0}=2.8$ (dashed line). \newline
\textbf{Right diagram:} for $m_{0}=2.8$, $q_{M}=0$ (continuous line), $%
q_{M}=1.25$ (dotted line) and $q_{M}=1.5$ (dashed line).}
\label{fr1}
\end{figure}

\begin{figure}[tbp]
$%
\begin{array}{ccc}
\epsfxsize=5.5cm \epsffile{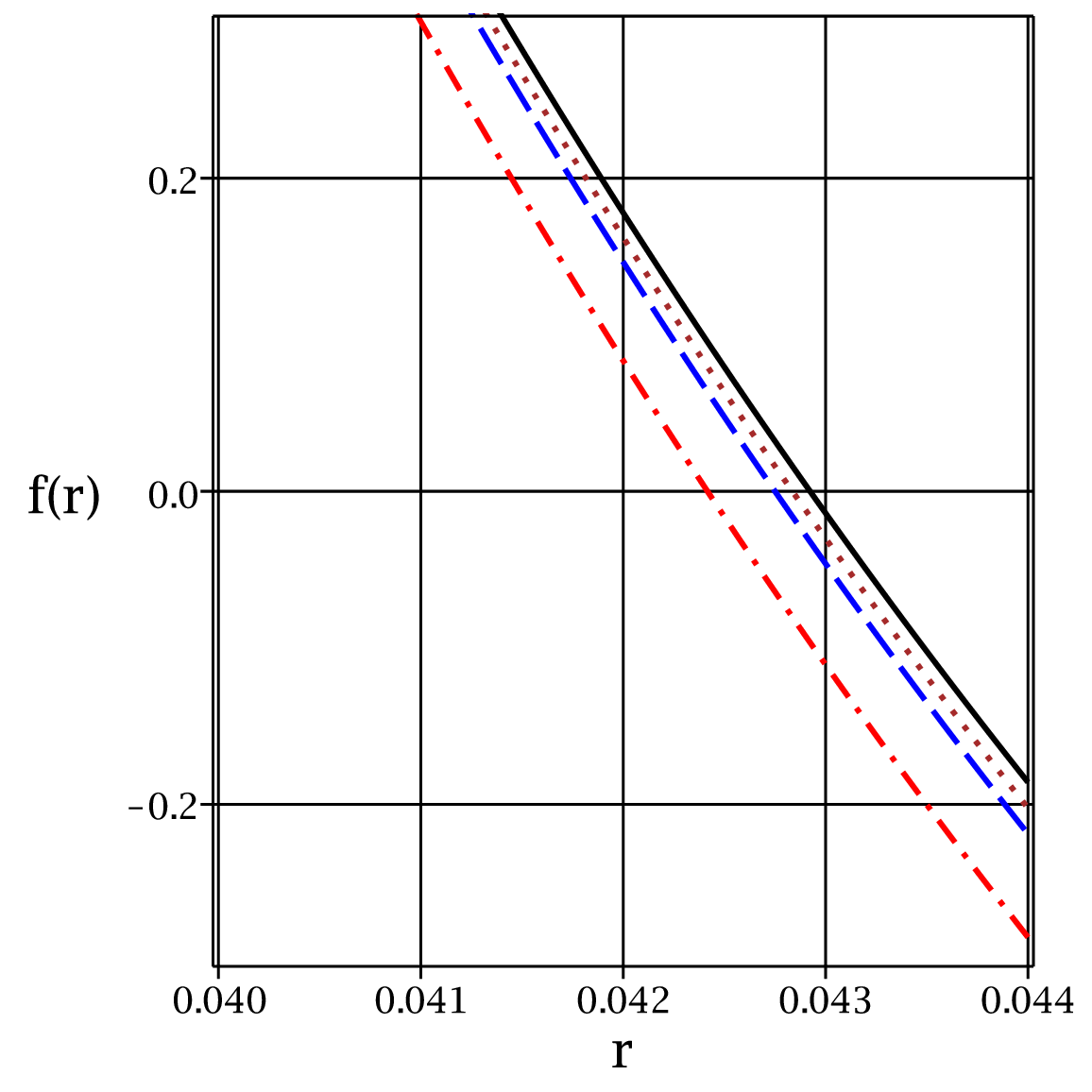} & \epsfxsize=5.5cm %
\epsffile{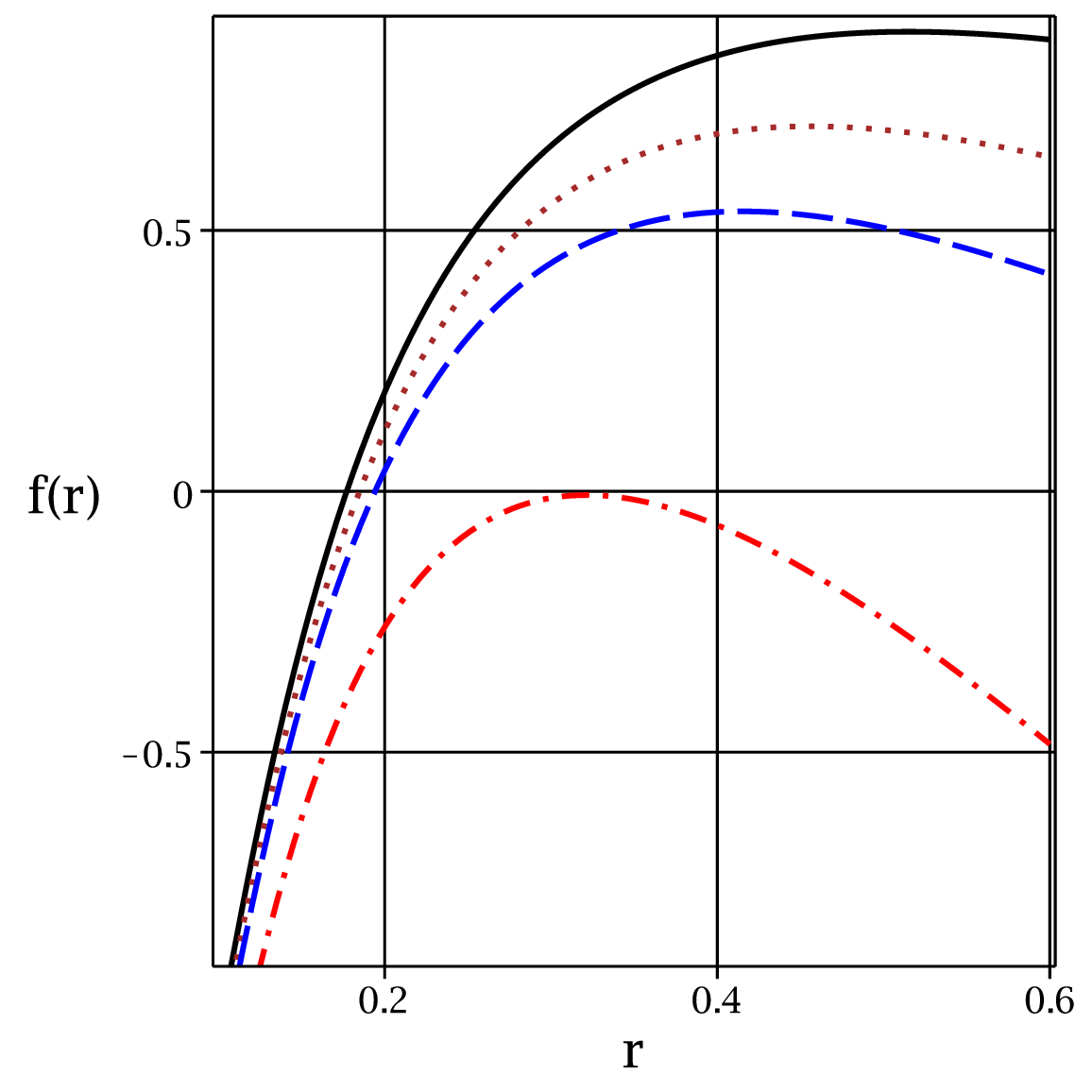} & \epsfxsize=5.5cm \epsffile{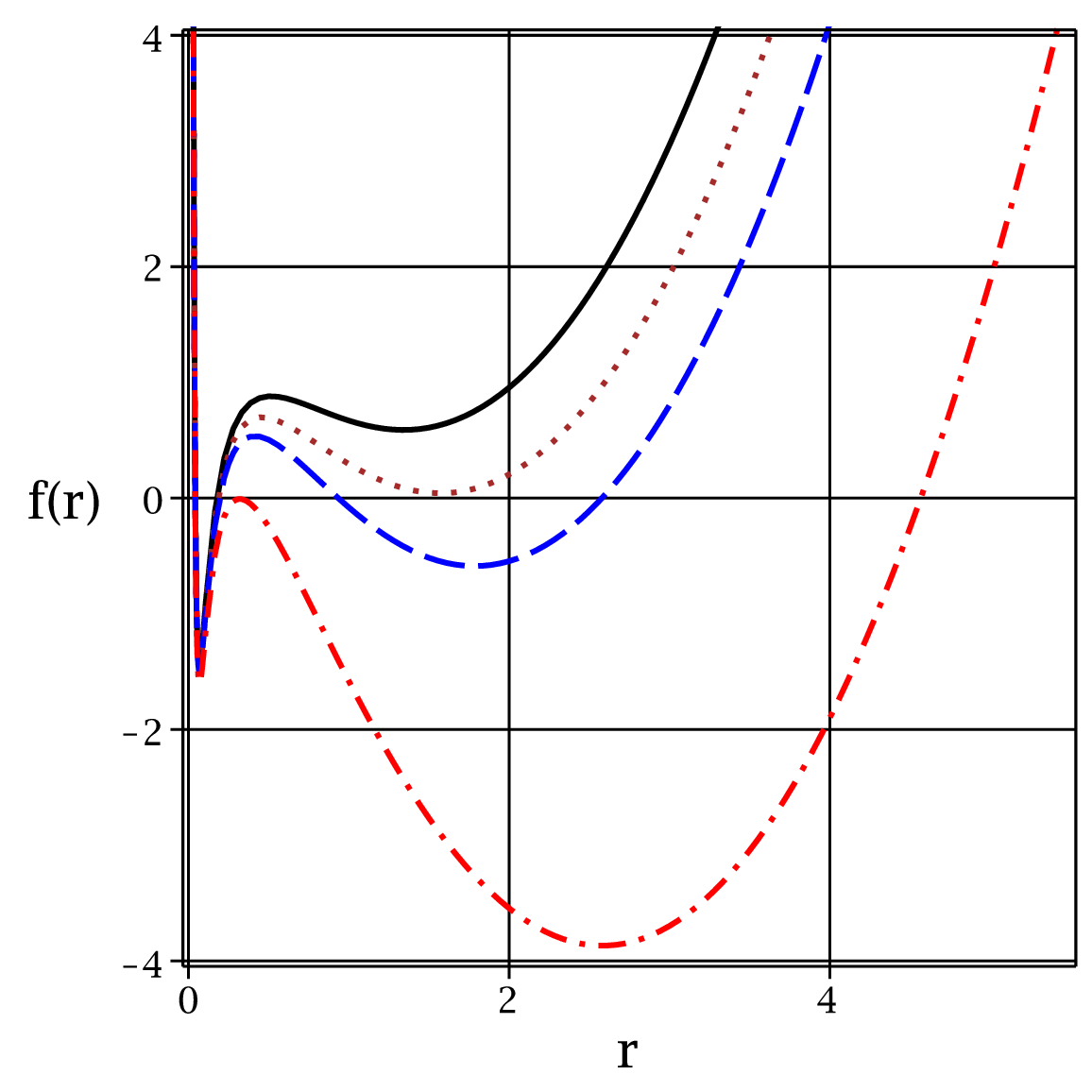}%
\end{array}
$%
\caption{\textbf{For different scales:} $\protect\psi(r)$ versus $r$ for $%
b=1 $, $q_{E}=q_{M}=0.1$, $m=0.3$, $c=3$, $c_{2}=1$ and $k=1$;
\newline For $c_{1}=-8$ (continuous line), $c_{1}=-9$ (dotted
line), $c_{1}=-10$ (dashed line) and $c_{1}=-14$ (dashed-dotted
line).} \label{fr2}
\end{figure}

Now, we focus on the existence of horizon. Due to complexity of
the obtained metric function, it is not possible to study its
roots (which are the horizons) analytically. Therefore, we employ
numerical evaluation and plot
some diagrams for finding the possible roots for the metric function (Figs. %
\ref{fr1} and \ref{fr2}).

Evidently, the existence of horizon and their numbers are highly
sensitive to values of the massive parameters. By suitable choices
of different parameters, these black holes my have
Reissner-Nordstr\"{o}m like behavior (Fig. \ref{fr1}). Meaning,
that they may have: two horizons, one extreme horizon or no
horizon. On the other hand, for another set of values for
different parameters, the Reissner-Nordstr\"{o}m like behavior
will be
modified and existence of more than two horizons will be observed (see Fig. %
\ref{fr2}). The geometrical structure of the black holes depends
on the number and type of horizons. In other words, it is possible
to have usual horizon for outer horizon or an extreme one
depending on choices of massive
parameters. Since the existence of multi horizons was discussed before \cite%
{HendiMass3}, we ignore it and concentrate on thermodynamic
behavior. In addition, it is worthwhile to mention that the metric
function with no real positive root results into the absence of
horizon covering the singularity. This case is known as naked
singularity and is not of interest in this paper.

\section{Thermodynamical quantities \label{Thermo1}}

Our next step for studying holographical aspects of massive dyonic
black holes is obtaining thermodynamical quantities. The Hawking
temperature could be obtained by employing the concept of surface
gravity. Using this concept, the Hawking temperature on the outer
horizon of these black holes ($r_{+}$) is given by
\begin{equation}
T=\frac{1}{4\pi }\left[ \frac{k}{r_{+}}+\frac{3r_{+}}{l^{2}}-\frac{%
q_{E}^{2}+q_{M}^{2}}{r_{+}^{3}}+m^{2}\left( cc_{1}+\frac{c^{2}c_{2}}{r_{+}}%
\right) \right] .  \label{TotalTT}
\end{equation}

The obtained black hole solutions in bulk are asymptotically $AdS$
which are related to a $CFT$ theory living on the boundary of
bulk. The
electromagnetic field of bulk is dual to global $U(1)$ current operator $%
J_{\mu }$. It is a matter of calculation to show that the
conserved global charge of the $CFT$ related to this current is
given by \cite{dyonicmassless}
\begin{equation}
\left\langle J^{t}\right\rangle =\frac{q_{E}}{16\pi G_{4}},
\end{equation}%
which by using the holographic dictionary, one can obtain \cite%
{dyonicmassless}
\begin{equation}
\left\langle J^{t}\right\rangle =\frac{\sqrt{2}N^{\frac{3}{2}}}{24\pi l^{2}}%
q_{E},
\end{equation}%
in which $N$ is the degree of gauge group in $CFT$. Studying
asymptotic value of the bulk field strength (from Eq.
(\ref{gaugeP})) shows that the boundary $CFT$ has a magnetic field
strength
\begin{equation}
B=\frac{q_{M}}{l^{2}},
\end{equation}%
which may be employed to study holographical aspects of the
boundary theory.

The total mass of these black holes could be obtained by using
Arnowitt-Deser-Misner approach or on shell action method \cite%
{dyonicmassless}. The total mass is given by
\begin{equation}
M=\frac{r_{+}}{2}\left[ k+\frac{r_{+}^{2}}{l^{2}}+\frac{q_{E}^{2}+q_{M}^{2}}{%
r_{+}^{2}}+m^{2}\left( \frac{cc_{1}}{2}r_{+}+c^{2}c_{2}\right)
\right] . \label{Mass}
\end{equation}

Recent progresses in studying thermodynamical structure of the
black holes suggest that the cosmological constant plays the role
of a thermodynamical variable. In other words, this quantity is a
variable which can be interpreted as the thermodynamical pressure.
The relation between the cosmological constant and pressure for
the Einstein black holes can be written as
\begin{equation}
P=-\frac{\Lambda }{8\pi }=\frac{3}{8\pi l^{2}}.  \label{PEP}
\end{equation}

It is worthwhile to mention that generalization to scalar-tensor
theories modifies the mentioned relation between thermodynamical
pressure and cosmological constant (\ref{PEP}), while the massive
gravity does not have any effect on this relation. It should be
pointed out that it is possible to
construct massive theory of gravity using scalar tensor theories \cite%
{New1,New2}. In this case, the relation between pressure and
cosmological constant could be affected by massive nature of the
gravity. Replacing $l$ with its corresponding pressure in total
mass, one can obtain enthalpy for these black holes
\begin{equation}
H=\frac{4\pi P}{3}r_{+}^{3}+\frac{r_{+}}{2}\left[ k+\frac{q_{E}^{2}+q_{M}^{2}%
}{r_{+}^{2}}+m^{2}\left( \frac{cc_{1}}{2}r_{+}+c^{2}c_{2}\right)
\right] . \label{internal}
\end{equation}

The electric potential for our solution is given by
\begin{equation}
\Phi _{E}=\frac{q_{E}}{r_{+}}.  \label{PhiE}
\end{equation}

Hereon, we are interested in doing our study in an ensemble in
which its asymptotic value of $A_{t}$ (electric potential) is
constant. Therefore, we
will replace $q_{E}$ with its corresponding electric potential, $%
\Phi_{E}$, in thermodynamical quantities. Such case corresponds to
fixing the chemical potential in $CFT$. The mechanism here is to
consider the $A_{t} $ component of the electromagnetic potential
vector as a fixed value on the horizon (with respect to infinity)
and replace all the corresponding quantities related to electric
field with the electric potential.

Using enthalpy, we can extract other thermodynamical quantities
with the following forms
\begin{eqnarray}
V &=&\frac{\partial H}{\partial P}=\frac{4\pi }{3}r_{+}^{3},  \label{V} \\
q_{E} &=&\frac{\partial H}{\partial \Phi _{E}}=\frac{r_{+}\Phi
_{E}}{G_{4}},
\label{qE} \\
\Phi _{M} &=&\frac{\partial H}{\partial
q_{M}}=\frac{q_{M}}{G_{4}r_{+}},
\label{PhiM} \\
S &=&\frac{\partial H}{\partial T}=\frac{1}{4G_{4}}4\pi r_{+}^{2}.
\label{TotalS}
\end{eqnarray}

Here, $\Phi _{M}$ represents the intensive variable related to
magnetic charge.

The free energy can be written as
\begin{equation}
W=H-TS-\Phi _{E}q_{E},
\end{equation}%
where by using the obtained thermodynamical quantities, one can
extract the free energy of massive dyonic black hole in the
following form
\begin{equation}
W=-\frac{2\pi P}{3}r_{+}^{3}+\frac{r_{+}}{4}\left[ k+\frac{3q_{M}^{2}}{%
r_{+}^{2}}-\Phi _{E}^{2}+m^{2}c^{2}c_{2}\right] .  \label{free1}
\end{equation}

Now, we are in a position to study thermodynamic and holographic
structures of the massive dyonic black holes.

\subsection{Mass/Enthalpy}

The mass of black holes is the same as internal energy in usual
thermodynamics whereas by employing the relation between the
cosmological constant and thermodynamical pressure, it will be
interpreted as the enthalpy. Let us focus on it being enthalpy.
First of all, the horizon radius has polynomial-like behavior
which indicates that there exists an
extremum for the enthalpy. The dominant term for small black holes is the $%
q_{M}$ term, whereas, for large black holes, the $P$ term is
dominant. Since these parameters are positive valued, it is safe
to say that the extremum for these black holes is actually a
minimum.

\begin{figure}[tbp]
$%
\begin{array}{cc}
\epsfxsize=7cm \epsffile{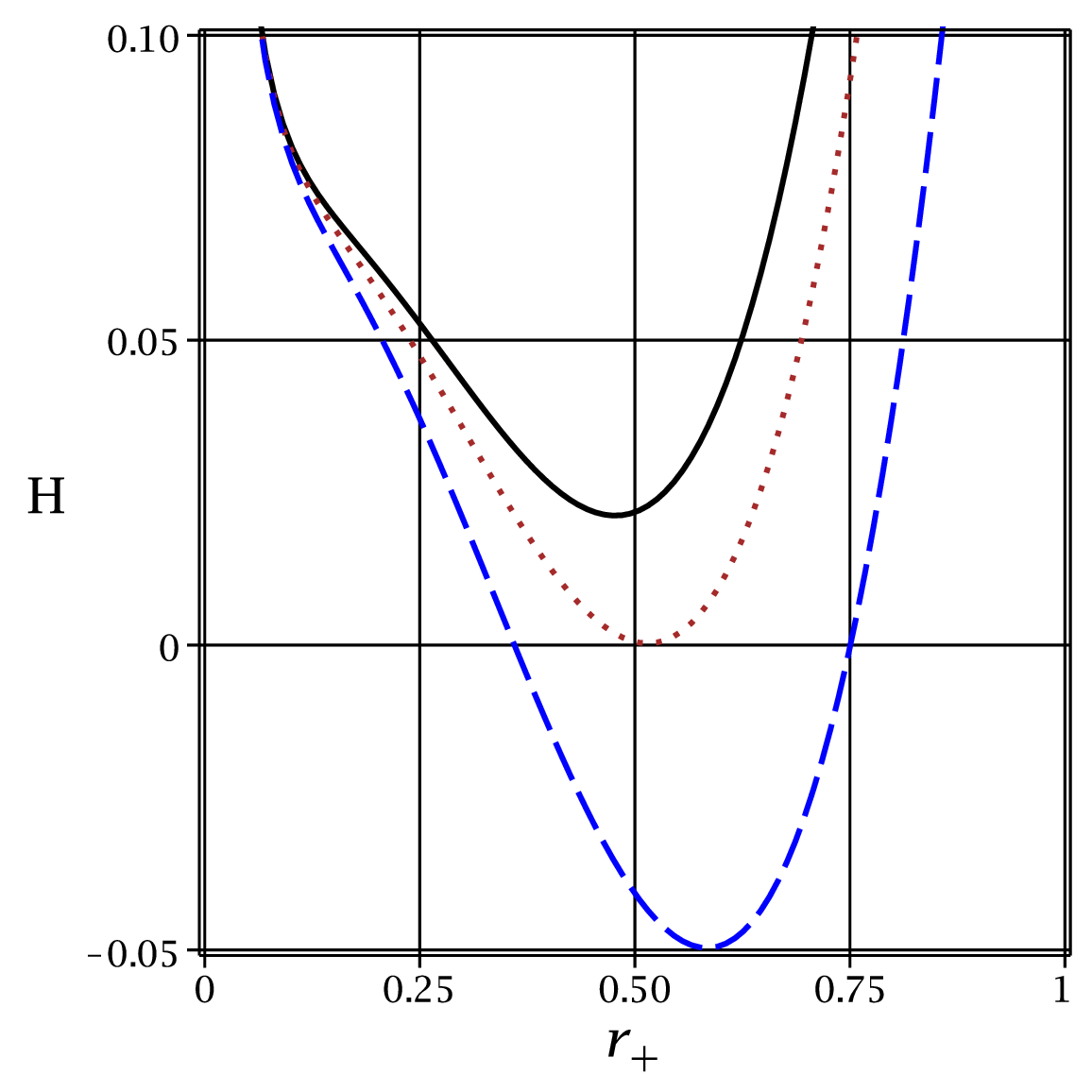} & \epsfxsize=7cm %
\epsffile{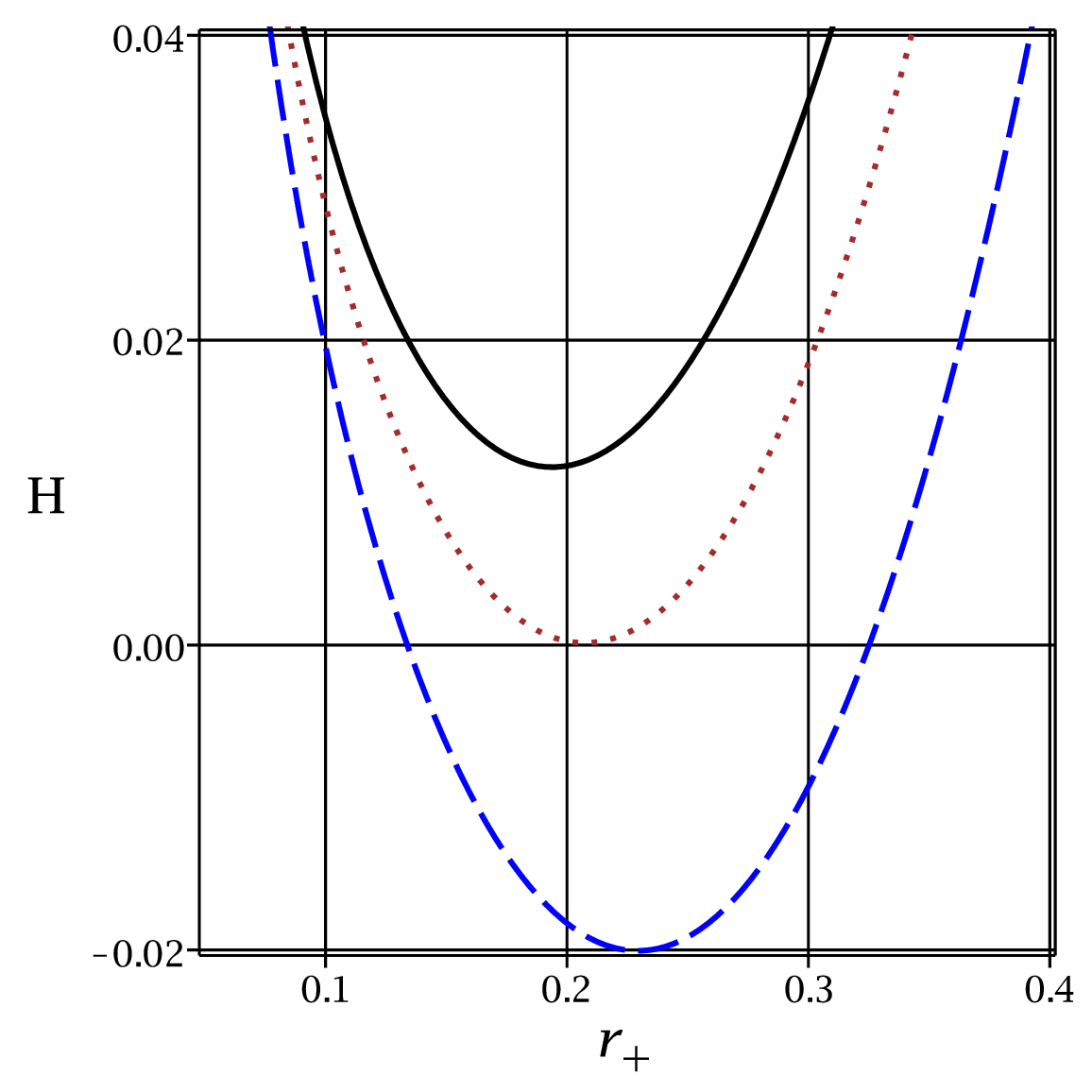}%
\end{array}
$%
\caption{$H$ versus $r_{+}$ for $P=0.5$, $\Phi _{E}=1$, $q_{M}=0.1$, $m=1$, $%
c=1$ and $k=1$; \newline
\textbf{Left diagram:} for $c_{2}=1$, $c_{1}=-8$ (continuous line), $%
c_{1}=-8.34$ (dotted line) and $c_{1}=-9$ (dashed line). \newline
\textbf{Right diagram:} for $c_{1}=1$, $c_{2}=-0.4$ (continuous line), $%
c_{2}=-0.515$ (dotted line) and $c_{2}=-0.7$ (dashed line).}
\label{Fig1}
\end{figure}

Due to the nature of $P$, $\Phi_{E}$, $q_{M}$, $m$ and $c$, these
parameters are positive and they have positive contributions to
the values of enthalpy. On the other hand, $c_{1},$ $c_{2}$ and
$k$ could be negative or positive. Therefore, they could either
contribute negatively or positively to the enthalpy. If we
consider the latter parameters to be positive, the enthalpy will
be always positive without any root. But, if we take them to be
negative valued, it may be possible to obtain one or two roots for
the enthalpy depending on the choices of different parameters.
Existence of negative enthalpy indicates that the system in this
case is not a physical one. Therefore, in this case (presence of
two roots for enthalpy), small and large black holes will be
physical whereas the medium black holes will be non-physical.
Since we could not obtain the root of mass analytically, we employ
numerical calculations and present the results in various diagrams
(see Fig. \ref{Fig1}).

\subsection{Temperature}

Another thermodynamic quantity of interest is temperature. Taking
a closer look at the temperature, one can see that for large and
small horizon radii, $q_{M}$ and $P$ are dominant terms,
respectively. Since $q_{M}$ term is negative and $P$ term is
positive, one can conclude that at least one root exists for the
temperature. This root separates the non-physical solutions with
negative temperature from physical black holes with positive
temperature. Therefore, there exists a condition for having
physical black holes with positive temperature which is
\begin{equation}
8\pi P\ r_{+}^{4}+m^{2}c_{1}c\ r_{+}^{3}+(k^{2}-\Phi _{E}^{2} +
m^{2}c_{2}c^{2} )\ r_{+}^{2} -q_{M}^{2} >0.  \label{posT}
\end{equation}

It is worthwhile to mention that the presence of $q_{M}$ leads the
temperature to diverge toward -$\infty $ in the limit of $%
r_{+}\longrightarrow 0$, faster. This is one of the effects of the
presence of magnetic charge. In addition, one can see that $c_{1}$
in temperature is not coupled with horizon radius. Here, it merely
plays the role of a constant. Due to the coupling of different
parameters with varying orders of horizon radius, for every
specific ranges of horizon radius, one of these terms could be
dominant. This may lead to the presence of extrema for the
temperature. It is straightforward to show that the extrema of
temperature are given by
\begin{equation}
r_{Extremum-T}=\,\sqrt{{\frac{{m}^{2}c^{2}{c}_{2}-{\Phi }_{E}^{2}+k\pm \sqrt{%
{m}^{2}c^{2}{c}_{2}\left( 2k+{m}^{2}c^{2}{c}_{2}-2{\Phi }_{E}^{2}\right) +{%
\Phi }_{E}^{4}+k\left( k-2{\Phi }_{E}^{2}\right) -96\,P\pi \,q_{M}^{2}}}{%
16P\pi }}}.  \label{rootT}
\end{equation}

The above obtained extrema for the temperature indicate that by
suitable choices of different parameters, it may be possible to
have one of the following cases:

I) Temperature being only an increasing function of the horizon
radius without any extremum.

II) Existence of only one extremum with one root for the
temperature.

III) Presence of one minimum and one maximum with one root for the
temperature.

We will investigate these cases later in a separate section with
some diagrams for more clarifications.\newline

\textbf{$q_{M}\longrightarrow 0$\ limit:}\newline In the absence
of magnetic charge, it is possible to derive an interesting
property for these black holes. In the expression for temperature,
$k$, $\Phi _{E}$ and $c_{2}$ are coupled with horizon radius of
the same order. In the absence of the magnetic charge, the
dominant terms for small values of horizon radius will be
\begin{equation}
m^{2}c_{2}c^{2}+k-\Phi _{E}^{2}.
\end{equation}

Now, it is possible to tune this relation in a way so that it
vanishes. If we eliminate this relation, one can find the
following relation for temperature
\begin{equation}
T=2r_{+}P+\frac{m^{2}cc_{1}}{4\pi },
\end{equation}%
which for the limit of $r_{+}\longrightarrow 0$, it is non-zero.

Remembering that in the evaporation of black holes by the Hawking
radiation mechanism, the horizon radius eventually vanishes.
However, we see here that in this case, there will be a remnant
for the temperature of black holes. This indicates that all the
information regarding existence of the black holes is not vanished
completely despite the statement of paradox information. In fact,
a trace of existence of black holes will remain which presents
itself as a fluctuation in temperature of the background spacetime
where black hole was present. This specific property for the
temperature is due to the existence of massive gravitons. This
shows that generalization from massless gravitons to massive ones,
introduces new properties to the thermodynamics of black holes
which could solve and answer some long standing questions
regarding the physics of black holes such as the information
paradox. It is worthwhile to mention that existence of the remnant
for temperature of the black holes, to our knowledge, so far was
only reported for black holes in the context of massive gravity \cite%
{HendiMass5} and it is one of the unique properties of the massive
gravity which make it different from other modified theories of
the gravity.

\subsection{Heat Capacity}

Our next thermodynamical quantity of interest is the heat
capacity. This quantity is of interest because of two important
properties; sign and divergencies. Basically, the conditions
regarding the stability of black holes could be attained by
studying the sign of the heat capacity. Regardless of other
parameters of the black holes, the negativity of it shows that
black holes are thermally unstable while the vice versa is true
for stability. On the other hand, it has been argued that black
holes go under second order phase transitions where they meet
divergencies of the heat capacity. Since we are working in
extended phase space, the heat capacity is given by
\begin{equation}
C_{Q,P}=\frac{T}{\left( \frac{\partial ^{2}M}{\partial S^{2}}\right) _{Q,P}}%
=T{\left( \frac{\partial S}{\partial T}\right) _{Q,P}}=T\frac{\left( \frac{%
\partial S}{\partial r_{+}}\right) _{Q,P}}{\left( \frac{\partial T}{\partial
r_{+}}\right) _{Q,P}}.  \label{CQ}
\end{equation}

By using the temperature (\ref{TotalTT}) and entropy
(\ref{TotalS}), one can obtain the heat capacity of massive
charged dyonic black holes as
\begin{equation}
C_{Q,P}=2\pi r_{+}^{2}\,{\frac{{m}^{2}cr_{+}^{2}\left( \,{c}%
_{2}c+c_{1}r_{+}\right) -{\Phi }_{E}^{2}{r}_{+}^{2}-q_{M}^{2}+k{r}%
_{+}^{2}+8\pi \,{r}_{+}^{4}P}{{\Phi }_{E}^{2}{r}_{+}^{2}+3q_{M}^{2}+8\pi \,{r%
}_{+}^{4}P-{m}^{2}r_{+}^{2}\,{c}_{2}c^{2}-k{r}_{+}^{2}}.}
\label{CQQ}
\end{equation}

In order to have a positive heat capacity, the numerator and
denominator of heat capacity must be whether both negative or
positive. The numerator of heat capacity is exactly the same with
positive condition that we found for the temperature (see Eq.
(\ref{posT})). Since this relation is supposed to be positive, the
denominator must also be positive to have a positive heat
capacity, and hence, thermally stable solutions. Therefore, we
have
\begin{equation*}
{\Phi }_{E}^{2}{r}_{+}^{2}+3q_{M}^{2}+8\pi \,{r}_{+}^{4}P-{m}^{2}r_{+}^{2}\,{%
c}_{2}c^{2}-k{r}_{+}^{2}>0,
\end{equation*}%
in which by combining it with previous condition for having
positive numerator, we obtain
\begin{equation}
0<m^{2}c_{2}c^{2}r_{+}^{2}+r_{+}^{3}m^{2}c_{1}c-\Phi
_{E}^{2}r_{+}^{2}-q_{M}^{2}+k^{2}r_{+}^{2}+8\pi
Pr_{+}^{4}<r_{+}^{3}m^{2}c_{1}c-2q_{M}^{2}.  \label{posCQ}
\end{equation}

In conclusion, if Eq. (\ref{posCQ}) is satisfied, the heat
capacity will be positive and black holes are thermally stable.

On the other hand, divergencies of the heat capacity are obtained
according to the following relation
\begin{equation}
r_{\infty -C_{Q,P}}=\sqrt{{\frac{{m}^{2}c^{2}{c}_{2}-{\Phi
}_{E}^{2}+k\pm
\sqrt{{m}^{2}c^{2}{c}_{2}\left( 2k+{m}^{2}c^{2}{c}_{2}-2{\Phi }%
_{E}^{2}\right) +{\Phi }_{E}^{4}+k\left( k-2{\Phi }_{E}^{2}\right)
-96\pi Pq_{M}^{2}}}{16P\pi }}}.  \label{rcCQ}
\end{equation}

Evidently, depending on the choices of different parameters, the
heat capacity may come across one of the following cases:

I) Absence of divergency which is observed when it is not real
valued.

II) Presence of only one divergency which takes place only when
the following conditions for different parameters hold
simultaneously
\begin{equation*}
\begin{array}{c}
{m}^{2}c^{2}{c}_{2}\left( 2k+{m}^{2}c^{2}{c}_{2}-2{\Phi }_{E}^{2}\right) +{%
\Phi }_{E}^{4}+k\left( k-2{\Phi }_{E}^{2}\right) -96\,P\pi \,q_{M}^{2}=0, \\
\\
{m}^{2}c^{2}{c}_{2}-{\Phi }_{E}^{2}+k>{0}.%
\end{array}%
\end{equation*}

III) Existence of two divergencies. This case happens if both
positive and negative branches of the obtained solutions are
positive and real valued. Therefore, another set of conditions are
imposed for negative branch as well which are
\begin{equation*}
\begin{array}{c}
{m}^{2}c^{2}{c}_{2}\left( 2k+{m}^{2}c^{2}{c}_{2}-2{\Phi }_{E}^{2}\right) +{%
\Phi }_{E}^{4}+k\left( k-2{\Phi }_{E}^{2}\right) -96\pi Pq_{M}^{2}>0, \\
\\
{m}^{2}c^{2}{c}_{2}-{\Phi }_{E}^{2}+k\pm \sqrt{{m}^{2}c^{2}{c}_{2}\left( 2k+{%
m}^{2}c^{2}{c}_{2}-2{\Phi }_{E}^{2}\right) +{\Phi
}_{E}^{4}+k\left( k-2{\Phi
}_{E}^{2}\right) -96\pi Pq_{M}^{2}}>0.%
\end{array}%
\end{equation*}

Returning to the obtained divergency, let's examine different
limiting cases. First of all, interestingly, the magnetic charge,
$q_{M}$ is coupled with pressure and the only trace of the
pressure in numerator is due to this term. In the absence of the
magnetic charge, divergencies will be only decreasing functions of
the pressure. This highlights the contribution and the effect of
the magnetic charge in thermodynamical phase transition of these
black holes. It is worthwhile to mention that magnetic charge term
is negative and its contribution is toward decreasing the value of
divergencies. On the other hand, for the electric part of
solutions, except for ${\Phi }_{E}^{4}$, all the other terms
including electric part contribute to negativity of the obtained
divergencies. In the absence of electric part, for positive
branches of the solutions, only magnetic charge term contributes
to negativity. Finally, the massive terms only contribute to
positivity of the divergencies. This shows that thermodynamical
phase transitions more likely take place in larger horizon radius
for massive graviton cases compared to massless ones. This is one
of the contributions of the massive gravity in thermodynamical
structure of the black holes. Later, we will present some diagrams
and give interpretations regarding the thermodynamical behavior of
black holes for mentioned divergencies, and hence, phase
transitions. It is worthwhile to mention that obtained
divergencies for the heat capacity are the same as extrema that
were extracted for temperature (\ref{rootT}).

\subsection{Free energy}

The free energy is considered to be one of the thermodynamical
potentials which could be used to derive other thermodynamical
quantities. The free energy could also be employed to extract
information regarding the phase transitions of the black holes and
their chemical equilibrium. Thermodynamically speaking, when a
system reaches chemical equilibrium at constant temperature and
pressure, its free energy minimizes. This means that the presence
of an extremum in free energy diagrams indicates that system
undergoes a second order phase transition in that extremum. This
point is known as equilibrium point and derivative of the free
energy with respect to thermodynamical coordinate vanishes there.

Using the obtained free energy (\ref{free1}), it is a matter of
calculation to show that roots of free energy are

\begin{equation}
r_{root-W}=\,\sqrt{3{\frac{k+{c}^{2}\,{m}^{2}c_{2}-{\Phi }_{E}^{2}\pm \sqrt{%
32\,P\pi \,q_{M}^{2}+{m}^{2}c^{2}{c}_{2}\left(
2k+{m}^{2}c^{2}{c}_{2}-2{\Phi }_{E}^{2}\right) +{\Phi
}_{E}^{4}+k\left( k-2{\Phi }_{E}^{2}\right) }}{16\pi \,P}}}.
\label{rootW}
\end{equation}

Obtained roots for the free energy are almost identical with
obtained divergencies for the heat capacity with a few
differences. Here, magnetic charge term coupled with pressure is
positive and therefore, its contribution is toward increasing the
values of root. Otherwise, same arguments that were mentioned for
divergencies of the heat capacity, regarding the effects of
different parameters, are valid for the roots of free energy as
well.

The equilibrium points for these black holes could be obtained by
using the first derivative of the free energy with respect to
thermodynamical coordinates. Here, we choose horizon radius as our
thermodynamical coordinate which leads to following equilibrium
points
\begin{equation}
r_{Equiblirium-W}=\,\sqrt{{\frac{{m}^{2}c^{2}{c}_{2}-{\Phi
}_{E}^{2}+k\pm
\sqrt{{m}^{2}c^{2}{c}_{2}\left( 2k+{m}^{2}c^{2}{c}_{2}-2{\Phi }%
_{E}^{2}\right) +{\Phi }_{E}^{4}+k\left( k-2{\Phi }_{E}^{2}\right)
-96\,P\pi \,q_{M}^{2}}}{16P\pi }}}.  \label{rcW}
\end{equation}

A simple comparison shows that obtained equilibrium points are
identical to extrema and divergencies which were extracted for the
temperature and the heat capacity, respectively. Now, since we
have two roots and two extrema for the free energy, one of these
cases may happen:

I) Existence of two roots and one extremum.

II) Presence of two roots and two extrema.

III) Absence of extremum and existence of only one root.

IV) Absence of extremum and root.

We will plot some diagrams to elaborate mentioned cases later. So,
we temporarily postpone the discussions here.

\section{van der Waals like behavior}

\label{vdW}

Replacing the cosmological constant with its corresponding
pressure provides the possibility of studying the critical
behavior of black holes in the context of van der Waals like
characteristic. Here, by replacing the cosmological constant in
the temperature, we will obtain the equation of the state in the
following form
\begin{equation}
P=\frac{1}{8\pi }\left[ \,{\frac{{\Phi }_{E}^{2}}{{r}_{+}^{2}}}+\,{\frac{%
q_{M}^{2}}{{r}_{+}^{4}}}-{\frac{k}{{r}_{+}^{2}}}+{\frac{4T}{r_{+}}}-{m}%
^{2}c\left( {\frac{\,{cc}_{2}}{{r}_{+}^{2}}}-{\frac{{c}_{1}}{r_{+}}}\right) %
\right] .  \label{Pr}
\end{equation}

As it was pointed out earlier, the total volume of the black hole,
conjugate to pressure, is obtainable in terms of the horizon
radius. Therefore, it is possible to obtain the equation of state
as
\begin{equation}
P={\frac{{\Phi }_{E}^{2}}{12}}\sqrt[3]{\frac{6}{\pi V^{2}}}+{\frac{T}{6}}%
\sqrt[3]{\frac{36\pi }{V}}-{\frac{k}{12}}\sqrt[3]{\frac{6}{\pi V^{2}}}-\,{%
\frac{{m}^{2}c\left( 2\sqrt[3]{\pi }{cc}_{2}+\sqrt[3]{6V}{c}_{1}\right) }{{24%
}}}\sqrt[3]{\frac{6}{\pi ^{2}V^{2}}}+\,{\frac{q_{M}^{2}}{18}}\sqrt[3]{\frac{%
36\pi }{V^{4}}}.  \label{PV1}
\end{equation}

Before we study the possible existence of the van der Waals like
critical behavior, let us examine the effects of different
parameters on the pressure.

Evidently, the massive terms have negative effect on the value of
the pressure. This indicates that generalization from massless to
massive gravitons leads to decrease the thermodynamical pressure
of black holes. Contrary to massive terms, electric and magnetic
parts of the solutions have positive contributions into values of
the pressure. Especially for small values of horizon radius, the
contribution of magnetic charge term becomes highly significant.

As for topological term ($k-$term), its contribution depends on
the choices of topology. For spherical case, it will decrease the
value of the pressure while the opposite is observed for
hyperbolic horizon. Regarding $T-$term, we should note that since
positive values of the temperature are valid for the black holes,
its contribution is always positive.

It is worthwhile to mention that due to specific coupling of
different factors of horizon radius with different terms, by
suitable choices, it is possible to cancel the effects of some of
terms. In other words, if the following equations hold
simultaneously,
\begin{eqnarray*}
4\pi T-m^{2}cc_{1} &=&0, \\
&& \\
{\Phi }_{E}^{2}-m^{2}c^{2}c_{1}-k &=&0,
\end{eqnarray*}%
the pressure will be only a function of the magnetic charge.

In order to study the critical behavior of the system, we use
specific volume which is related to horizon radius, $r_{+}=v/2$.
Therefore, hereafter we work with the following equation of state
\begin{equation}
P=\frac{1}{2\pi }\left[ \,{\frac{{\Phi }_{E}^{2}}{{v}^{2}}}+\,{\frac{%
4q_{M}^{2}}{{v}^{4}}}-{\frac{k}{{v}^{2}}}+{\frac{2T}{v}}-{m}^{2}c\left( {%
\frac{\,{cc}_{2}}{{v}^{2}}}-{\frac{{c}_{1}}{2v}}\right) \right] .
\label{Pv}
\end{equation}

We are in now a position to study possible phase transition and
van der Waals like behavior for these black holes.

\subsection{Van der Waals like phase transition}

The first step for studying van der Waals like behavior of the
black holes
is obtaining their critical points. To do so, we employ the properties of $%
P-v$ diagrams which is known as inflection point with the
following properties
\begin{equation}
\left( \frac{\partial P}{\partial v}\right) _{T}=\left( \frac{\partial ^{2}P%
}{\partial v^{2}}\right) _{T}=0.  \label{infel}
\end{equation}

Using obtained equation of the state (\ref{Pv}) with properties of
inflection point (\ref{infel}), one can obtain the following
relation for calculating critical volume
\begin{equation}
\left( {m}^{2}\,{c}^{2}c_{2}-{\Phi }_{E}^{2}+k\right)
{v}^{2}-24q_{M}^{2}=0. \label{critEq}
\end{equation}

It is a matter of calculation to obtain critical volume as

\begin{equation}
v_{c}=\,{\frac{12q_{M}}{\sqrt{6{m}^{2}\,{c}^{2}c_{2}-6{\Phi
}_{E}^{2}+6k}}.} \label{vc}
\end{equation}

By employing calculated critical volume, one can find the critical
values of temperature, pressure and free energy which take the
following forms
\begin{equation}
T_{c}=\frac{{m}^{2}\,cc_{1}}{4\pi }+\frac{\sqrt{6}\left( {m}^{2}\,{c}%
^{2}c_{2}-\,{\Phi_{E} }^{2}+\,k\right) ^{\frac{3}{2}}}{12\pi q_{M}}-{\frac{%
\sqrt{6}\,\left( {m}^{2}\,{c}^{2}c_{2}-\,{\Phi_{E} }^{2}+\,k\right) ^{\frac{3%
}{2}}}{216\pi q_{M}},}  \label{Tc}
\end{equation}
\begin{equation}
P_{c}=\,{\frac{\left( {m}^{2}\,{c}^{2}c_{2}-{\Phi }_{E}^{2}+k\right) ^{2}}{%
96\pi q_{M}^{2}}}  \label{Pc}
\end{equation}%
\begin{equation}
W_{c}=\,\frac{{2q_{M}}}{\sqrt{6}}{\sqrt{\left( {m}^{2}\,{c}^{2}c_{2}-{\Phi }%
_{E}^{2}+k\right) }.}  \label{Wc}
\end{equation}

In addition, the universal ratio of $\frac{P_{c}v_{c}}{T_{c}}$ for
these black holes will be obtained as
\begin{equation}
\frac{P_{c}v_{c}}{T_{c}}=\,{\frac{\frac{3}{4}\sqrt{6}}{2\sqrt{6}+9\,{m}%
^{2}cc_{1}q_{M}^{2}\left( {m}^{2}\,{c}^{2}c_{2}-{\Phi
}_{E}^{2}+k\right) ^{-3/2}}},  \label{PcvcTc}
\end{equation}%
which is modified in the presence of massive term. In the absence
of massive
terms, Eq. (\ref{PcvcTc}) reduces to the universal ration of Reissner Nordstr%
\"{o}m black holes \cite{kubuiznak}.

Before we proceed, it is necessary to study the obtained critical
values. The critical volume is an increasing function of the
magnetic charge and electric part of the solutions while it is a
decreasing function of the massive parameter. The effect of the
topological term is depending on the choice of topology. For
spherical horizon, critical horizon radius is a decreasing
function of the topological term while the opposite is observed
for hyperbolic horizon. As one can see, generalization to massive
gravity has interesting effects on critical behavior of the
system. The most important contribution is that critical behavior
could be observed for non-spherical horizons as well. Previously,
it was shown that van der Waals like behavior is only observable
for spherical black holes whereas black holes with other type of
horizons suffer the absence of such behavior in their phase
diagrams. Here, the generalization from massless gravitons to
massive ones, provides the possibility of van der Waals like
behavior for black holes with different horizon. In other words,
the topological restriction for having van der Waals like behavior
for black holes is relaxed. Here, the existence of real valued
critical volume is restricted by the following condition
\begin{equation*}
{m}^{2}\,{c}^{2}c_{2}-{\Phi }_{E}^{2}+k>0.
\end{equation*}

In other words, it is possible to eliminate critical behavior for
these black holes with suitable choices of different parameters.
Another interesting fact is that critical volume does not depend
on the variations of $c_{1}$. In other words, generalization to
massive gravity has no effect on the critical volume. Finally, it
is worth mentioning again that critical horizon radius is a
decreasing function of the massive gravity. These are the effects
of massive parameter on the critical volume.

As for the critical temperature, contrary to the critical volume,
it is a decreasing function of the magnetic charge and electric
part while it is an increasing function of the massive parameter.
The effects of different topologies are also opposite to what were
observed for critical volume. Interestingly, the critical
temperature is $c_{1}$ dependent and it is an increasing function
of this parameter. The existence of positive critical temperature
is restricted by the following condition
\begin{equation*}
{m}^{2}\,cc_{1}v_{c}^{3}+\left( 4\,{m}^{2}\,{c}^{2}c_{2}-4\,{\Phi }%
_{E}^{2}+4\,k\right) v_{c}^{2}-32\,q_{M}^{2}>0,
\end{equation*}%
which shows that it is possible to have positive critical volume
with negative critical temperature (by suitable choice of magnetic
charge).

Due to specific structure of the critical pressure, it is always
positive. The critical pressure is also independent of variation
of the $c_{1}$ parameter which could be employed to make previous
comments for this critical value as well. In addition, here too,
critical pressure is a decreasing function of the magnetic charge
and $\Phi_{E}$, while it is an increasing function of the massive
parameter and topological factor.

Finally, obtained critical free energy shows that its value is
always positive (if previous condition for having real valued
critical horizon radius is satisfied). In addition, it is an
increasing function of the massive parameter, magnetic charge and
topological factor while it is a decreasing function of the
electric charge.

\begin{figure}[tbp]
$%
\begin{array}{ccc}
\epsfxsize=5.5cm \epsffile{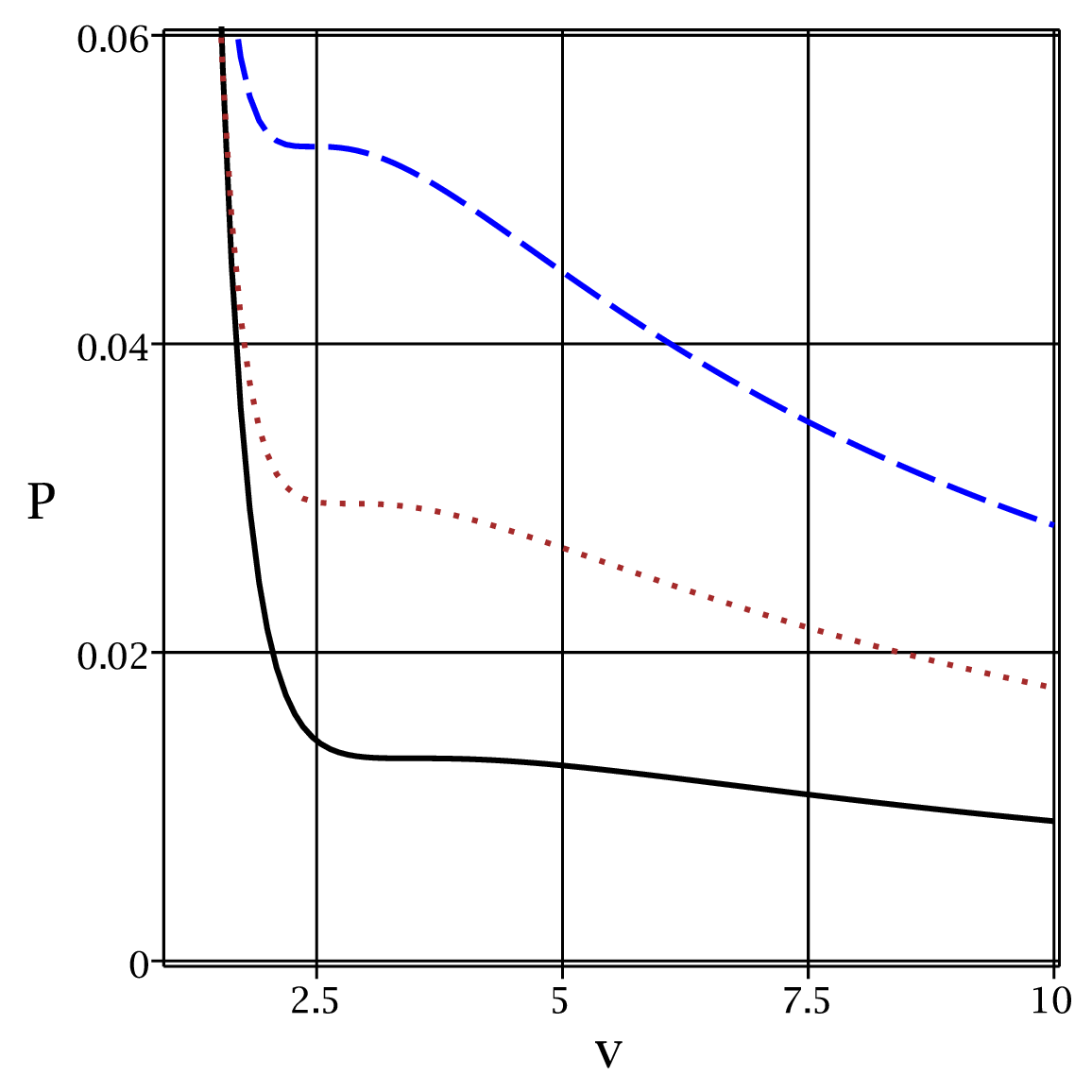} & \epsfxsize=5.5cm %
\epsffile{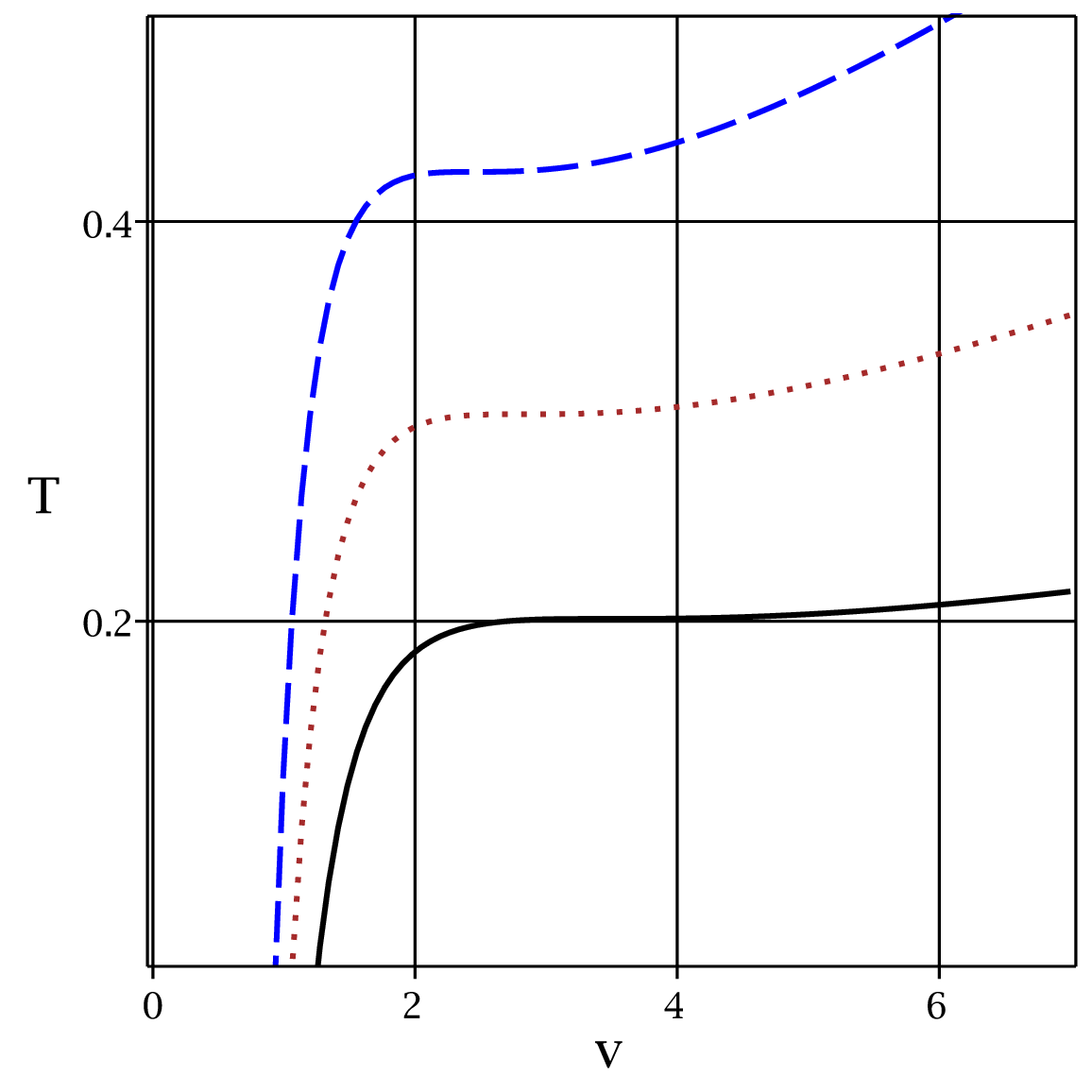} & \epsfxsize=5.5cm \epsffile{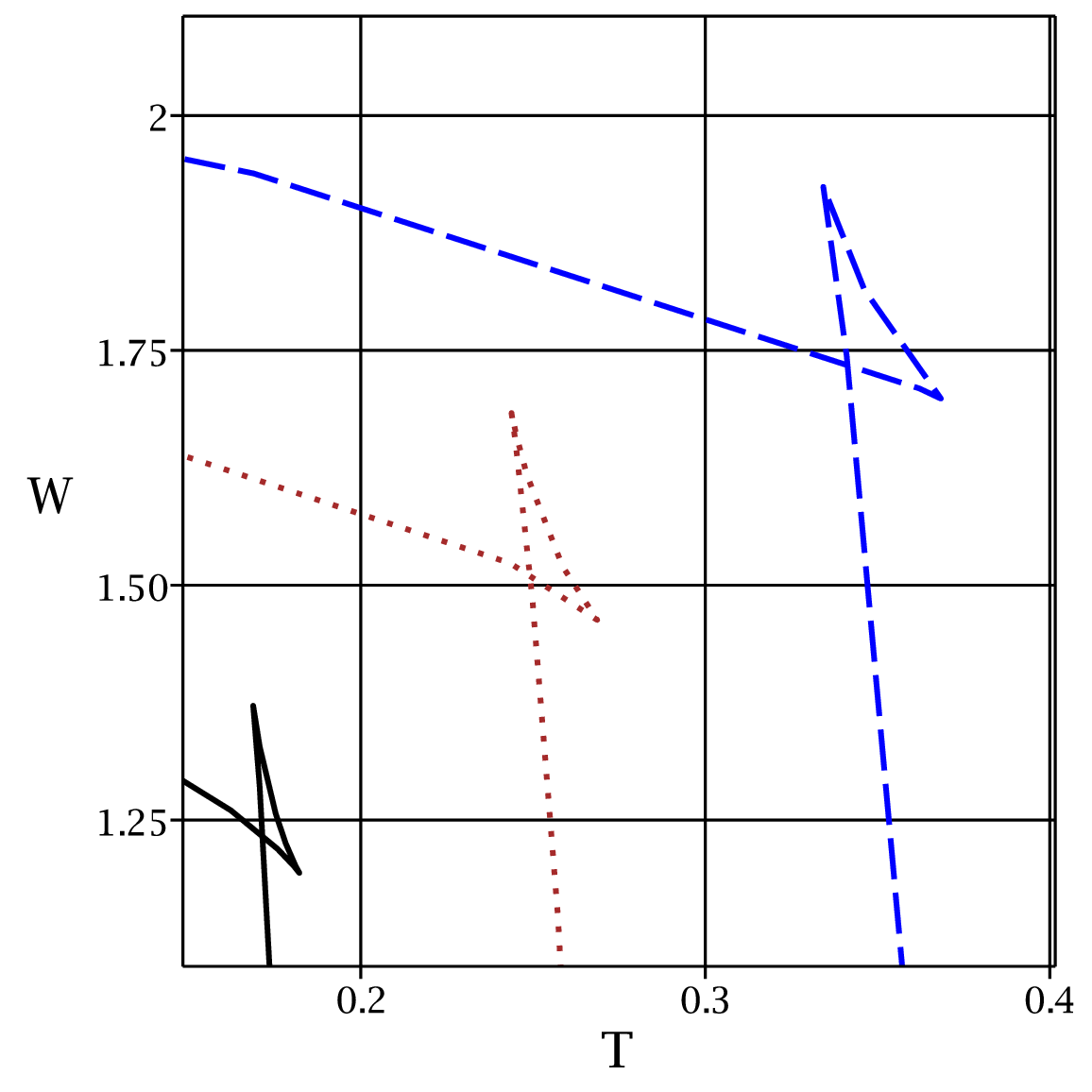}%
\end{array}
$%
\caption{ $P-v$ (left), $T-v$ (middle) and $W-T$ (right) diagrams for $m=0.5$%
, $c=c_{1}=2$, $c_{2}=3$, $\Phi _{E}=0.1$ and $q_{M}=0.1$; $k=-1$
(continuous line), $k=0$ (dotted line) and $k=1$ (dashed line).
\newline $P-v$ diagram for $T=T_{c}$, $T-v$ diagram for $P=P_{c} $
and $W-T$ diagram for $P=0.5P_{c}$. } \label{Fig3}
\end{figure}

To elaborate the existence of van der Waals like behavior for
different topological black holes, we plot phase diagrams for
obtained critical values (see Fig. \ref{Fig3}).

Formation of swallow-tail shape for $W-T$ (right panel of Fig.
\ref{Fig3}), existence of inflection point for $P-v$ (left panel
of Fig. \ref{Fig3}) and
presence of subcritical isobars in $T-v$ diagrams (middle panel of Fig. \ref%
{Fig3}), indicate that these black holes with different
topological structures, enjoy second order phase transition and
van der Waals like behavior in their phase diagrams. This is a new
achievement which is seen in massive gravity context
\cite{PRLwithMann}.

\subsection{Limiting cases and van der Waals like behavior for non-spherical
black holes}

In previous studies regarding the dyonic black holes, it was
pointed out that planar black holes suffer the absence of van der
Waals like phase transition in their phase diagrams. Here, in
previous section, it was shown that generalization to massive
gravity solves this problem and introduces van der Waals like
phase transition into phase structure of the black holes with
hyperbolic and flat horizons. This possibility opens up new
avenues in studying different aspects of black holes with
application in gauge/gravity duality. In the AdS/CFT
correspondence, $l$ is interpreted as a measure of the number of
degrees of freedom, $N$, of the boundary field theory and its
relation is determined by the CFTs under consideration. That being
said, some authors consider that varying in pressure, hence
$\Lambda$, corresponds to vary the number of colors, $N$, in the
boundary Yang-Mills theory and the volume conjugating to pressure
in the boundary field theory is equivalent to an associated
chemical potential for color \cite{heat2,C2,C3}. Another approach
considers the variation in $\Lambda$ corresponds to variation of
the volume on which the field theory resides \cite{C4}. Some of
the applications of extended phase space and Van der Waals like
behavior in AdS/CFT are in the context of; Holographic
superconductors \cite{C5}, Holographic entanglement entropy
\cite{C6} and etc. Later, we will show that different approaches
in studying critical behavior of the black holes yield consistent
results. Here, the effects of the variation of the magnetic
charge, $q_{M}$, on van der Waals like behavior of these black
holes in fixed $\Phi_{E}$ ensemble are studied.

\begin{figure}[tbp]
$%
\begin{array}{ccc}
\epsfxsize=5.5cm \epsffile{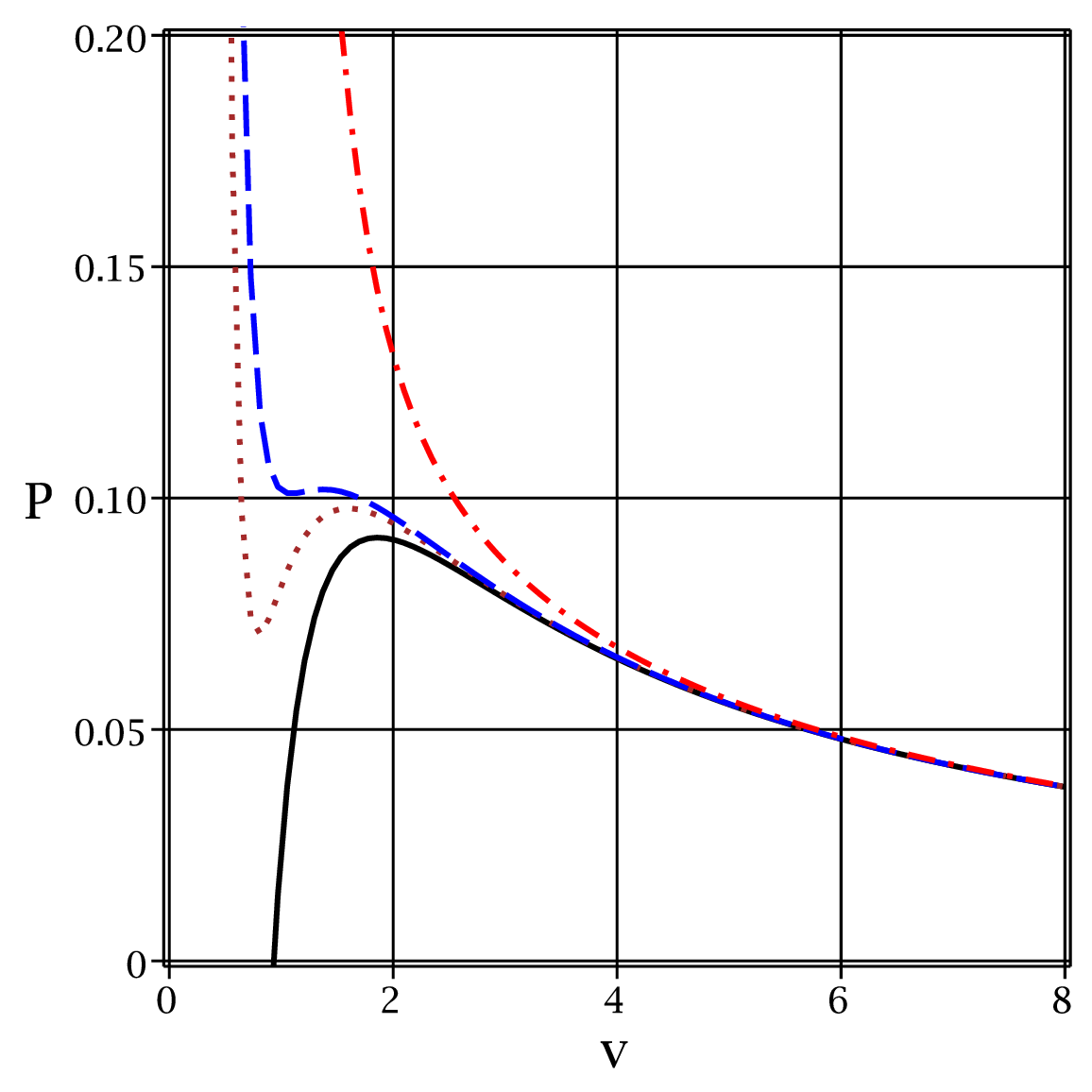} & \epsfxsize=5.5cm %
\epsffile{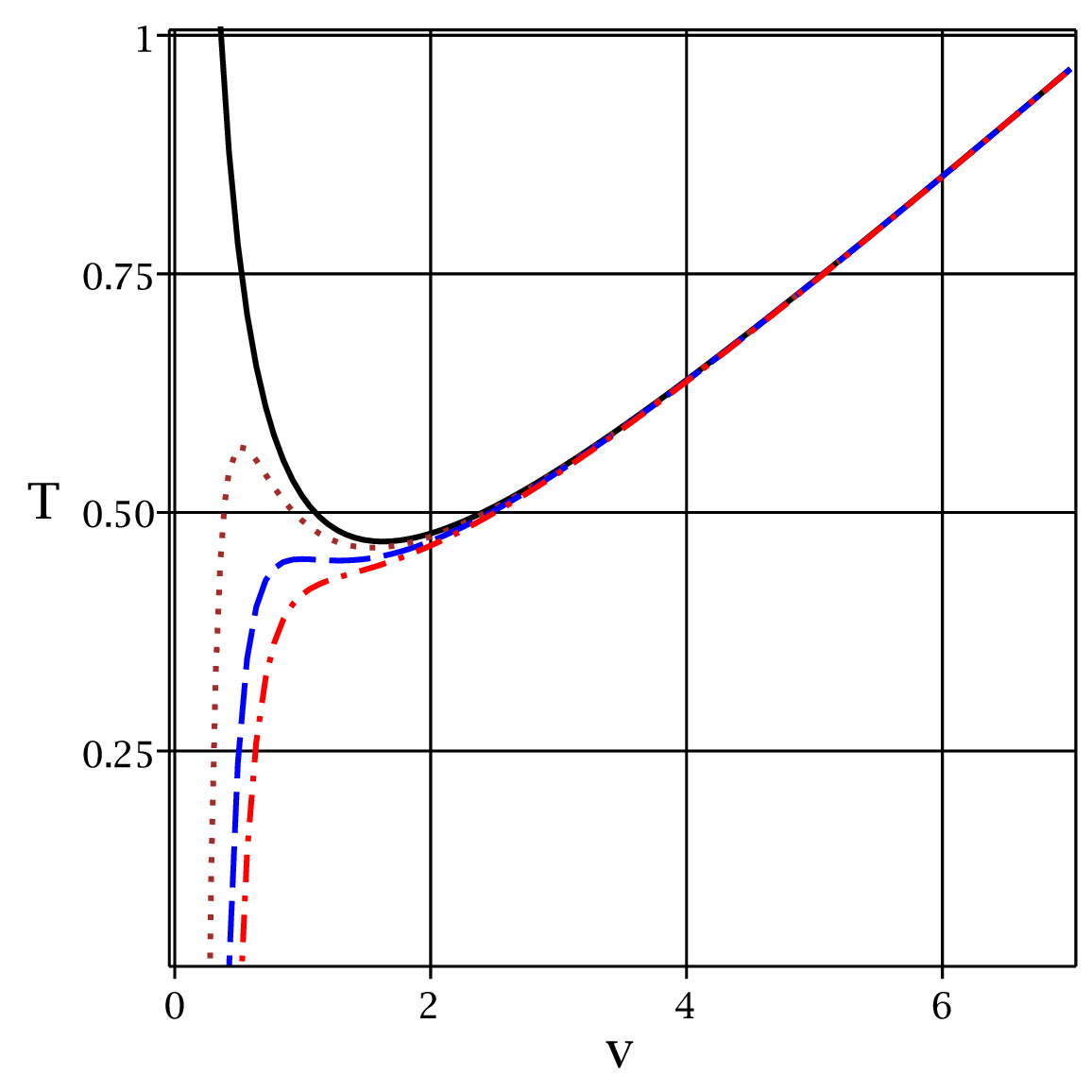} & \epsfxsize=5.5cm \epsffile{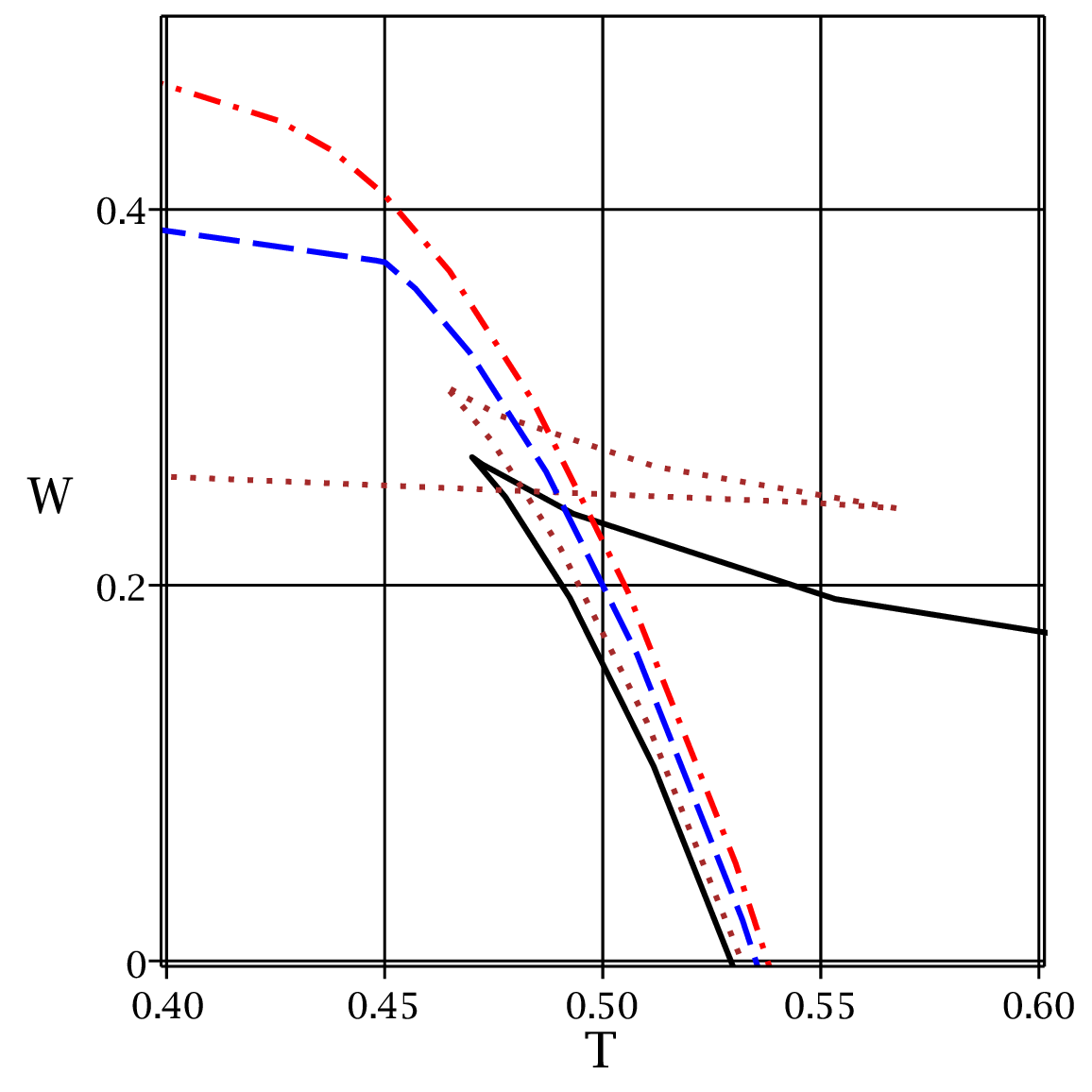}%
\end{array}
$%
\caption{ $P-v$ (left), $T-v$ (middle) and $W-T$ (right) diagrams for $m=0.5$%
, $c=c_{1}=2$, $c_{2}=3$, $\Phi _{E}=0.1$ and $k=-1$; \newline For
$T-v$ and $W-T$ panels: P=0.12, $q_{M}=0$ (continuous line),
$q_{M}=0.2$ (dotted line), $q_{M}=0.32$ (dashed line) and
$q_{M}=0.4$ (dashed-dotted line). \newline For $P-v$ panel:
T=0.42, $q_{M}=0$ (continuous line), $q_{M}=0.3$ (dotted line),
$q_{M}=0.35$ (dashed line) and $q_{M}=1$ (dashed-dotted line). }
\label{hyperbolic}
\end{figure}

\begin{figure}[tbp]
$%
\begin{array}{ccc}
\epsfxsize=5.5cm \epsffile{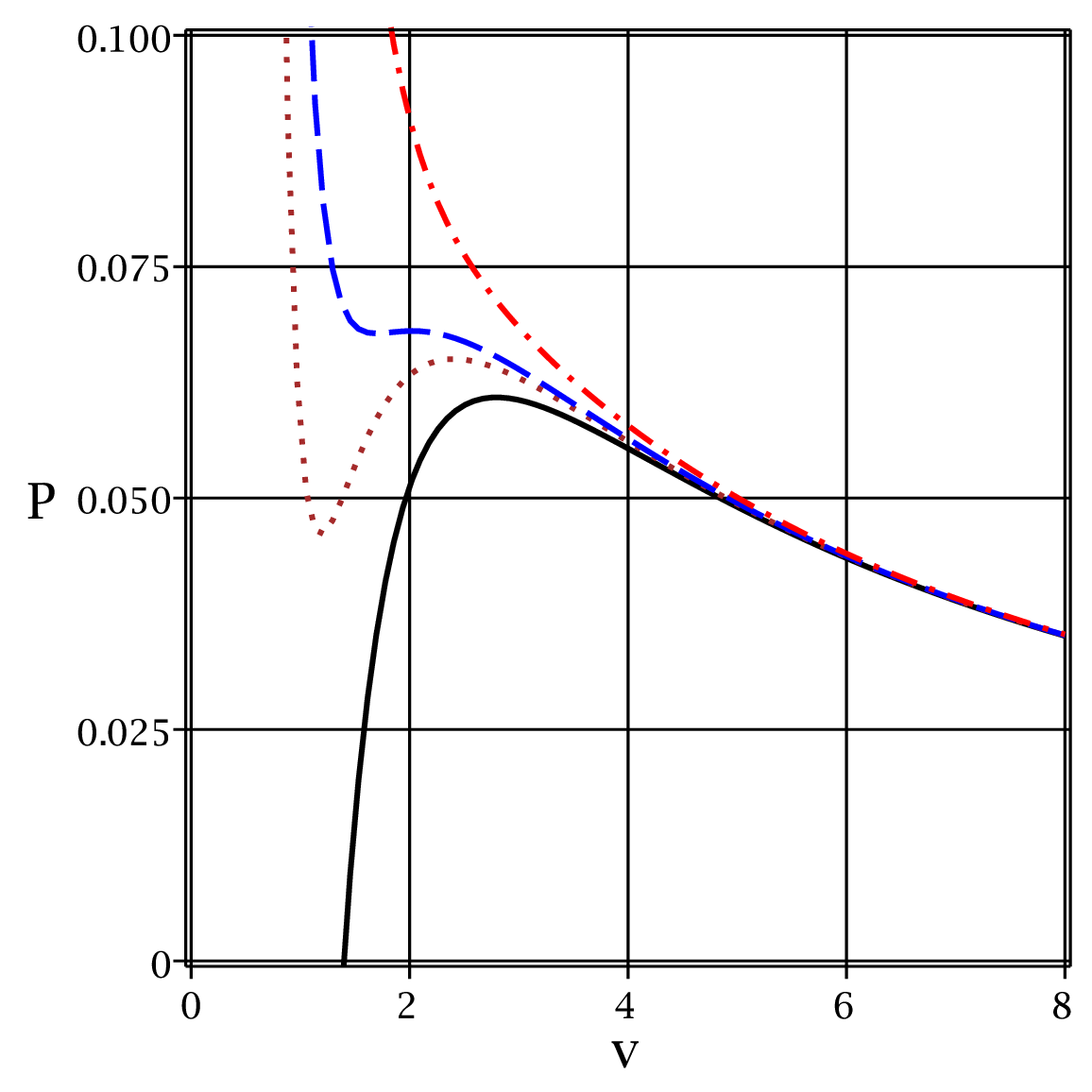} & \epsfxsize=5.5cm %
\epsffile{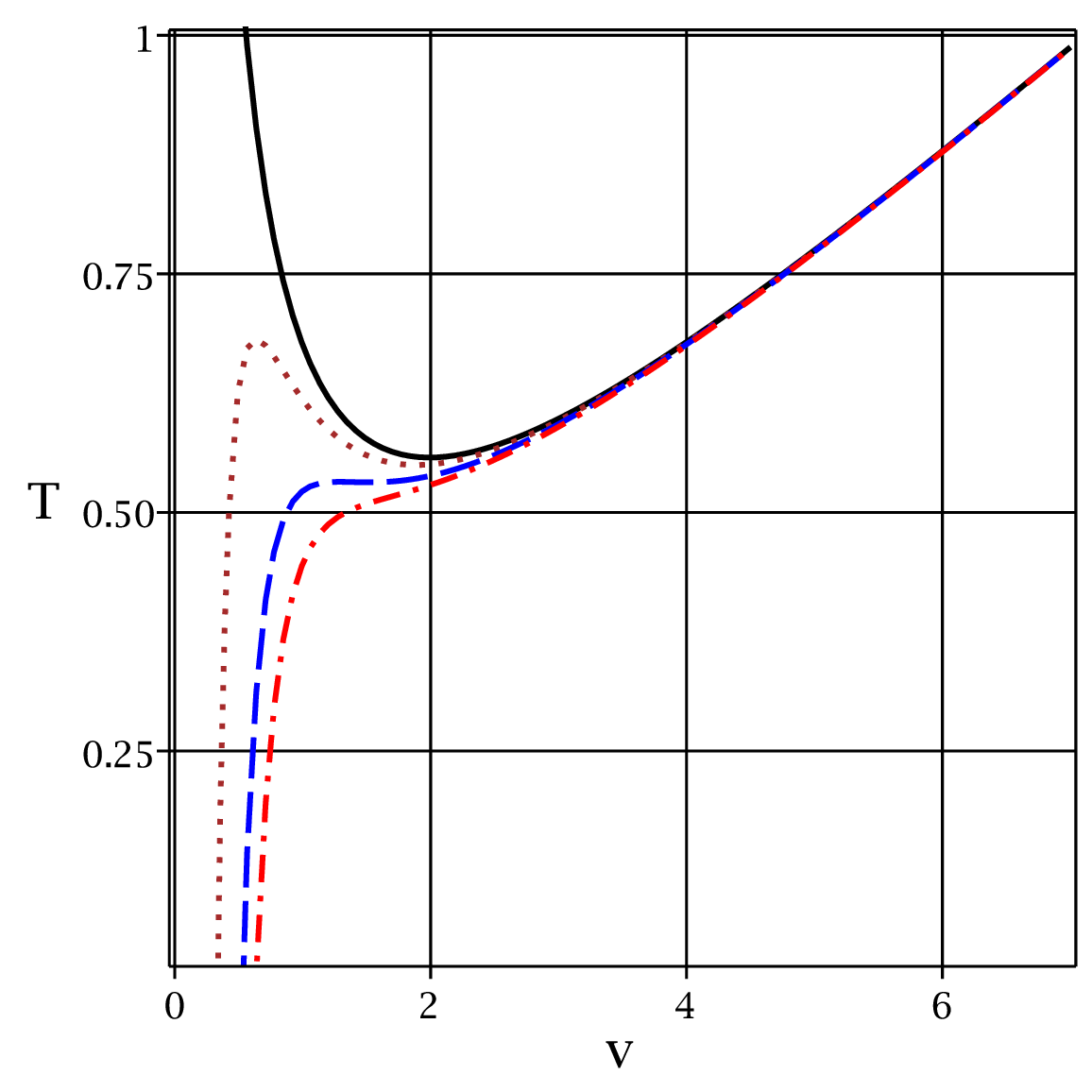} & \epsfxsize=5.5cm \epsffile{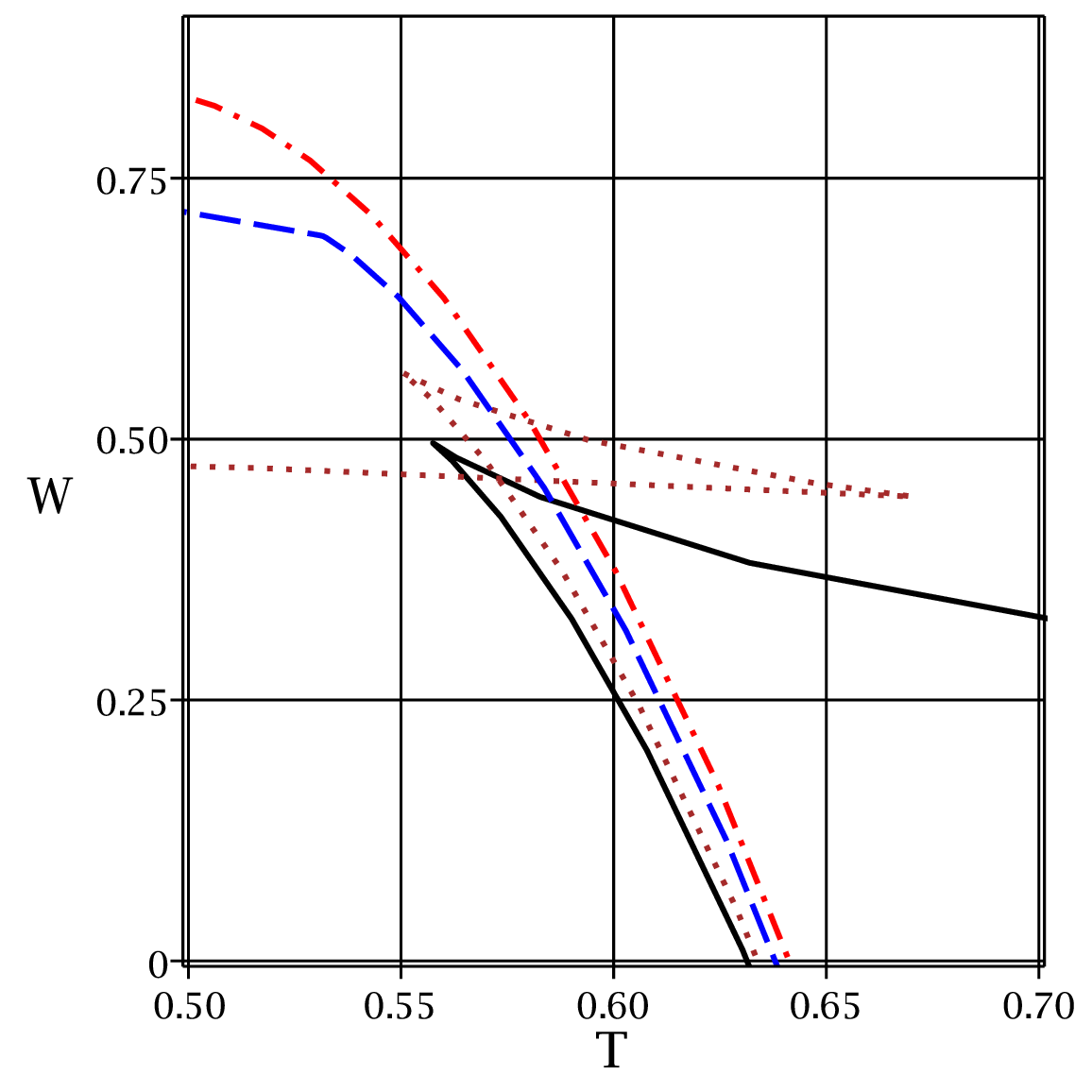}%
\end{array}
$%
\caption{ $P-v$ (left), $T-v$ (middle) and $W-T$ (right) diagrams for $m=0.5$%
, $c=c_{1}=2$, $c_{2}=3$, $\Phi _{E}=0.1$ and $k=0$; \newline For
$T-v$ and $W-T$ panels: P=0.12, $q_{M}=0$ (continuous line),
$q_{M}=0.3$ (dotted line), $q_{M}=0.49$ (dashed line) and
$q_{M}=0.6$ (dashed-dotted line). \newline For $P-v$ panel:
T=0.42, $q_{M}=0$ (continuous line), $q_{M}=0.55$ (dotted line),
$q_{M}=0.65$ (dashed line) and $q_{M}=1$ (dashed-dotted line). }
\label{flat}
\end{figure}

\begin{figure}[tbp]
$%
\begin{array}{ccc}
\epsfxsize=5.5cm \epsffile{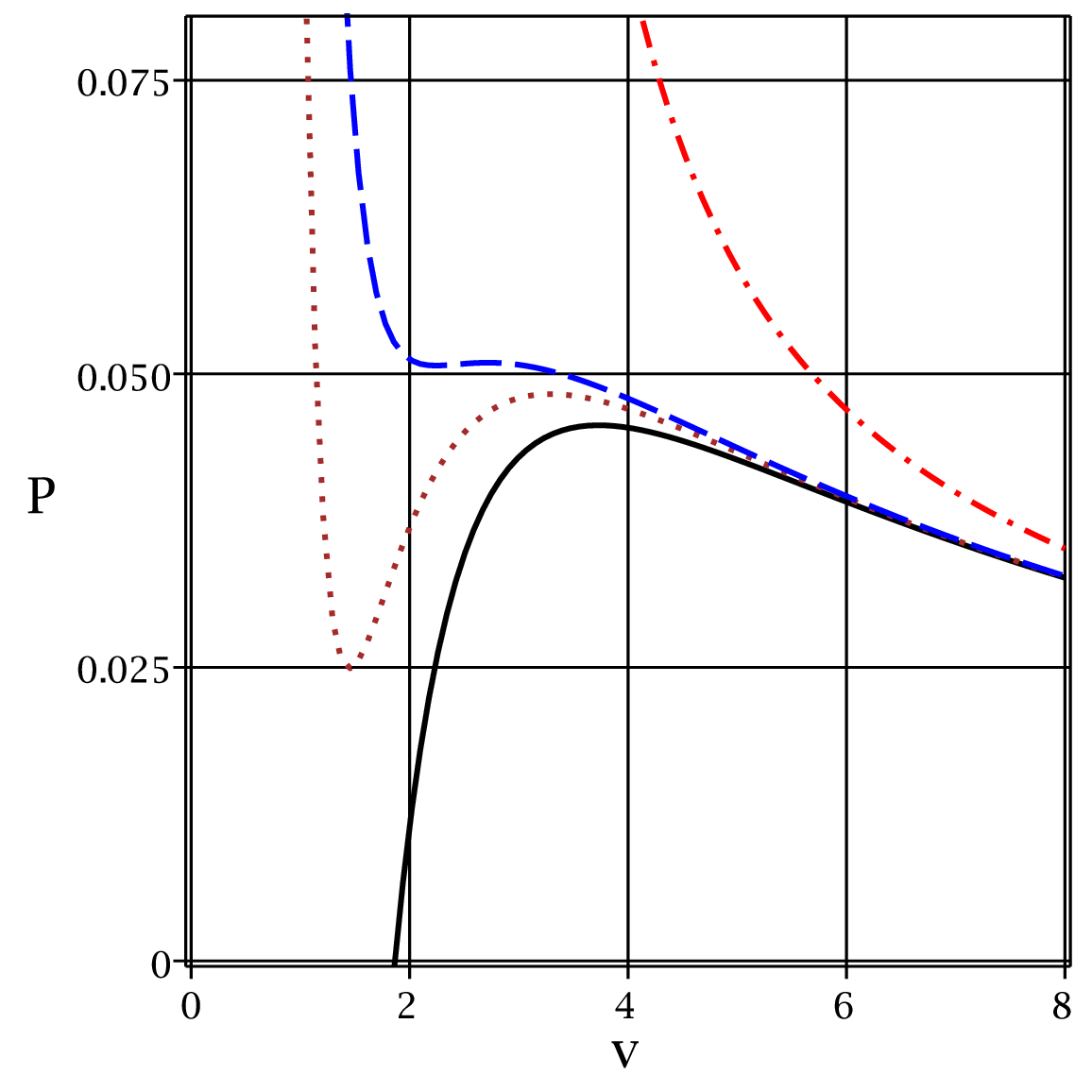} & \epsfxsize=5.5cm %
\epsffile{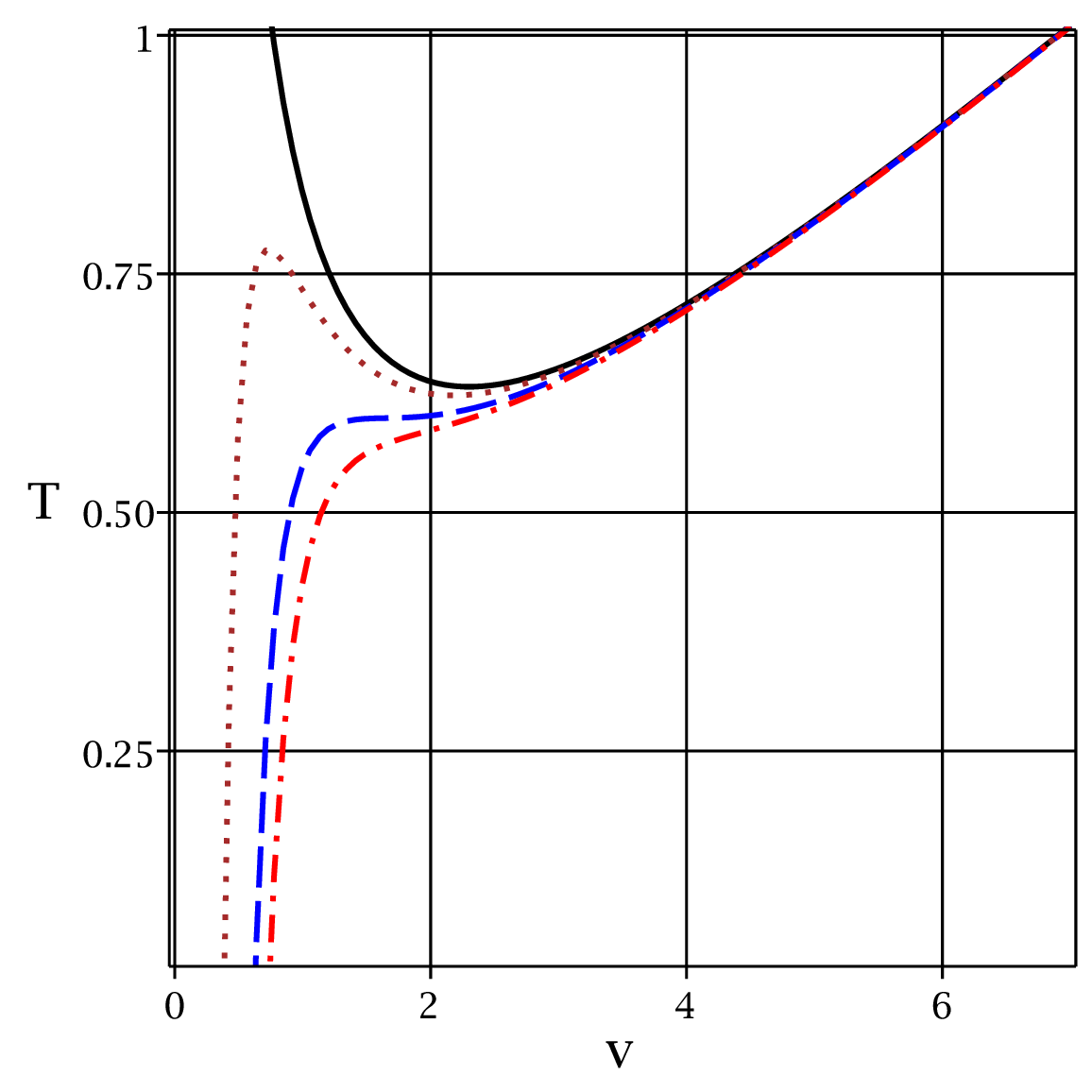} & \epsfxsize=5.5cm \epsffile{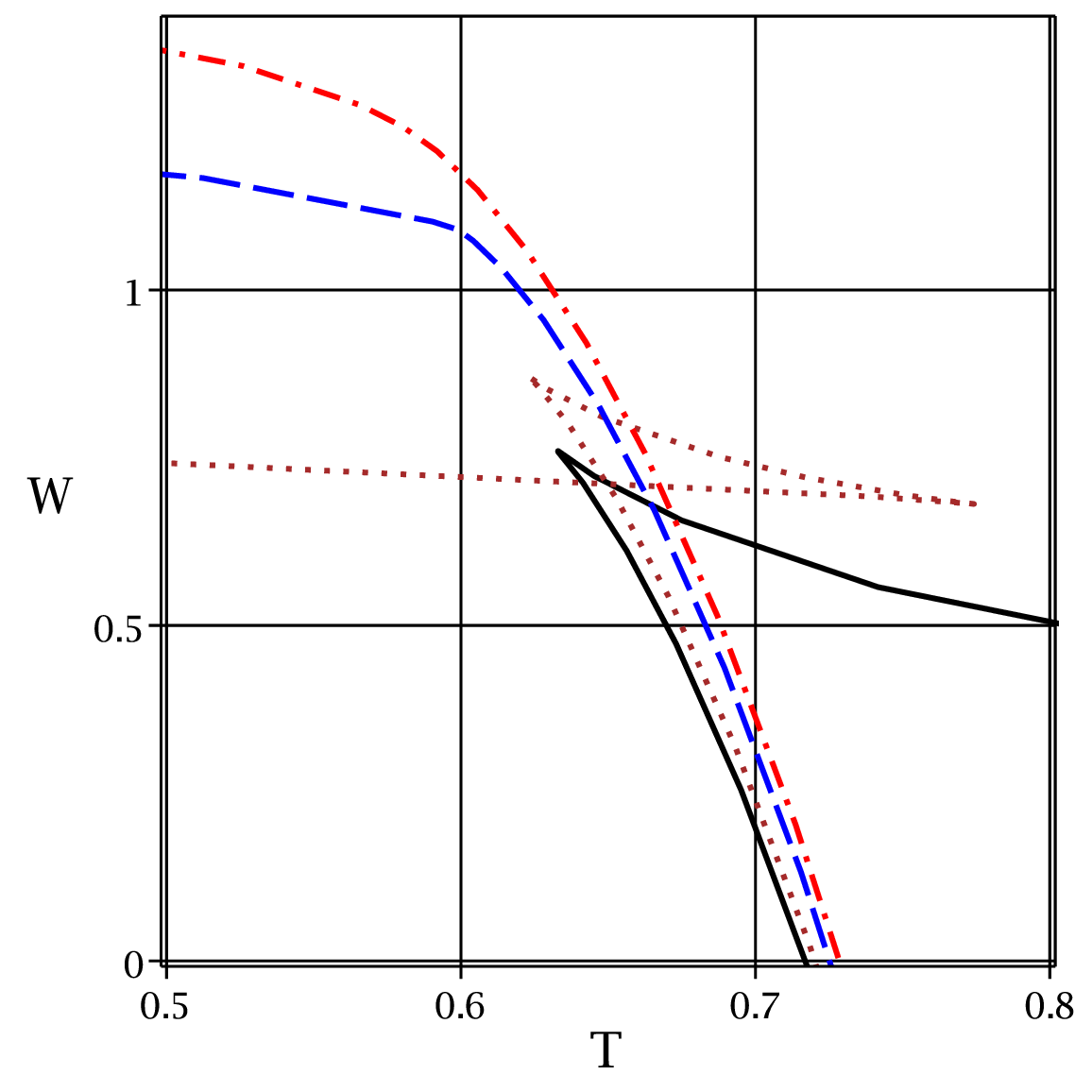}%
\end{array}
$%
\caption{ $P-v$ (left), $T-v$ (middle) and $W-T$ (right) diagrams for $m=0.5$%
, $c=c_{1}=2$, $c_{2}=3$, $\Phi _{E}=0.1$ and $k=1$; \newline For
$T-v$ and $W-T$ panels: P=0.12, $q_{M}=0$ (continuous line),
$q_{M}=0.4$ (dotted line), $q_{M}=0.67$ (dashed line) and
$q_{M}=0.8$ (dashed-dotted line). \newline For $P-v$ panel:
T=0.42, $q_{M}=0$ (continuous line), $q_{M}=0.8$ (dotted line),
$q_{M}=1$ (dashed line) and $q_{M}=4$ (dashed-dotted line). }
\label{spherical}
\end{figure}

We have plotted three sets of figures for hyperbolic (Fig. \ref{hyperbolic}%
), flat (Fig. \ref{flat}) and spherical (Fig. \ref{spherical})
cases. These figures contain $P-v$ (left panels), $T-v$ (middle
panels) and $W-T$ (right panels) diagrams. In these diagrams,
electric part, pressure and massive parameter are considered to be
fixed and variation of the magnetic charge is studied.


First, we take $W-T$ diagrams into account. Evidently, for large
values of electromagnetic charge, no phase transition exists and
only one phase is
observable for black holes (left panels of Figs. \ref{hyperbolic}-\ref%
{spherical}, dashed-dotted lines). The formation of two phases and
van der Waals like behavior could be observed by decreasing
magnetic charge till it
meets a critical magnetic charge, $q_{M_{c}}$ (left panels of Figs. \ref%
{hyperbolic}-\ref{spherical}, dashed lines). Here, a phase
transition of small/large black holes takes place. For magnetic
charges smaller than critical magnetic charge, an unstable phase
is formed. Therefore, in this case, three phases exist which are
small, medium and large black holes. This could be observed in
$W-T$ diagrams as formation of the swallow-tail (left panels of
Figs. \ref{hyperbolic}-\ref{spherical}, dotted lines). Finally,
for the limit $q_{M}\longrightarrow 0$, all the three phases
reduces to two phases for black holes (left panels of Figs. \ref{hyperbolic}-%
\ref{spherical}, continuous lines) in which the smaller black
holes are unstable while the larger ones are stable.

Now, we study $T-v$ diagrams. Evidently, here too, for the
magnetic charges larger than the critical magnetic charge, the
temperature is only an increasing function of the volume without
any subcritical isobar and extremum. Here, only one stable phase
is possible for black holes (middle panels of Figs.
\ref{hyperbolic}-\ref{spherical}, dashed-dotted lines). For
critical magnetic charge, an extremum is observed and a
subcritical isobar is formed (middle panels of Figs.
\ref{hyperbolic}-\ref{spherical}, dashed
lines). The extremum separates small and large phases from each other. For $%
0<q_{M}<q_{M_{c}}$, two extrema are formed for $T-v$ diagrams. The
unstable phase of medium black holes is located between these two
extrema and small and large phases are located before maximum and
after minimum, respectively (middle panels of Figs.
\ref{hyperbolic}-\ref{spherical}, dotted lines). Interestingly,
for $q_{M}\longrightarrow 0$, the behavior of the temperature is
modified. Here, a minimum is observable for the temperature in
which the
phase before this minimum is the unstable one (middle panels of Figs. \ref%
{hyperbolic}-\ref{spherical}, continuous lines). At the minimum a
phase transition to larger stable black holes takes place.

For $P-v$ diagrams, in case of $q_{M}>q_{M_{c}}$, the pressure is
only a decreasing function of the volume without any extremum
which indicates that
only a single stable phase exists (left panels of Figs. \ref{hyperbolic}-\ref%
{spherical}, dashed-dotted lines). For $q_{M}=q_{M_{c}}$,
inflection point is formed and two phases of small and large
stable black holes are separated (left panels of Figs.
\ref{hyperbolic}-\ref{spherical}, dashed lines). The case
$0<q_{M}<q_{M_{c}}$ leads to appearance of two extrema where the
unstable medium black hole phase is located between them (left
panels of Figs. \ref{hyperbolic}-\ref{spherical}, dotted lines).
Finally, in the absence of magnetic charge, the thermodynamical
behavior of the pressure is modified and a maximum is formed. The
maximum marks the phase transition between unstable small black
holes and stable large black holes.

Here, we see that modifications in thermodynamical structure of
the black holes for non-spherical black holes are similar to those
that are observed for spherical ones. This similarity shows that
non-spherical black holes also enjoy van der Waals like behavior
in their phase diagrams. It is worthwhile to mention that critical
values and the places of mentioned modification depends on choices
of the topology. Later, we will present more details regarding
different phases in various comparative diagrams.

\subsection{Alternative method for studying van der Waals like behavior}

Instead of using the equation of state and properties of
inflection point for studying critical behavior of the black
holes, it is possible to use another method which uses the slope
of temperature versus entropy. This method was first introduced in
Ref. \cite{int} and employed in several other papers. It was shown
that the results of this method are consistent with other methods.
Here we examine the validity of this method in dyonic massive
gravity.

The method employs the slope of temperature versus entropy to
derive a new relation for the pressure which is independent of the
equation of state. If the new relation admits a maximum, the black
holes enjoy second order phase transition and van der Waals like
behavior in their phase space. The phase transition point is
located at the maximum and its pressure corresponds to the
critical pressure.

For these black holes, the new relation for pressure by using the
heat capacity (\ref{CQQ}) is
\begin{equation}
P_{New}={\frac{\left( \,{m}^{2}\,{c}^{2}c_{2}-4\,{\Phi
}_{E}^{2}+4k\right) v_{c}^{2}-12\,q_{M}^{2}}{2\pi {v}^{4}}.}
\label{Pnew}
\end{equation}

We should note that the subscript \emph{"NEW"}, simply, stands for
the new method that we have used to obtain pressure and
temperature. The maximum pressure and its corresponding volume are
obtained, respectively, in following forms
\begin{equation}
v_{Maximum}={\frac{12q_{M}}{\sqrt{6{m}^{2}\,{c}^{2}c_{2}-6{\Phi }_{E}^{2}+6k}%
},}  \label{vMax}
\end{equation}%
\begin{equation}
P_{Maximum}={\frac{\left( {m}^{2}\,{c}^{2}c_{2}-{\Phi }_{E}^{2}+k\right) ^{2}%
}{96\pi q_{M}^{2}}},  \label{PMax}
\end{equation}%
which are exactly critical horizon volume and pressure that were
extracted in previous section (see Eqs. (\ref{vc}) and
(\ref{Pc})). This shows that the results of this method are
consistent with those derived in previous section and this method
provides satisfactory results. Now, replacing pressure in the
temperature (\ref{TotalTT}) with Eq. (\ref{Pnew}), one can find a
relation independent of pressure for the temperature
\begin{equation}
T_{New}={\frac{m^{2}cc_{1}{v}^{3}+4\left( {m}^{2}\,{c}^{2}c_{2}-{\Phi }%
_{E}^{2}+k\right) {v}^{2}-32\,q_{M}^{2}}{4\pi {v}^{3}}}.
\label{Tnew}
\end{equation}

Using manual calculations, it
can be shown that the new relations for the temperature and
pressure coincide with usual $T$ and $P$, which is in
agreement with the existence of a single temperature (and also
pressure in an extended phase space) in the dual CFT. Replacing
volume with obtained maximum volume, one can find the following
(maximum) temperature which is essentially the critical
temperature
\begin{equation}
T_{Maximum}={\frac{27m^{2}cc_{1}q_{M}\,+\left(
6{m}^{2}\,{c}^{2}c_{2}-6{\Phi }_{E}^{2}+6k\right)
^{\frac{3}{2}}}{108\pi q_{M}}}.  \label{TMax}
\end{equation}

Numerical evaluation shows that this temperature and critical
temperature which was obtained in previous section are the same
and therefore, we are able to extract critical temperature using
this method.

\begin{figure}[tbp]
$%
\begin{array}{cc}
\epsfxsize=5.5cm \epsffile{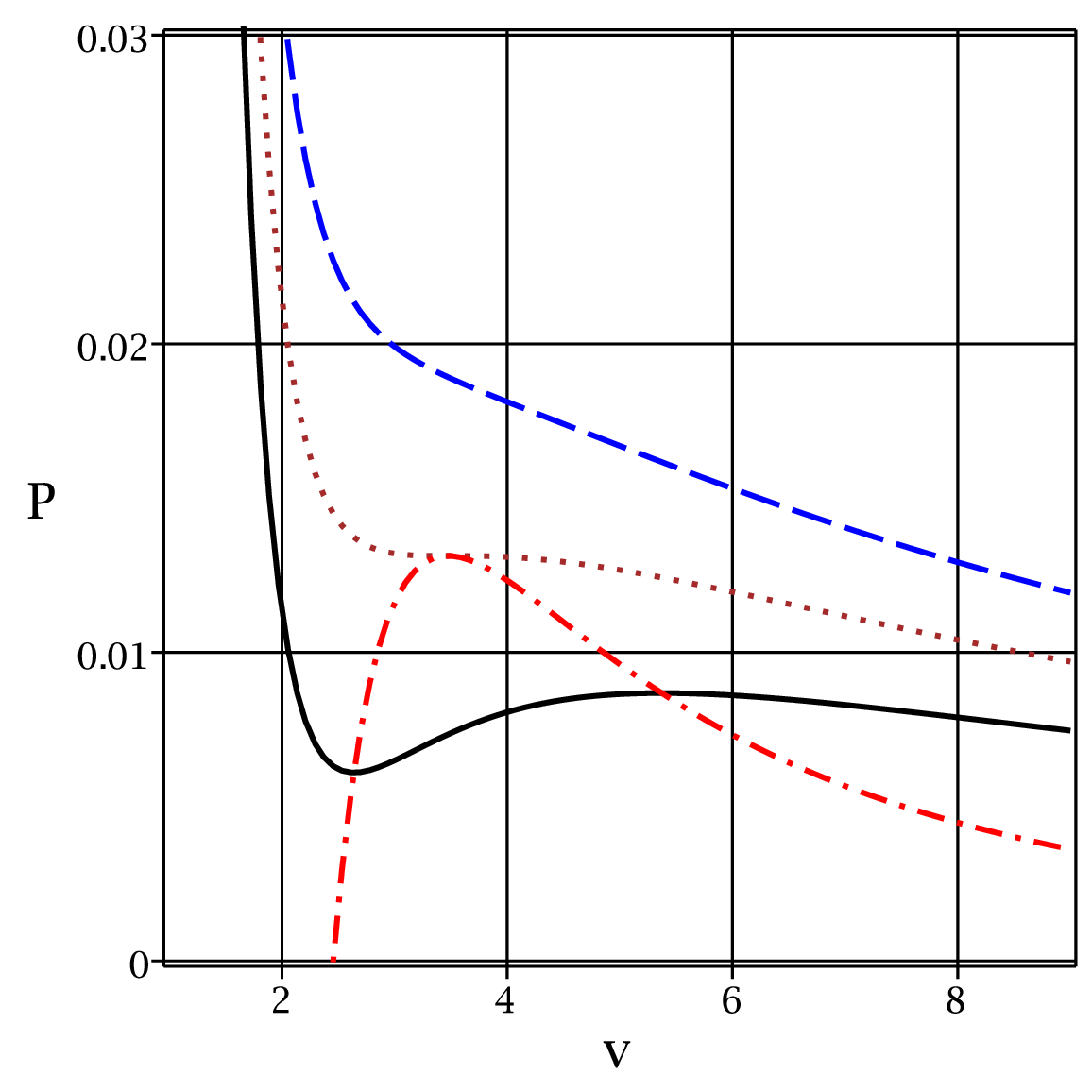} & \epsfxsize=5.5cm %
\epsffile{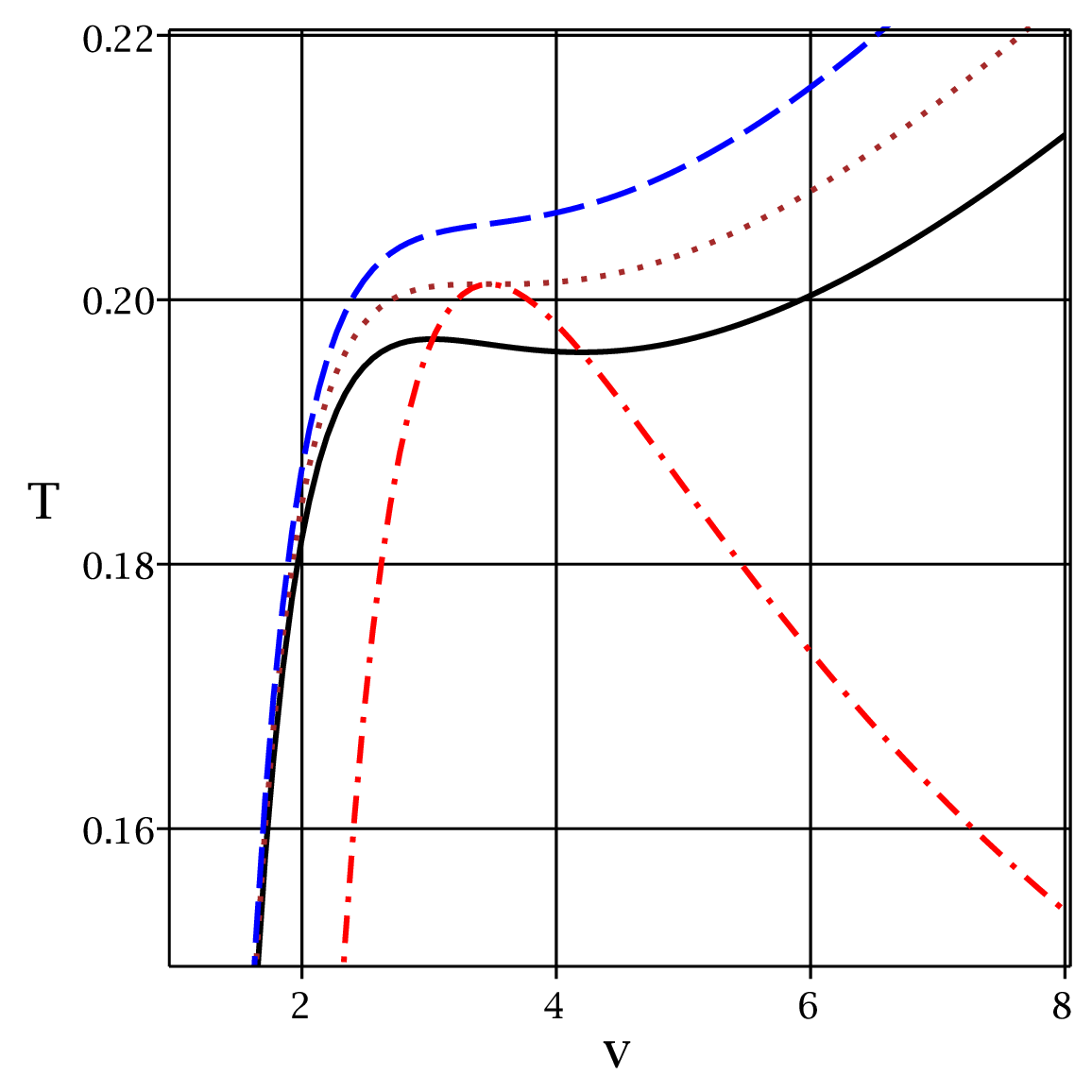}%
\end{array}
$%
\caption{For $m=0.5$, $c=c_{1}=2$, $c_{2}=3$, $\Phi _{E}=0.1$,
$q_{M}=0.1$ and $k=-1$; \newline
\textbf{Left panel:} $P_{new}$ (dashed-dotted line) and $P$ diagrams versus $%
v$ for $T=0.9T_{c}$ (continuous line), $T=T_{c}$ (dotted line) and $%
T=1.1T_{c}$ (dashed line). \newline \textbf{Right panel:}
$T_{new}$ (dashed-dotted line) and $T$ diagrams versus
$v$ for $P=0.9P_{c}$ (continuous line), $P=P_{c}$ (dotted line) and $%
P=1.1P_{c}$ (dashed line). } \label{Fig4}
\end{figure}

It is evident that by using the method which was employed in this
section, one can extract relations for temperature and pressure
which are independent of each other and it is different from
equation of the state. In addition, these new relations could be
employed to plot new $T-v$, $P-v$ and $P-T$ diagrams which are
different from previous phase diagrams. This enables one to study
other thermodynamical properties of the black holes which were not
possible to investigate due to limitations of the previous
methods. In order
to elaborate this matter, we will plot some diagrams (Figs. \ref{Fig4}-\ref%
{Fig7}).

\begin{figure}[tbp]
$%
\begin{array}{cc}
\epsfxsize=5.5cm \epsffile{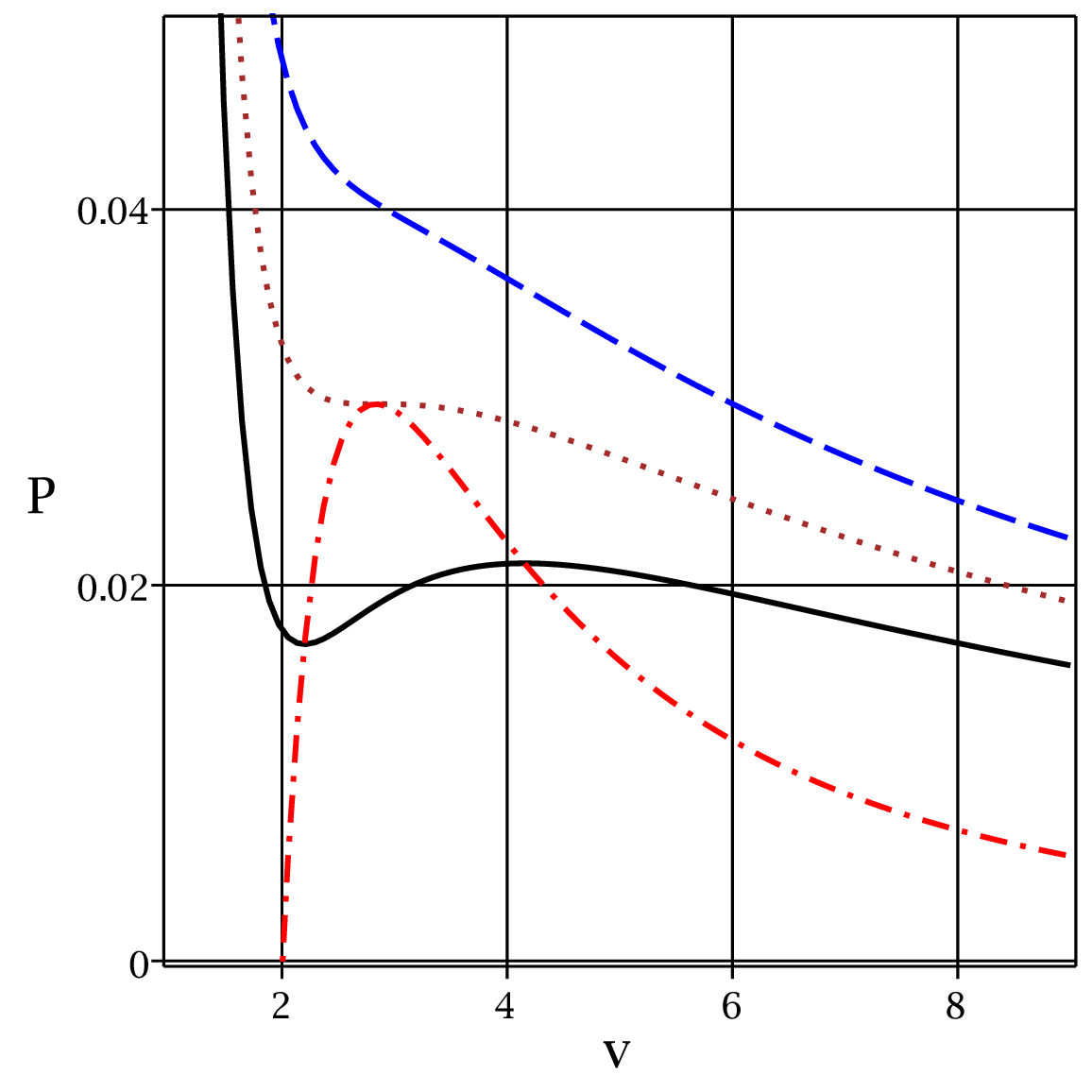} & \epsfxsize=5.5cm %
\epsffile{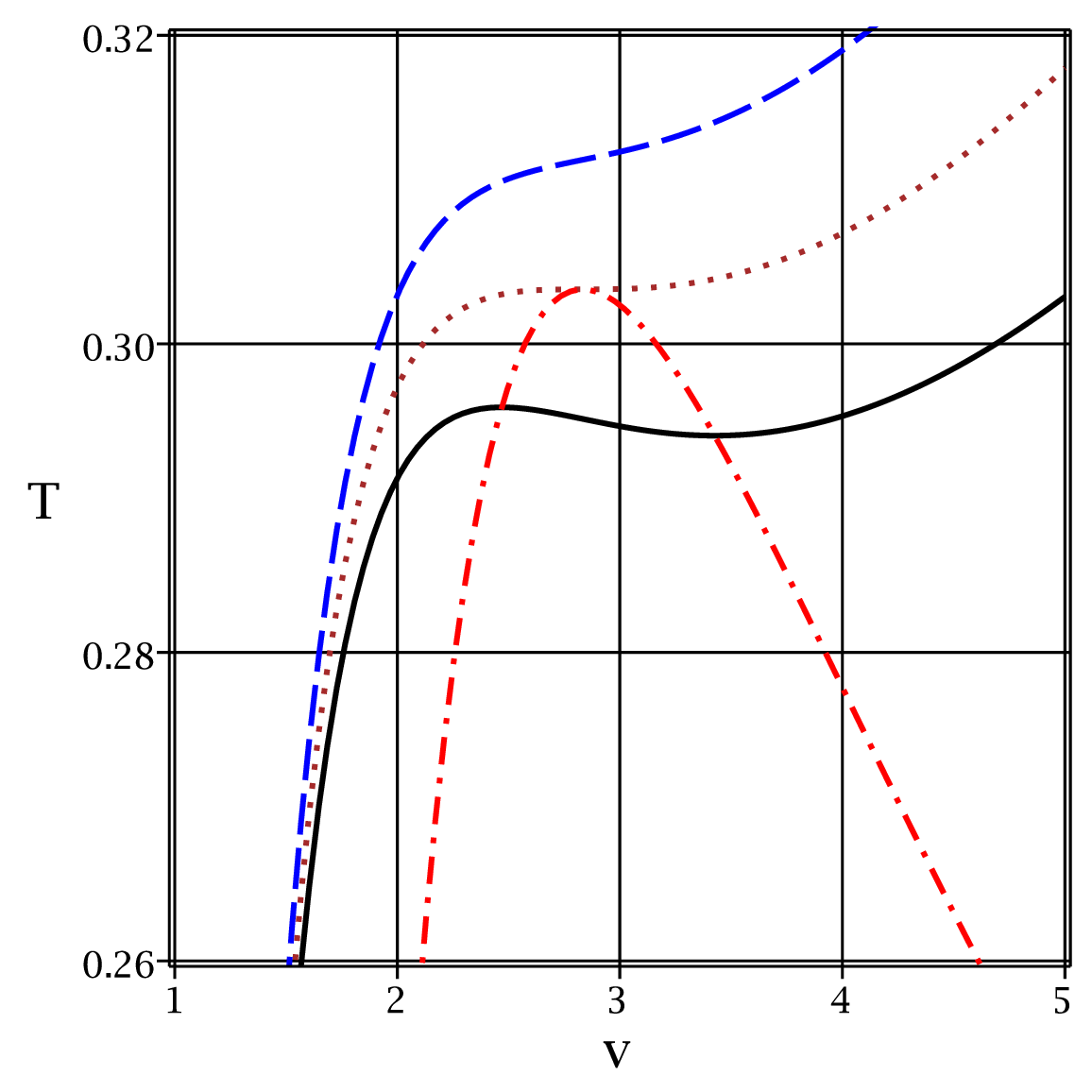}%
\end{array}
$%
\caption{For $m=0.5$, $c=c_{1}=2$, $c_{2}=3$, $\Phi _{E}=0.1$,
$q_{M}=0.1$ and $k=0$; \newline
\textbf{Left panel:} $P_{new}$ (dashed-dotted line) and $P$ diagrams versus $%
v$ for $T=0.9T_{c}$ (continuous line), $T=T_{c}$ (dotted line) and $%
T=1.1T_{c}$ (dashed line). \newline \textbf{Right panel:}
$T_{new}$ (dashed-dotted line) and $T$ diagrams versus
$v$ for $P=0.9P_{c}$ (continuous line), $P=P_{c}$ (dotted line) and $%
P=1.1P_{c}$ (dashed line). } \label{Fig5}
\end{figure}
\begin{figure}[tbp]
$%
\begin{array}{cc}
\epsfxsize=5.5cm \epsffile{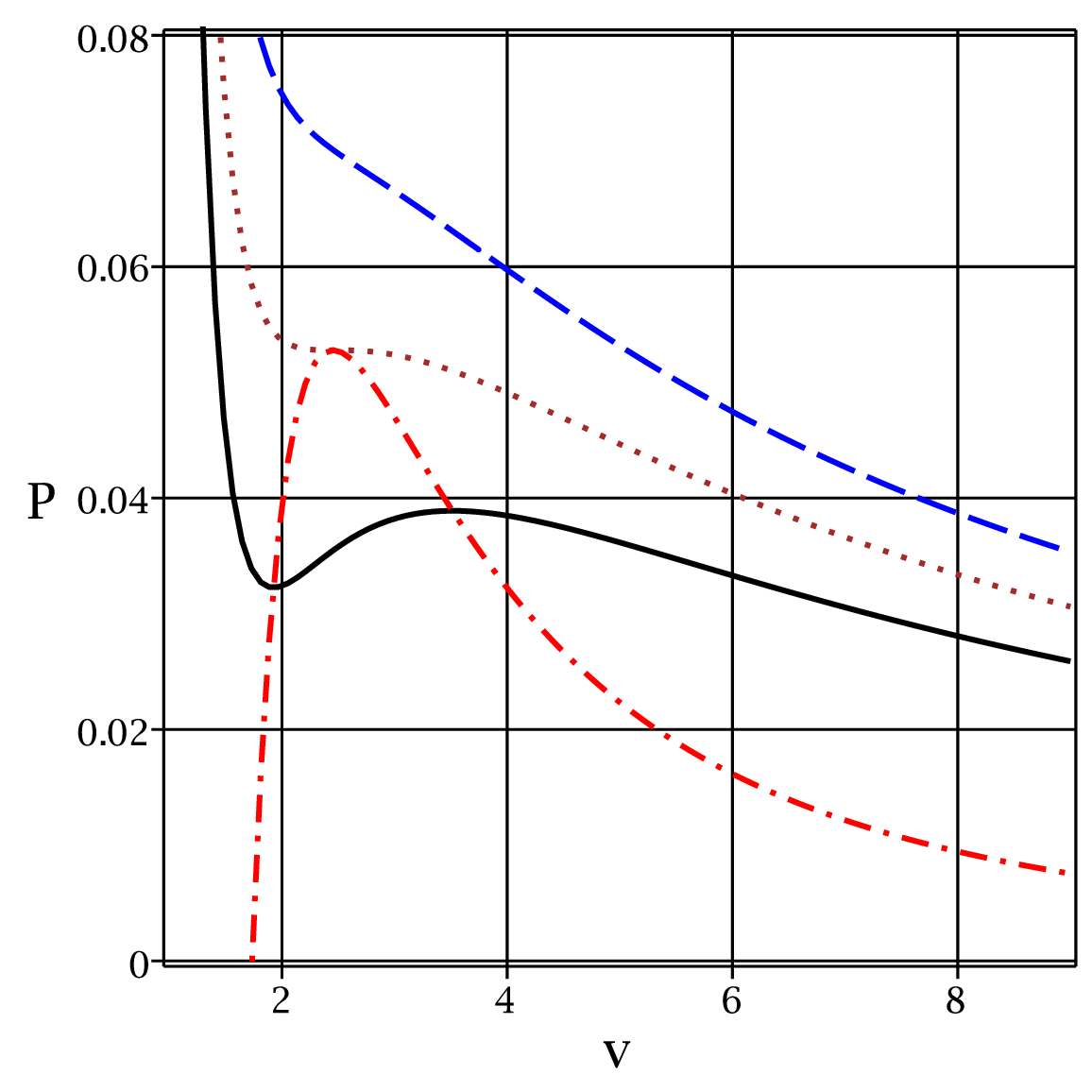} & \epsfxsize=5.5cm %
\epsffile{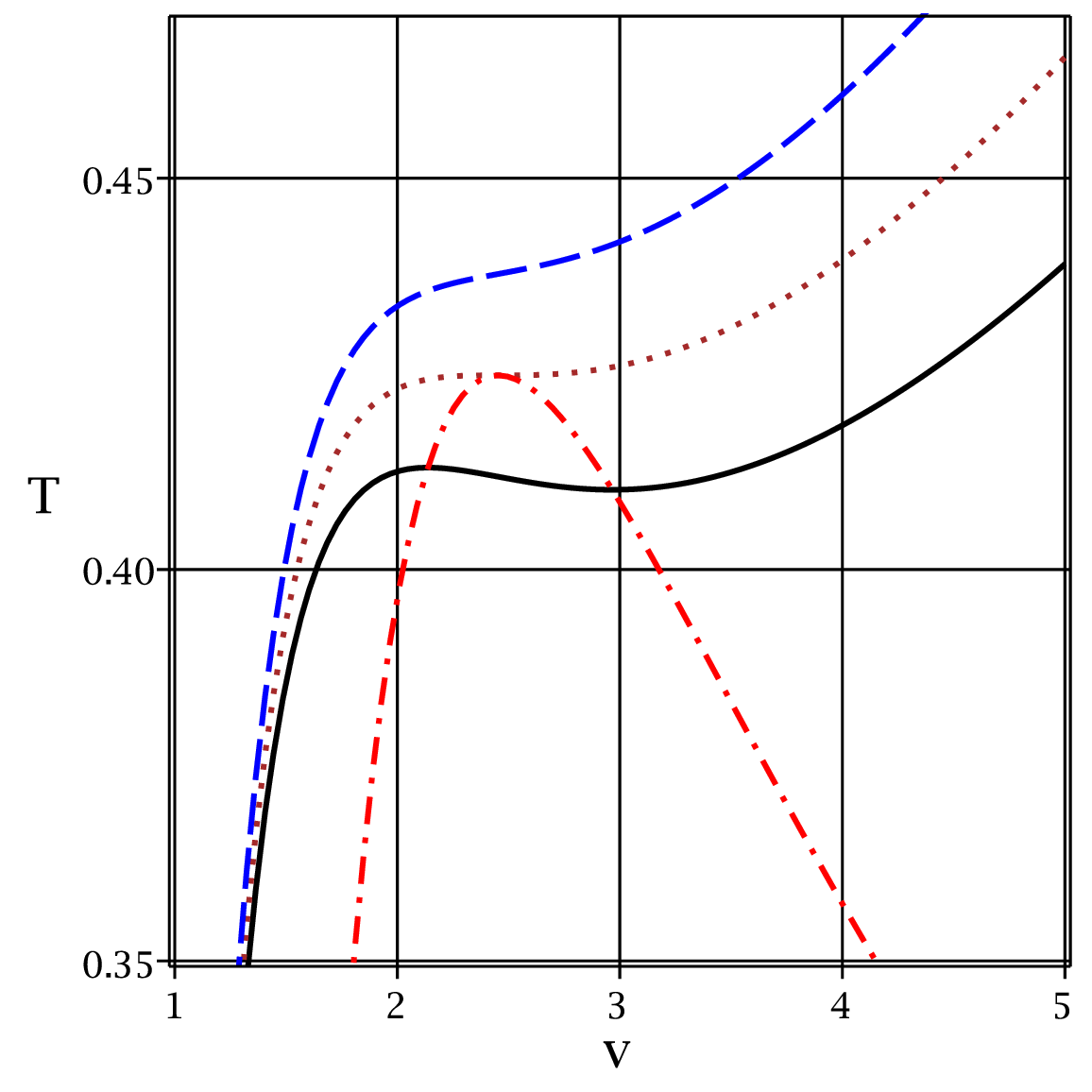}%
\end{array}
$%
\caption{For $m=0.5$, $c=c_{1}=2$, $c_{2}=3$, $\Phi _{E}=0.1$,
$q_{M}=0.1$ and $k=1$; \newline
\textbf{Left panel:} $P_{new}$ (dashed-dotted line) and $P$ diagrams versus $%
v$ for $T=0.9T_{c}$ (continuous line), $T=T_{c}$ (dotted line) and $%
T=1.1T_{c}$ (dashed line). \newline \textbf{Right panel:}
$T_{new}$ (dashed-dotted line) and $T$ diagrams versus
$v$ for $P=0.9P_{c}$ (continuous line), $P=P_{c}$ (dotted line) and $%
P=1.1P_{c}$ (dashed line). } \label{Fig6}
\end{figure}
\begin{figure}[tbp]
$%
\begin{array}{ccc}
\epsfxsize=5.5cm \epsffile{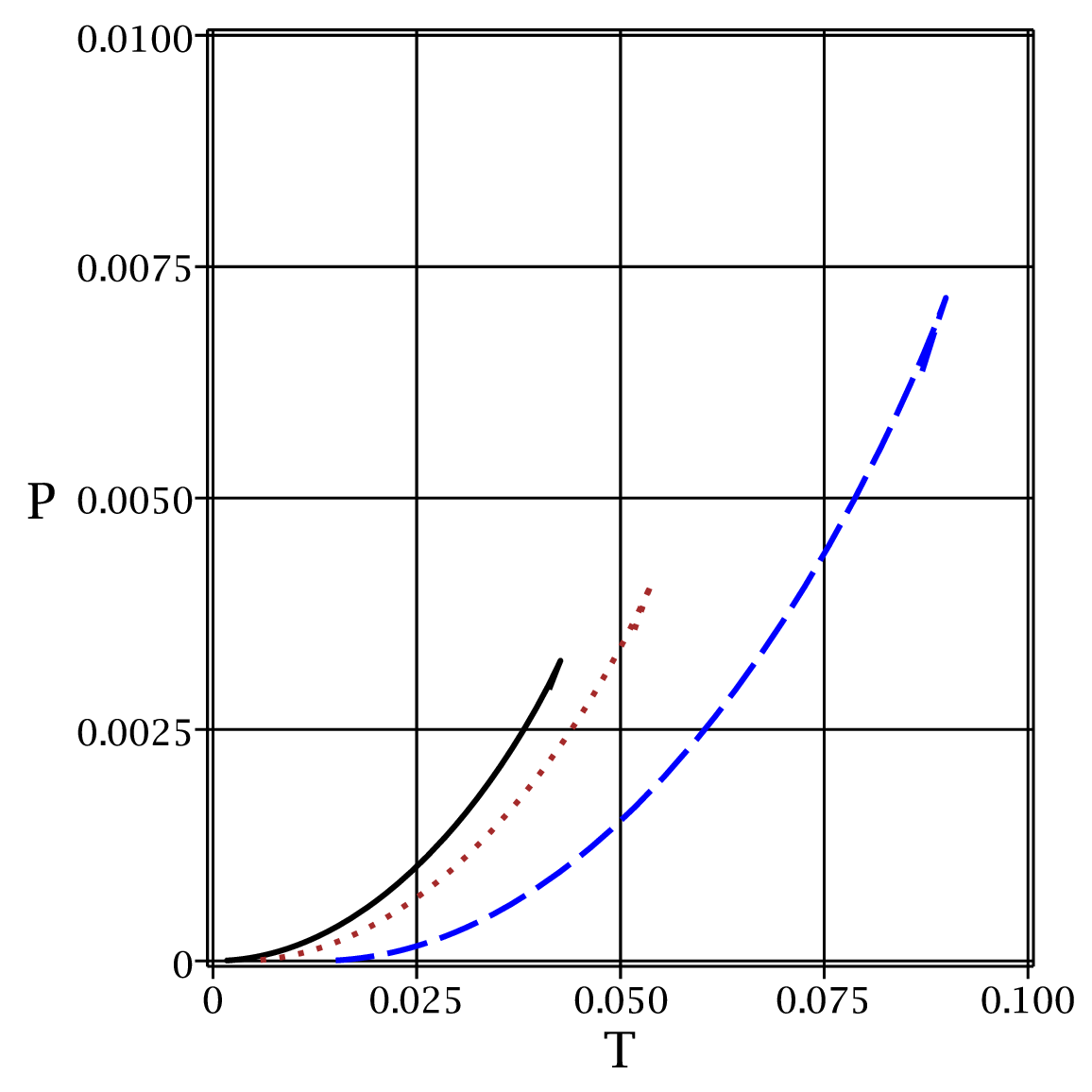} & \epsfxsize=5.5cm %
\epsffile{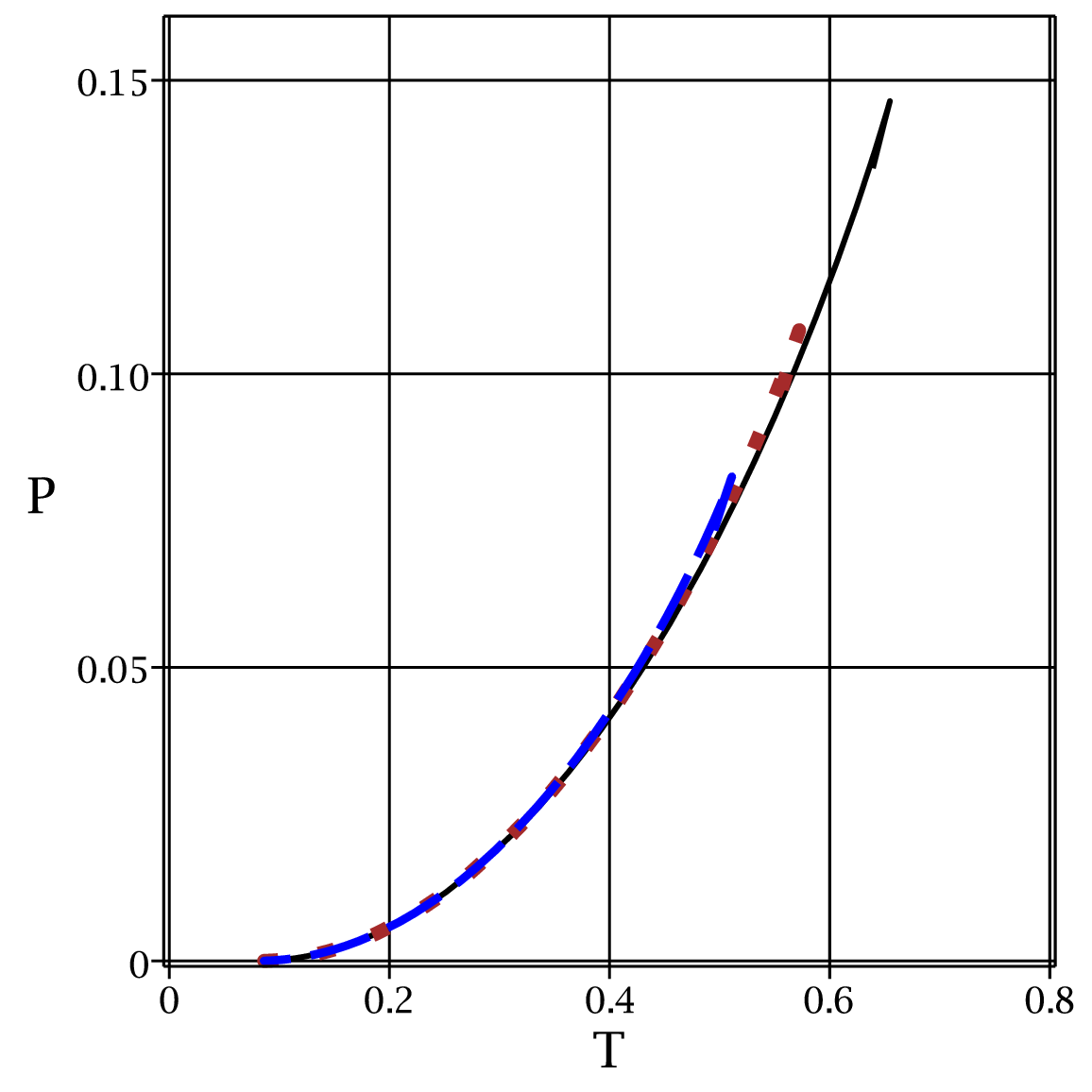} & \epsfxsize=5.5cm \epsffile{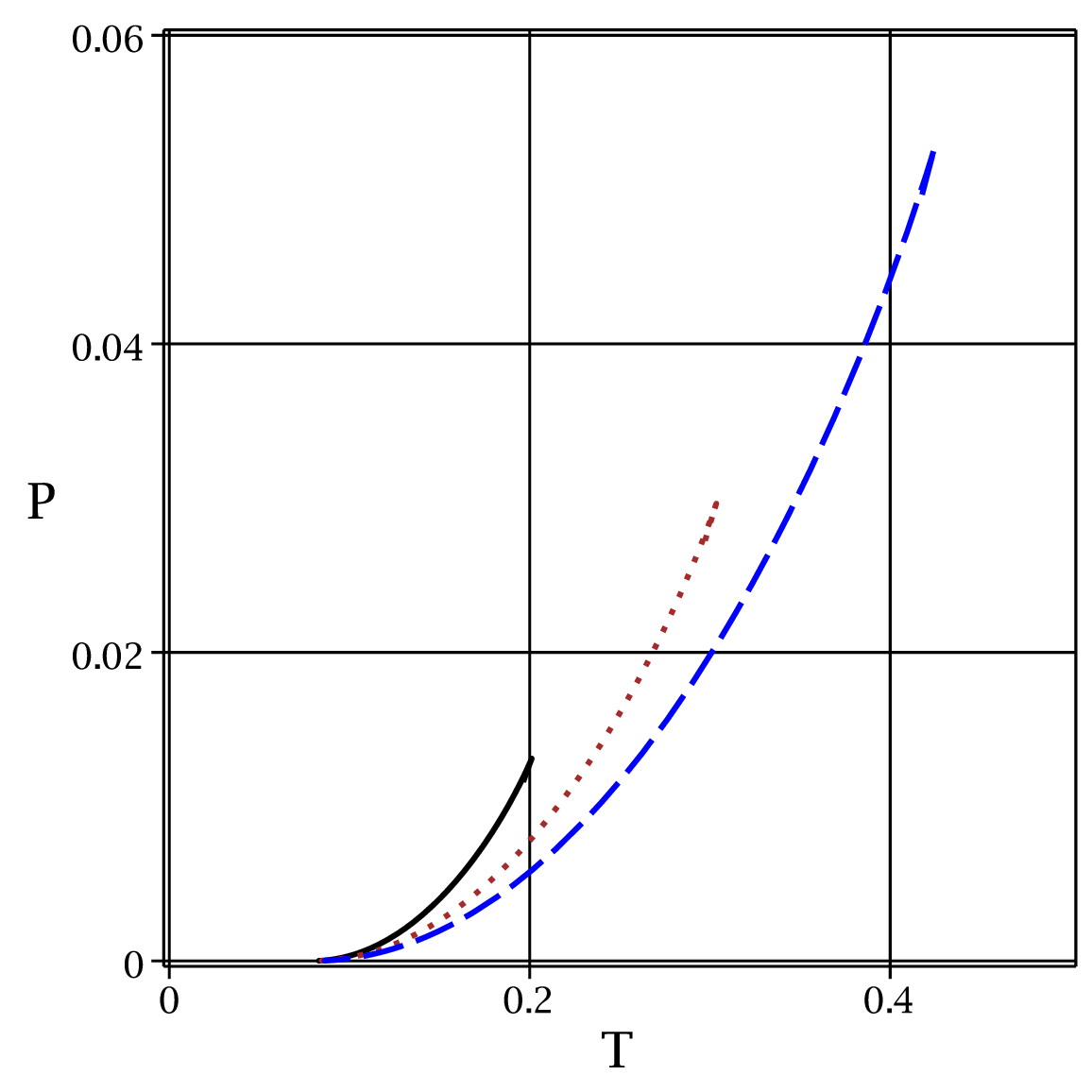}%
\end{array}
$%
\caption{ $P$ versus $T$ diagrams for $c=c_{1}=2$, $c_{2}=3$ and
$\Phi _{E}=0.1$; \newline \textbf{Left panel:} $k=1$, $q_{M}=1$,
$m=0$ (continuous line), $m=0.1$ (dotted line) and $m=0.2$ (dashed
line). \newline
\textbf{Middle panel:} $k=1$, $m=0.5$, $q_{M}=0.6$ (continuous line), $%
q_{M}=0.7$ (dotted line) and $q_{M}=0.8$ (dashed line). \newline
\textbf{Left panel:} $q_{M}=1$, $m=0.5$, $k=-1$ (continuous line),
$k=0$ (dotted line) and $k=1$ (dashed line).} \label{Fig7}
\end{figure}

First of all, as one can see, the maxima of the new relations for
the temperature and pressure are the critical temperature and
pressure, respectively. For the pressures and temperatures smaller
than the critical pressure and temperature, the plotted diagrams
for new relations include all the possible phase transition
points that these black holes can acquire and the results of the
new relations are consistent with those derived in previous
section. In addition, we see that for the pressures and
temperatures larger than the critical pressure and temperature, no
phase transition exists.

On the other hand, $P-T$ diagrams are also plotted for more
clarifications. These diagrams contain information regarding the
region of equilibrium between two different phases which phase
transitions takes place between them. These two phases are small
and large black holes. The critical pressure and temperature are
located at the end of these diagrams and beyond these two critical
values, no phase transition takes place. Evidently, the
equilibrium region is an increasing function of the massive
parameter (left
panel of Fig. \ref{Fig7}) and topological factor (right panel of Fig. \ref%
{Fig7}) while it is a decreasing function of magnetic charge
(middle panel of Fig. \ref{Fig7}).

\subsection{Volume expansion coefficient, isothermal compressibility
coefficient and speed of sound}

Our final study in this section is obtaining specific properties
of the
critical systems known as the length/area/volume expansion coefficients, ($%
\alpha_{L}$/$\alpha_{A}$/$\alpha_{V}$), the isothermal
compressibility coefficient ($\kappa_{T}$), and the speed of sound
($c_{s}$).

The volume expansion coefficient for the black holes at constant
pressure represents the change in the volume of black holes
through the heat transferring. Here, the heat transferring could
be detected by variation in the temperature of black holes. The
length and area expansion coefficients for these black holes are
given by
\begin{equation}
\alpha _{L}=\frac{1}{L}\left( \frac{\partial L}{\partial T}\right)
_{P}\text{
\ \ }\&\text{ \ \ }\alpha _{A}=\frac{1}{A}\left( \frac{\partial A}{\partial T%
}\right) _{P}.
\end{equation}

It is worthwhile mentioning that the length and area of a black
hole recognize by the horizon radius (specific volume) and the
total entropy, respectively. Therefore, one can rewrite
$\alpha_{L}$ and $\alpha_{A}$ as
\begin{equation}
\alpha _{L}=\frac{1}{v}\left( \frac{\partial v}{\partial T}\right)
_{P}\text{
\ \ }\&\text{ \ \ }\alpha _{A}=\frac{1}{S}\left( \frac{\partial S}{\partial T%
}\right) _{P},
\end{equation}%
which for the obtained black hole solutions we can find
\begin{equation}
\alpha_{A}=2\alpha_{L},
\end{equation}
and
\begin{equation}
\alpha _{V}=3\alpha_{L}=\frac{1}{V}\left( \frac{\partial V}{\partial T}%
\right) _{P}={\frac{6{v}^{3}\pi }{2\,\pi P{v}^{4}-{v}^{2}\left( {m}^{2}\,{c}%
^{2}c_{2}-{\Phi }_{E}^{2}+k\right) +12q_{M}^{2}}},
\end{equation}
which are consistent with the results of usual thermodynamic
systems.

Besides, the isothermal compressibility coefficient represents
variation in the volume of black holes corresponding to change in
pressure. The isothermal term comes from the fact that we are
dealing with system in the fixed temperature. For massive dyonic
black holes, the isothermal compressibility coefficient can be
written as \cite{comp}
\begin{equation}
\kappa _{T}=-\frac{1}{V}\left( \frac{\partial V}{\partial P}\right) _{T}={%
\frac{6{v}^{4}\pi }{2\,\pi P{v}^{4}-{v}^{2}\left( {m}^{2}\,{c}^{2}c_{2}-{%
\Phi }_{E}^{2}+k\right) +12q_{M}^{2}}},
\end{equation}%
in which we have used the following relation
\begin{equation}
\left( \frac{\partial V}{\partial P}\right) _{T}\left( \frac{\partial P}{%
\partial T}\right) _{V}\left( \frac{\partial T}{\partial V}\right) _{P}=-1.
\end{equation}

Our next property of interest is the speed of sound. Here, the
speed of sound represents a breathing mode for variation of the
volume versus pressure while the area of the black holes remains
fixed. In order to calculate this property of the black hole,
first, we find the homogenous density of black holes by
\cite{dolan1,dolan2}
\begin{equation}
\rho =\frac{M}{V}=\,{\frac{8\,\pi
P{v}^{4}+3\,{m}^{2}{v}^{3}\,cc_{1}+\left(
12\,{\Phi }_{E}^{2}+12\,k+12\,{m}^{2}\,{c}^{2}c_{2}\right) {v}%
^{2}+48\,q_{M}^{2}}{8\pi {v}^{4}}}.
\end{equation}

Then, by using the following relation \cite{dolan1,dolan2},
\begin{equation}
\frac{1}{c_{s}^{2}}=\frac{\partial \rho }{\partial P},
\end{equation}%
one can find that the speed of sound for massive dyonic black
holes is
\begin{equation}
c_{s}^{-2}=1+\rho \kappa _{T},
\end{equation}%
which is in permitted region of $0\leq c_{s}^{2}\leq 1$.

In order to finalize the discussions in this section, we will give
some details regarding the obtained volume expansion and
isothermal compressibility coefficients. Since the denominator of
both these quantities are the same, we will limit our discussions
to the volume expansion coefficient. First of all, it is a matter
of calculation to show that divergencies of this quantity are
located at
\begin{equation}
v_{\infty -\alpha }=\sqrt{{\frac{{m}^{2}c^{2}{c}_{2}-{\Phi
}_{E}^{2}+k\pm
\sqrt{{m}^{2}c^{2}{c}_{2}\left( 2k+{m}^{2}c^{2}{c}_{2}-2{\Phi }%
_{E}^{2}\right) +{\Phi }_{E}^{4}+k\left( k-2{\Phi }_{E}^{2}\right)
-96\pi Pq_{M}^{2}}}{4\pi P}}}.
\end{equation}

A simple comparison between obtained divergencies for the volume
expansion coefficient and phase transition points (divergencies)
that were extracted for the heat capacity (\ref{rcCQ}) with
$r=v/2$, one can conclude that they are identical. In other words,
phase transition points of heat capacity match with divergencies
of the volume expansion coefficient. This shows that divergencies
of the volume expansion coefficient are actually phase transition
points. Therefore, studying this quantity would also provide the
possibility of obtaining phase transition points.

The leading orders in near horizon and asymptotic limits for the
volume expansion coefficient are in following forms, respectively,
\begin{equation}
\left. {\alpha_{V}}\right\vert _{very\;\; small\;\; v} ={\frac{\pi {v}^{3}}{%
\,2q_{M}^{2}}}-{\frac{\pi \,\left( {\Phi }_{E}^{2}-{m}^{2}\,{c}%
^{2}c_{2}-k\right) {v}^{5}}{24\,q_{M}^{4}},}
\end{equation}%
\begin{equation}
\left. {\alpha_{V}}\right\vert _{very\;\; large\;\; v}={\frac{3}{Pv}}-\frac{3%
}{2}\,{\frac{{\Phi }_{E}^{2}-{m}^{2}\,{c}^{2}c_{2}-k}{{P}^{2}\pi \,{v}^{3}}}%
+3\,\frac{{\frac{\left( {\Phi }_{E}^{2}-{m}^{2}\,{c}^{2}c_{2}-k\right) ^{2}}{%
4{P}^{2}\pi }-\frac{6\,q_{M}^{2}}{P}}}{{P\pi v}^{5}}.
\end{equation}

Evidently, the near horizon limit is dominated by the magnetic
charge. In the absence of magnetic charge, the near limit horizon
will be modified as
\begin{equation}
\left. {\alpha_{V}}\right\vert _{very\;\; small\;\; v} ={\frac{6\pi \,v}{{%
\Phi }_{E}^{2}-{m}^{2}\,{c}^{2}c_{2}-k}}-{\frac{12{\pi
}^{2}P{v}^{3}}{\left( {\Phi
}_{E}^{2}-{m}^{2}\,{c}^{2}c_{2}-k\right) ^{2}}.}
\end{equation}

This highlights another effect of the magnetic charge in
thermodynamical structure of the massive dyonic black holes. On
the other hand, the leading order in asymptotic behavior of these
black holes consists of pressure which is a decreasing function of
it. Here, we see that in the presence of magnetic charge, the
effects of the massive gravity present themselves in second
leading terms. The exception is in the absence of the magnetic
charge, in which the leading order of the near horizon limit
consists of massive terms.

\section{Connection between different phase diagrams}

In this section, we would like to see how different
thermodynamical quantities that were studied are related to each
other and how we should interpret phase transitions and thermal
stability conditions.

In studying temperature, heat capacity and free energy, it was
shown that extrema of the temperature, divergencies of the heat
capacity and equilibrium points of the free energy were located
exactly at the same points. Remembering that divergencies of the
heat capacity are points at which black holes suffer second order
phase transitions, one can conclude that extrema in temperature
and equilibrium points are also places in which black holes go
under second order phase transitions. Therefore, studying these
three quantities provides a uniform picture regarding
thermodynamical behavior of the black holes. On the other hand, it
was shown that different methods in studying van der Waals like
behavior of these black holes lead to consistent results. In other
words, the critical points that were extracted through different
methods were identical.

Now, in order to show that positions (and number) of divergencies
of the heat capacity are consistent with phase transition points
(and number) in
van der Waals like behavior, we have plotted various diagrams (see Figs. \ref%
{Fig8} and \ref{Fig9}).

\begin{figure}[tbp]
$%
\begin{array}{ccc}
\epsfxsize=5.5cm \epsffile{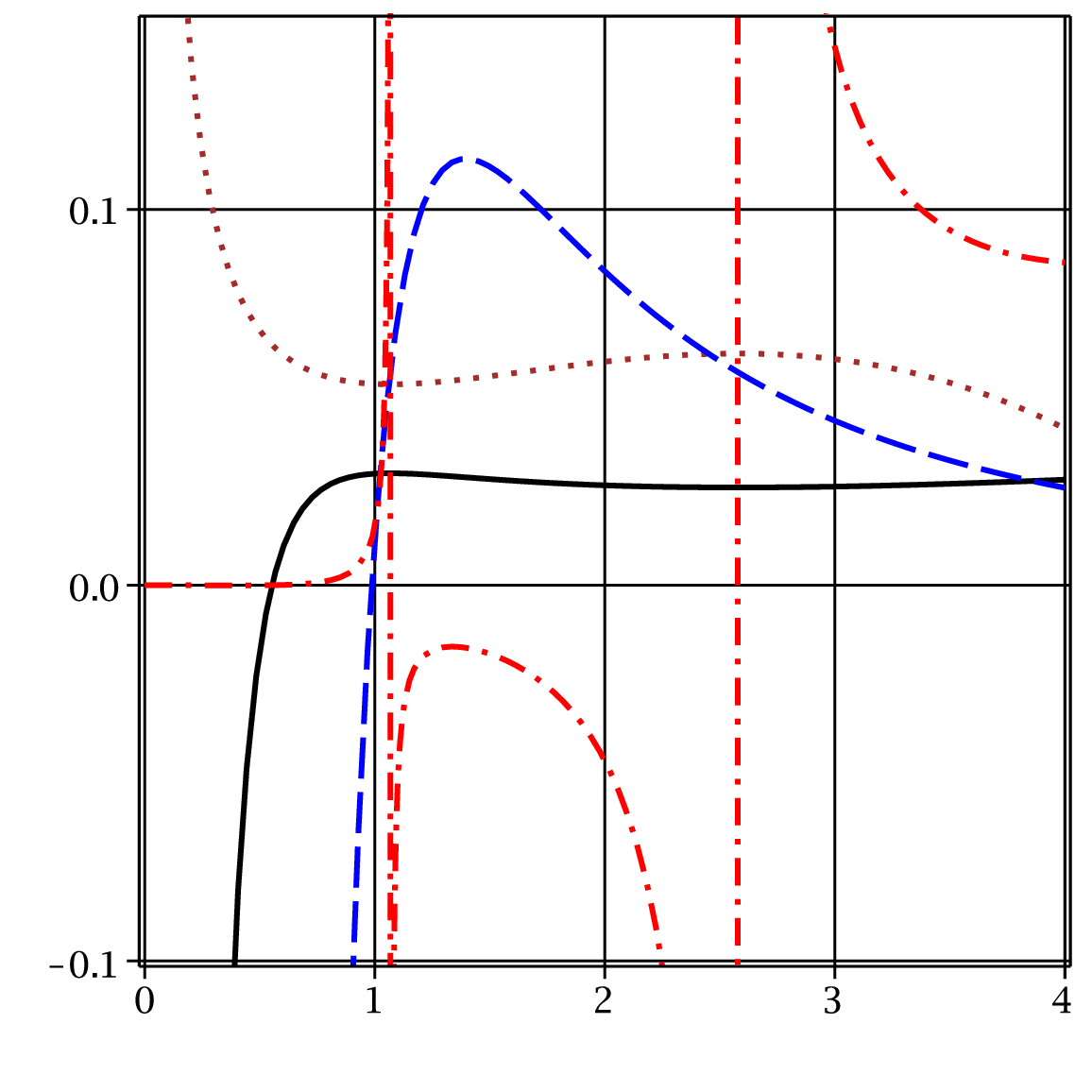} & \epsfxsize=5.5cm %
\epsffile{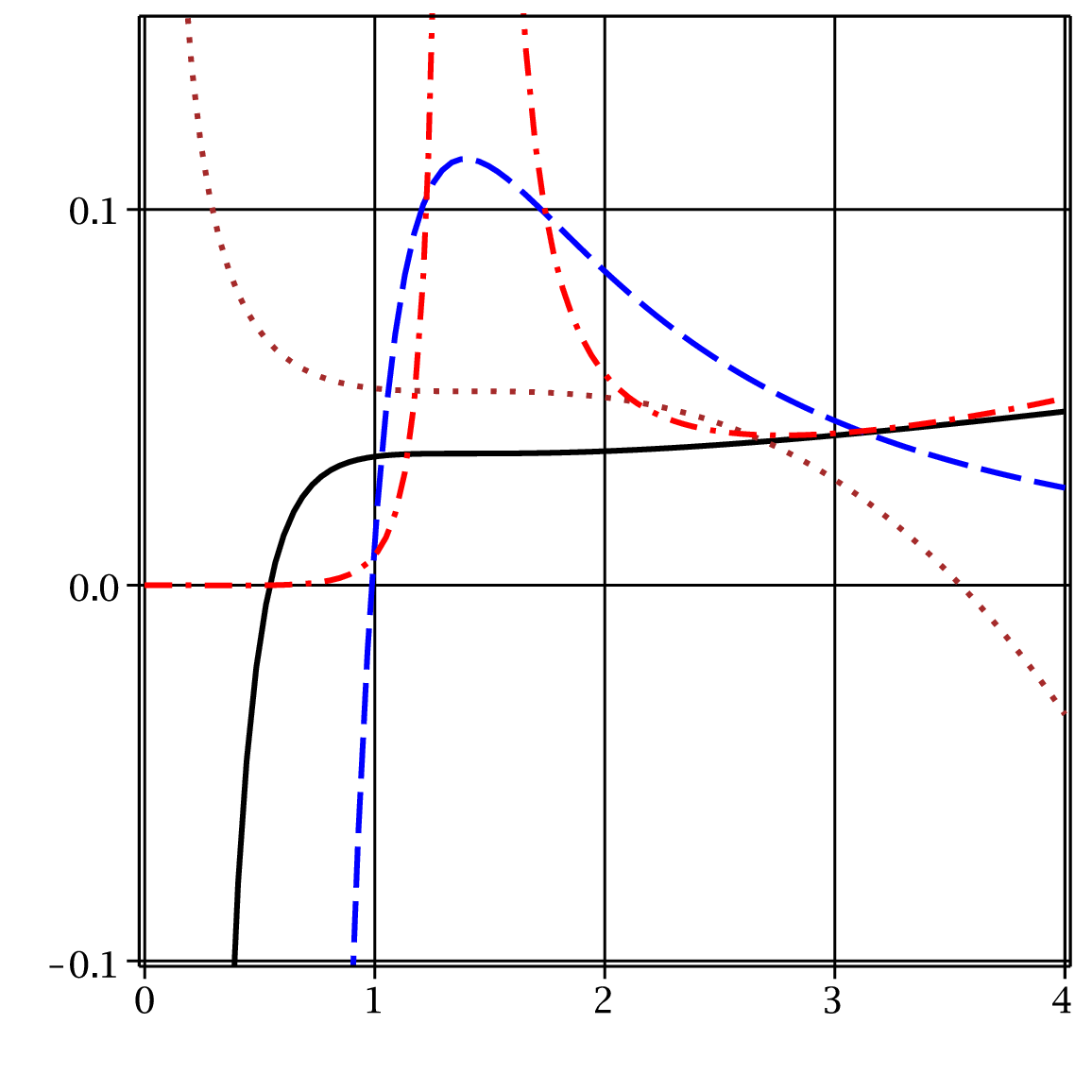} & \epsfxsize=5.5cm \epsffile{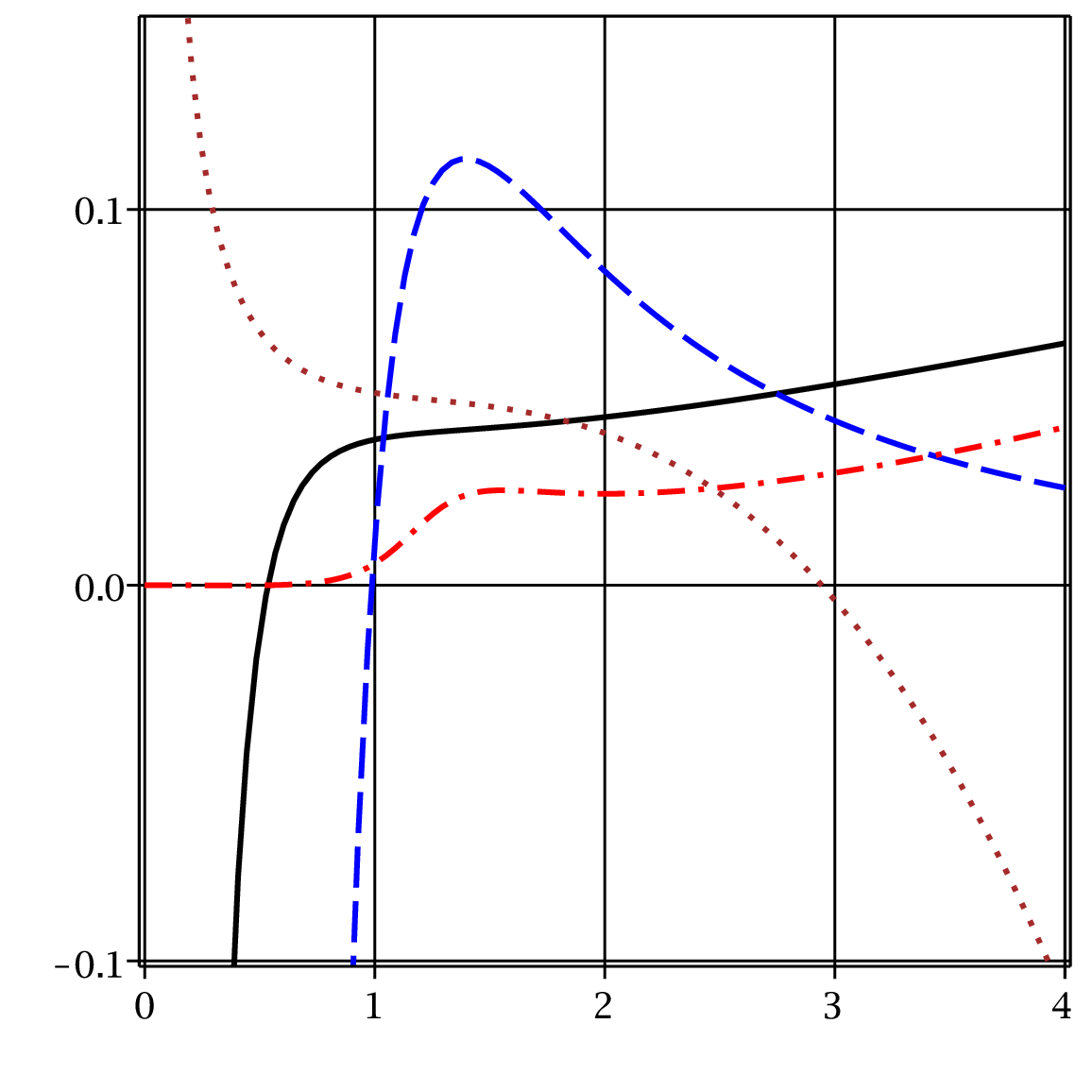}%
\end{array}
$%
\caption{$T$ (continuous line), $W$ (dotted line), $P_{New}$
(dashed line)
and $C_{P}$ (dashed-dotted line) versus $r_{+}$ diagrams for $c=c_{1}=2$, $%
c_{2}=3$, $\Phi _{E}=0.1$, $k=1$, $m=0.1$ and $q_{M}=0.6$;
\newline
\textbf{Left panel:} $P=0.5P_{c}$; \textbf{Middle panel:} $P=P_{c}$; \textbf{%
Right panel:} $P=1.5P_{c}$.} \label{Fig8}
\end{figure}
\begin{figure}[tbp]
$%
\begin{array}{ccc}
\epsfxsize=5.5cm \epsffile{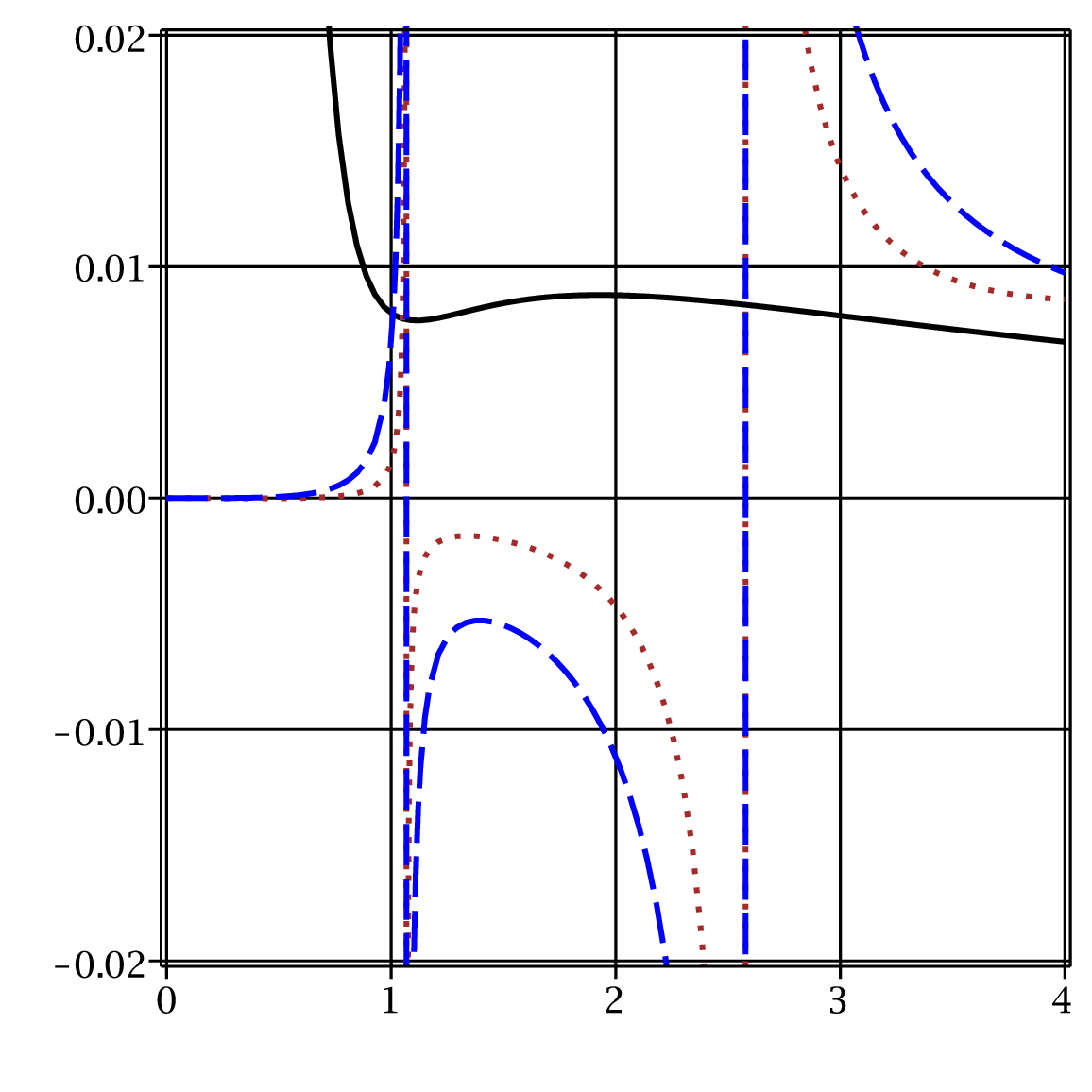} & \epsfxsize=5.5cm %
\epsffile{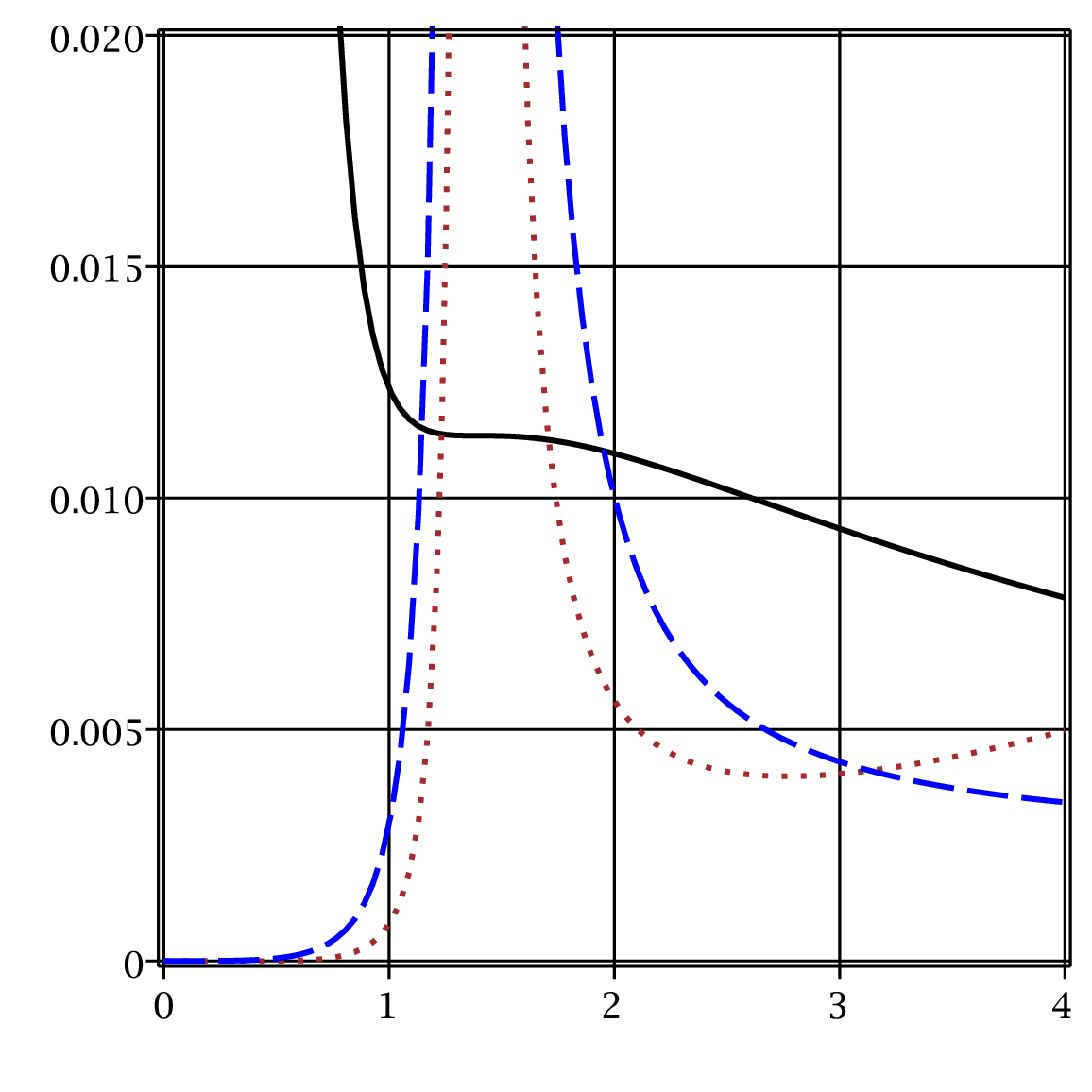} & \epsfxsize=5.5cm \epsffile{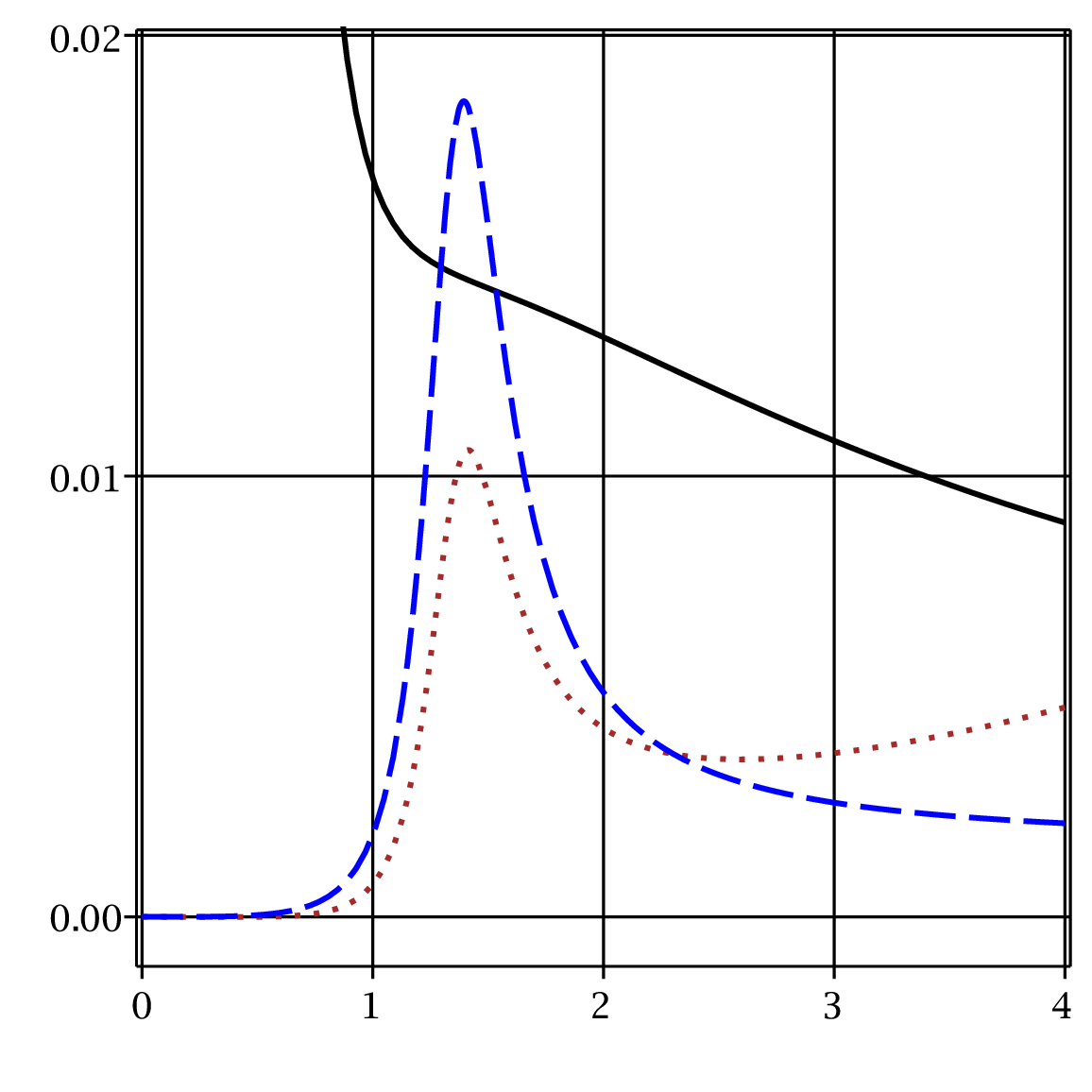}%
\end{array}
$%
\caption{$P$ (continuous line), $C_{P}$ (dotted line) and
$\protect\kappa$ (dashed line) versus $r_{+}$ diagrams for
$c=c_{1}=2$, $c_{2}=3$, $\Phi _{E}=0.1$, $k=1$, $m=0.1$ and
$q_{M}=0.6$; \newline \textbf{Left panel:} $P-r_{+}$ diagram for
$T=0.9T_{c}$ and $C_{P}-r_{+}$ for $P=0.5P_{c}$. \newline
\textbf{Middle panel:} $P-r_{+}$ diagram for $T=T_{c}$ and
$C_{P}-r_{+}$ for $P=P_{c}$. \newline \textbf{Right panel:}
$P-r_{+}$ diagram for $T=1.1T_{c}$ and $C_{P}-r_{+}$ for
$P=1.1P_{c}$.} \label{Fig9}
\end{figure}
Interestingly, for the cases where van der Waals like diagrams
($P-r_{+}$) present an inflection point (pressure and temperature
are the same as critical pressure and temperature), similarly, the
heat capacity enjoys only
one divergency in its diagrams (middle panels of Figs. \ref{Fig8} and \ref%
{Fig9}). The places of this divergency and inflection point
coincide which indicate that these two pictures are identical.
Here, there is a coexistence state between two black holes
(small/large black holes) with positive heat capacity.

On the other hand, when pressure and temperature are smaller than
the critical pressure and temperature, a phase transition point is
observed in the van der Waals diagrams. This phase transition
point matches with that of heat capacity (left panels of Figs.
\ref{Fig8} and \ref{Fig9}). In other words, the places of unstable
states in the van der Waals like diagrams are characterizing with
the divergencies of the heat capacity. Interestingly, in this
case, a region of instability (negative heat capacity) is formed
between two divergence points. In previous case, only two stable
phases existed which were small and large ones without any
unstable region. Here, we have three regions which are small,
medium and large black holes. In this case, medium black holes are
unstable while small and large black holes are thermally stable.
In previous case, there was a coexistence state between both
phases, whereas here, a phase transition take place over a region
which corresponds to medium black holes. If one considers the
growth of the black holes, it could be pointed that upon growing
to the size of smaller divergency, a phase transition takes place
and size of the black holes changes to the size of the larger
divergency whereas vice versa could be observed for Hawking
radiation and evaporation of the black holes. Such sudden changes
in the size of the black holes would leave its traces in
gravitational properties of the black holes, such as gravitational
wave and quasinormal modes. Recently, it was shown that quasi
normal modes of the black holes acquire a dramatic change in the
slopes of their frequencies in small and large black holes near
the critical point where the Van der Waals like thermodynamic
phase transition takes place \cite{van14,quasiB}. In fact, it was
argued that the quasinormal mode can be a dynamic probe of the
thermodynamic phase transition. Therefore, it might provide an
observational potential to study thermodynamical structure of the
black holes.

Finally, in the absence of phase transition in van der Waals like
diagrams,
heat capacity is smooth without any divergency (right panels of Figs. \ref%
{Fig8} and \ref{Fig9}). Here, we see that the critical points that
are extracted through different methods, are consistent and the
obtained critical points in the heat capacity are places in which
system has van der Waals like phase transition. This shows that
the thermodynamical picture which is drawn by different methods
for black holes are consistent with each other. It should be
pointed out that, although the phase transition points for
different methods are identical, every phase diagram describes a
part of thermodynamical properties of the black holes. In order to
have a complete picture regarding thermodynamical structure of the
black holes, it is necessary to study different phase diagrams in
more details.

Before, finishing this part, we will give more details regarding
the possibility of existence of black holes for different phases.
Previously, we observed that speed of sound must not exceed the
speed of light. This could be acquired if isothermal
compressibility coefficient is positive valued. In plotted
diagrams for this quantity (Fig. \ref{Fig9}), one can see that the
divergencies of isothermal compressibility coefficient are
identical to those obtained for the heat capacity. This shows that
divergencies of the isothermal compressibility coefficient are
marking a phase transition point which agrees with our earlier
results. For pressures smaller than critical pressure, isothermal
compressibility coefficient has two divergencies. Between these
two divergencies (medium black holes), isothermal compressibility
coefficient is negative valued which leads to the fact that speed
of sound will be more than unity, hence, more than the speed of
light. This shows that the medium black holes, which are thermally
unstable, have speed of sound larger than the speed of light.
Therefore medium black holes do not have physical properties. In
other words, although medium black holes have positive temperature
and geometrically well behaved, they are not physical and between
two divergencies, black holes solutions does not exists. The only
physical cases are small black holes (between root and smaller
divergency of the heat capacity) and large black holes (after
larger divergency of the heat capacity). It is worthwhile to
mention that for the case of $P=P_{c}$, isothermal compressibility
coefficient is positive valued with one divergency which marks the
phase transition between small/large black holes. For pressures
larger than critical pressure, isothermal compressibility
coefficient is positive valued without any root and divergency
which agrees with results of other methods.

\section{Conformal field theory and its magnetic properties}

\label{CFTsec}

\subsection{Magnetization and magnetic susceptibility of boundary theory}

The structure of the dyonic black holes admits a constant magnetic
field in dual theory of conformal field theory on the boundary.
This constant magnetic field is due to the presence of a magnetic
charge on the bulk. It is a matter of calculation to show that the
magnetic field is
\begin{equation}
B=\frac{q_{M}}{l^{2}}.  \label{B}
\end{equation}

Considering this magnetic field, it is possible to introduce the
corresponding conjugating quantity which is the magnetization.
Thermodynamically speaking, different phases living on the
boundary of the solutions could be recognized by this quantity.
Using the definition of the internal energy (\ref{internal}), one
can find magnetization for these black holes as
\begin{equation}
M=-\left( \frac{\partial H}{\partial B}\right)
=-\frac{l^{2}q_{M}}{r_{+}}, \label{M}
\end{equation}%
which shows that generalization to massive gravity has no direct
effect on magnetization. In addition, since $b$, $q_{M}$ and
$r_{+}$ are positive values, magnetic field and magnetization have
opposite sign. Now, by using
the obtained internal energy (\ref{internal}) and temperature (\ref{TotalTT}%
) with the magnetic field (\ref{B}) and magnetization (\ref{M}),
one can rewrite internal energy and temperature in following
forms, respectively,
\begin{equation}
H={\frac{B^{2}l^{8}{m}^{2}\,cc_{1}}{4M^{2}}}-{\frac{B^{3}l^{10}}{2M^{3}}}-{%
\frac{\,\left( k+{m}^{2}\,{c}^{2}c_{2}+{\Phi }_{E}^{2}\right) Bl^{4}}{2M}}-%
\frac{BM}{2},  \label{massM}
\end{equation}%
\begin{equation}
T={\frac{{m}^{2}\,cc_{1}}{4\pi }}-{\frac{\,\left( k+{m}^{2}\,{c}^{2}c_{2}-{%
\Phi }_{E}^{2}\right) M}{4\pi Bl^{4}}}-\,{\frac{3Bl^{2}}{4\,\pi M}}+\,{\frac{%
M^{3}}{4\pi Bl^{8}\pi }.}  \label{TM}
\end{equation}

Remembering that phase transition points are marked as extrema of
temperature, we obtain the magnetization at the extrema of
temperature as
\begin{equation}
M_{Extremum-T}=\frac{l^{2}\sqrt{6}}{6}\sqrt{\,{m}^{2}\,{c}^{2}c_{2}-{\Phi }%
_{E}^{2}+\,k\pm \,\sqrt{\left( k+{m}^{2}\,{c}^{2}c_{2}-{\Phi }%
_{E}^{2}+6Bl\right) \left( k+{m}^{2}\,{c}^{2}c_{2}-{\Phi }%
_{E}^{2}-6Bl\right) }}\,.  \label{rootTM}
\end{equation}

Evidently, depending on the choices of different parameters, the
temperature may be categorized under one the following cases:

I) Absence of extremum. This case happens if the following
conditions are not satisfied, simultaneously,
\begin{eqnarray*}
\left( k+{m}^{2}\,{c}^{2}c_{2}-{\Phi }_{E}^{2}+6Bl\right) \left( k+{m}^{2}\,{%
c}^{2}c_{2}-{\Phi }_{E}^{2}-6Bl\right) &>&0, \\
&& \\
{m}^{2}\,{c}^{2}c_{2}-{\Phi }_{E}^{2}+\,k\pm \,\sqrt{\left( k+{m}^{2}\,{c}%
^{2}c_{2}-{\Phi }_{E}^{2}+6Bl\right) \left( k+{m}^{2}\,{c}^{2}c_{2}-{\Phi E}%
^{2}-6Bl\right) } &>&0.
\end{eqnarray*}

II) Existence of only one extremum which takes place if
\begin{equation*}
\left( k+{m}^{2}\,{c}^{2}c_{2}-{\Phi }_{E}^{2}+6Bl\right) \left( k+{m}^{2}\,{%
c}^{2}c_{2}-{\Phi }_{E}^{2}-6Bl\right) =0
\end{equation*}

III) The presence of two extrema which happens if the mentioned
conditions of first case are satisfied.\newline

The existence of magnetic field and magnetization provides the
possibility of studying magnetic properties of the solutions.
Hereafter, we will focus on diamagnetic and paramagnetic behaviors
of the solutions. In ordinary materials, diamagnetic behavior is
due to quantized orbital motion of the electrons in the presence
of an external magnetic field. On the other hand, paramagnetism
originates from alignment of the spin of electrons parallel to
external magnetic field. Depending on the dominance of either
these two behaviors, materials could be categorized into
diamagnetic and paramagnetic ones. The magnetic property which
enables one to determine diamagnetic and paramagnetic nature of
the material is magnetic susceptibility, $\chi $. The positivity
of the susceptibility, $\chi >0$, indicates that system is
diamagnetic while, the opposite ($\chi <0$) is valid for
paramagnetic systems. The susceptibility is defined as
\begin{equation}
\chi =\left( \frac{\partial M}{\partial B}\right) ,
\end{equation}%
in which, for these black holes, we can find
\begin{equation}
\chi =\frac{\left( \frac{\partial T}{\partial B}\right) }{\left( \frac{%
\partial T}{\partial M}\right) }=-\frac{M}{B}{\frac{\left( {m}^{2}\,{c}%
^{2}c_{2}-{\Phi }_{E}^{2}+k\right) M^{2}l^{4}-M^{4}-3\,B^{2}l^{10}}{\left( {m%
}^{2}\,{c}^{2}c_{2}-{\Phi }_{E}^{2}+k\right) M^{2}l^{4}-3M^{4}-3\,B^{2}l^{10}%
}}.  \label{chi}
\end{equation}

If following condition is satisfied
\begin{equation*}
0<\left( {m}^{2}\,{c}^{2}c_{2}-{\Phi }_{E}^{2}+k\right)
M^{2}l^{4}-M^{4}-3\,B^{2}l^{10}<2M^{4},
\end{equation*}%
the susceptibility will be positive valued and system has
diamagnetic behavior, while the opposite is observed for
paramagnetic behavior. The roots of susceptibility are obtained in
following forms
\begin{equation}
M_{root-\chi }=\frac{l^{2}\sqrt{2}}{2}\sqrt{\,{m}^{2}\,{c}^{2}c_{2}-{\Phi }%
_{E}^{2}+\,k\pm \,\sqrt{{m}^{2}c^{2}{c}_{2}\left( 2k+{m}^{2}c^{2}{c}_{2}-2{%
\Phi }_{E}^{2}\right) +{\Phi }_{E}^{4}+k\left( k-2{\Phi
}_{E}^{2}\right) -12B^{2}l^{2}}},  \label{rootchi}
\end{equation}%
which points to the following cases:

I) No root exists for the susceptibility. This happens if the
following conditions are not satisfied, simultaneously,
\begin{eqnarray*}
{m}^{2}c^{2}{c}_{2}\left( 2k+{m}^{2}c^{2}{c}_{2}-2{\Phi }_{E}^{2}\right) +{%
\Phi }_{E}^{4}+k\left( k-2{\Phi }_{E}^{2}\right) -12B^{2}l^{2} &>&0, \\
&& \\
{m}^{2}\,{c}^{2}c_{2}-{\Phi }_{E}^{2}+\,k\pm \,\sqrt{{m}^{2}c^{2}{c}%
_{2}\left( 2k+{m}^{2}c^{2}{c}_{2}-2{\Phi }_{E}^{2}\right) +{\Phi }%
_{E}^{4}+k\left( k-2{\Phi }_{E}^{2}\right) -12B^{2}l^{2}} &>&0.
\end{eqnarray*}

II) Only one root exists which will happen if
\begin{equation*}
{m}^{2}c^{2}{c}_{2}\left( 2k+{m}^{2}c^{2}{c}_{2}-2{\Phi }_{E}^{2}\right) +{%
\Phi }_{E}^{4}+k\left( k-2{\Phi }_{E}^{2}\right) -12B^{2}l^{2}=0.
\end{equation*}

III) Two roots exists if conditions for the first case are
satisfied. The roots usually determines the places where the sign
of the susceptibility switches. The phase transition between
different phases are marked by divergencies that may exist in the
susceptibility. It is a matter of calculation to show that
divergencies of the susceptibility are obtained as
\begin{equation}
M_{\infty -\chi
}=\frac{l^{2}\sqrt{6}}{6}\sqrt{\,{m}^{2}\,{c}^{2}c_{2}-{\Phi
}_{E}^{2}+\,k\pm \,\sqrt{\left( k+{m}^{2}\,{c}^{2}c_{2}-{\Phi }%
_{E}^{2}+6Bl\right) \left( k+{m}^{2}\,{c}^{2}c_{2}-{\Phi }%
_{E}^{2}-6Bl\right) }}.  \label{Mcchi}
\end{equation}

The obtained divergencies (phase transition) are identical to
extrema that were extracted for the temperature (\ref{rootTM}).
This shows that phase transition point in boundary theory are
extrema that are observed in the temperature. Therefore, depending
on satisfaction of conditions that were pointed out for the
temperature, the boundary theory is divided into one of the
following cases: I) Absence of phase transition and smooth
behavior for the susceptibility. II) Existence of only one phase
transition point. III) Existence of two phase transition points
for the boundary theory.

\subsection{$T$ vs $M$ and $\protect\chi$ vs $M$ diagrams and
diamagnetic/paramagnetic behaviors}

\begin{figure}[tbp]
$%
\begin{array}{cc}
\epsfxsize=5.5cm \epsffile{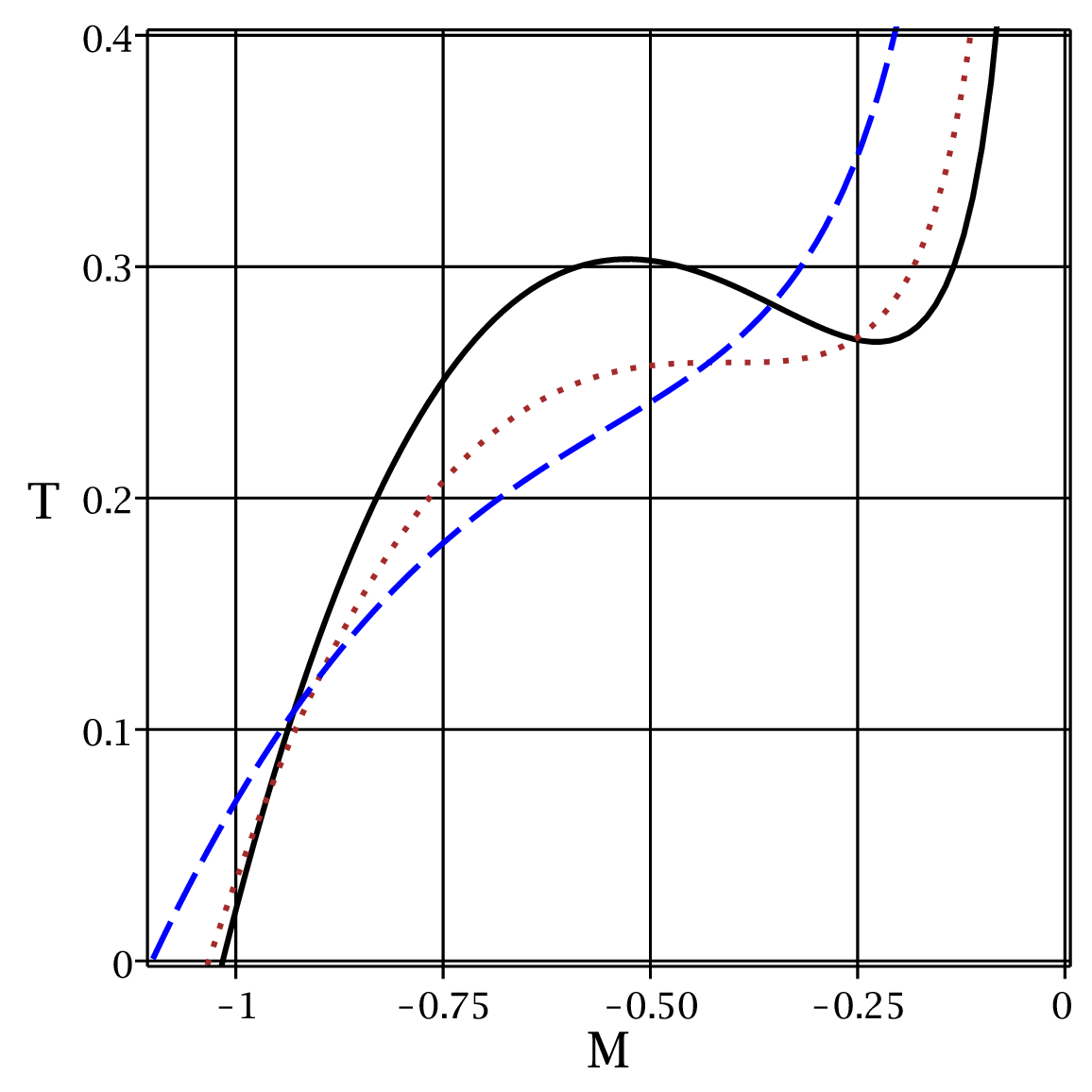} & \epsfxsize=5.5cm %
\epsffile{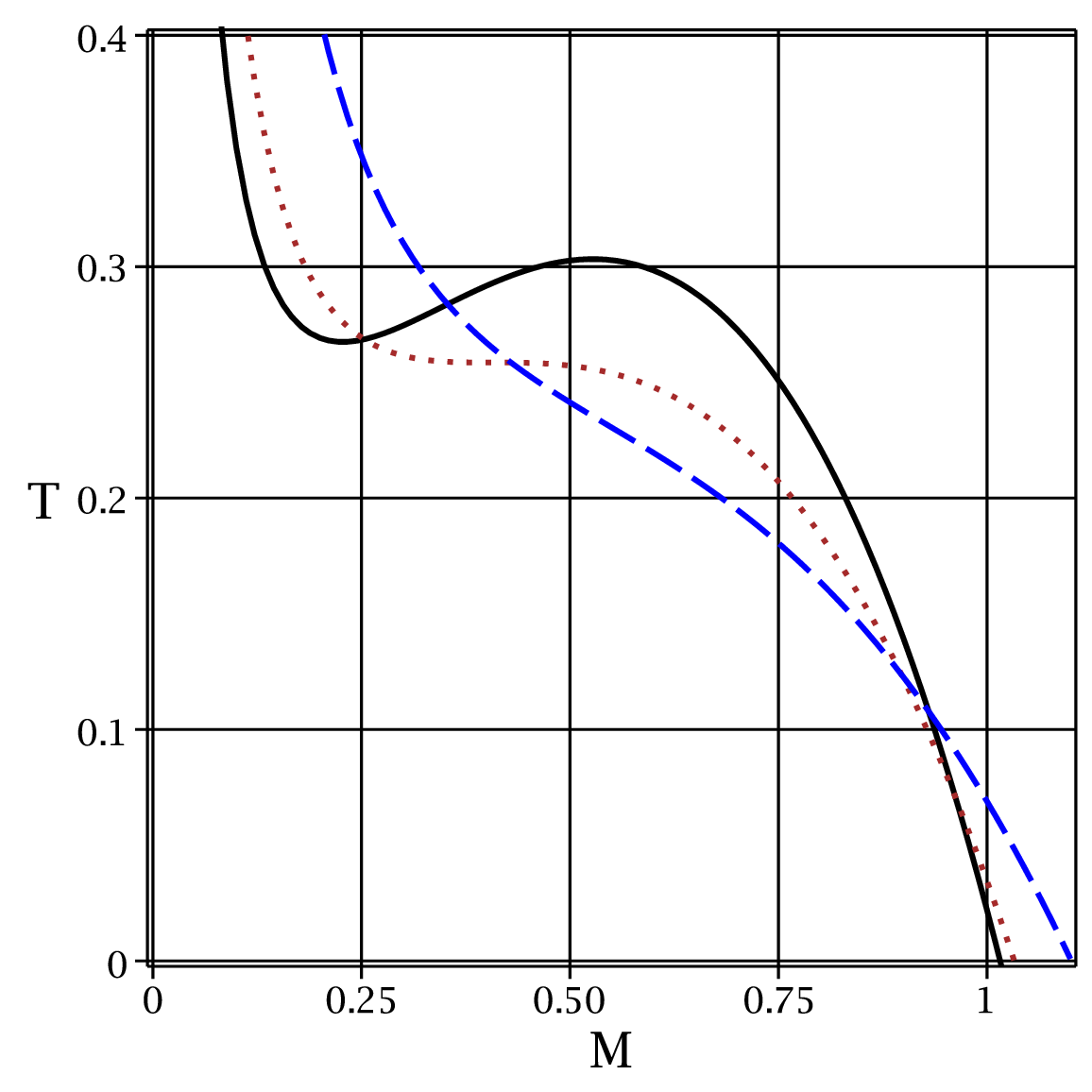} \\
\epsfxsize=5.5cm \epsffile{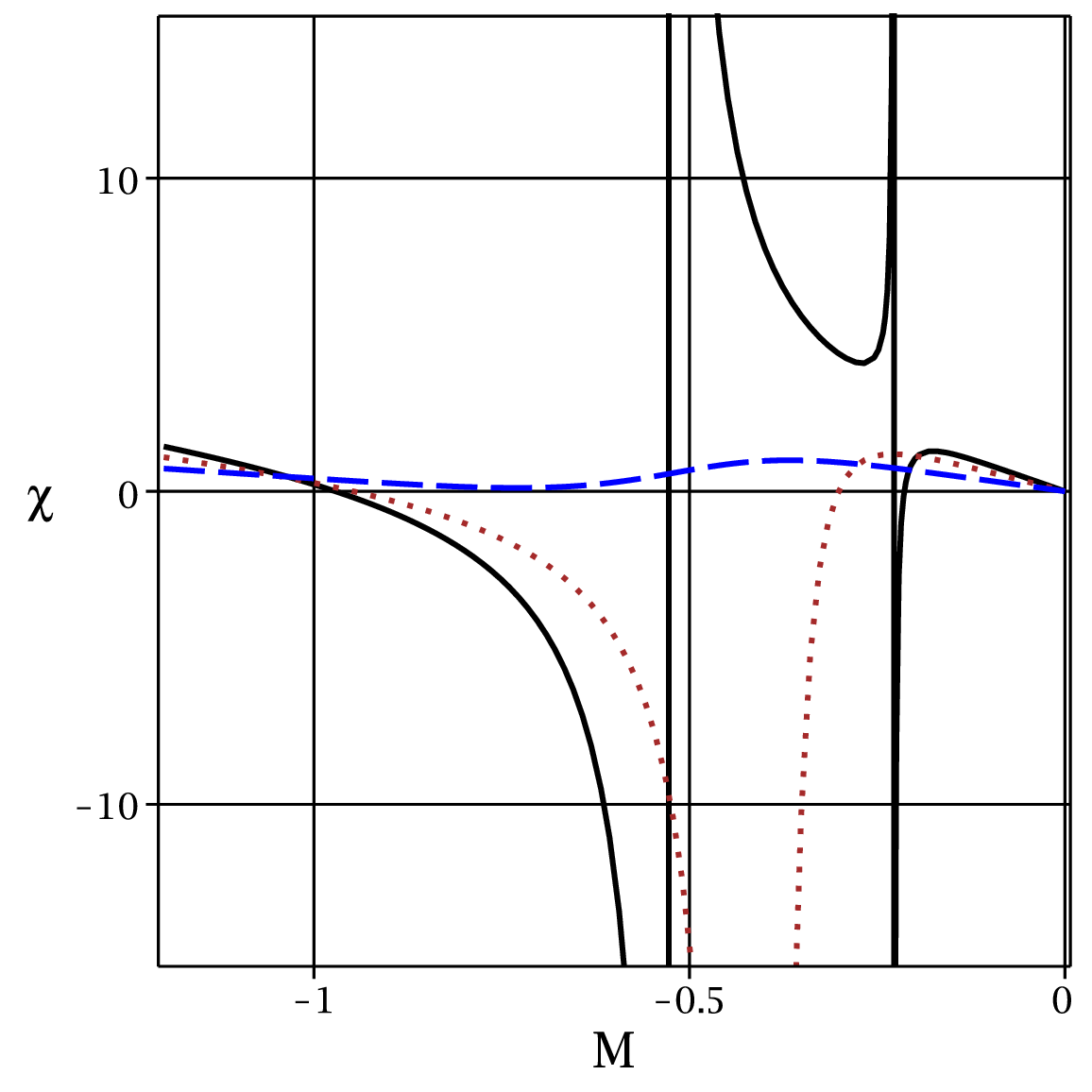} & \epsfxsize=5.5cm %
\epsffile{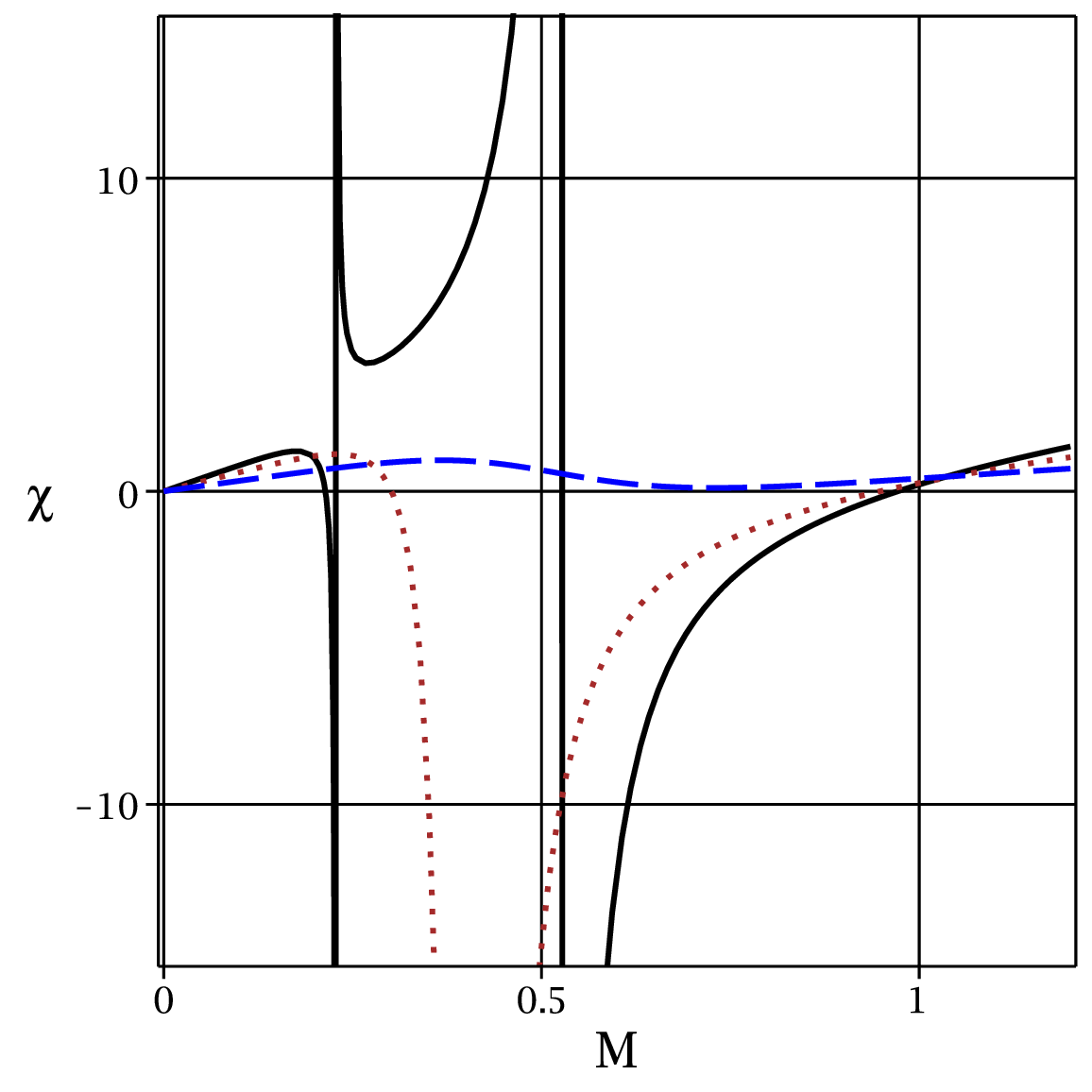}%
\end{array}
$%
\caption{$T$ (up panels) and $\protect\chi$ (down panels) versus
$M$ diagrams for $m=0$, $\Phi _{E}=0.1$, $k=1$ and $b=1$; \newline
\textbf{Left panels:} $B=0.12$ (continuous line), $B=0.165$
(dashed line) and $B=0.3$ (dashed-dotted line). \newline
\textbf{Right panels:} $B=-0.12$ (continuous line), $B=-0.165$
(dashed line) and $B=-0.3$ (dashed-dotted line).} \label{Fig10}
\end{figure}

\begin{figure}[tbp]
$%
\begin{array}{cc}
\epsfxsize=5.5cm \epsffile{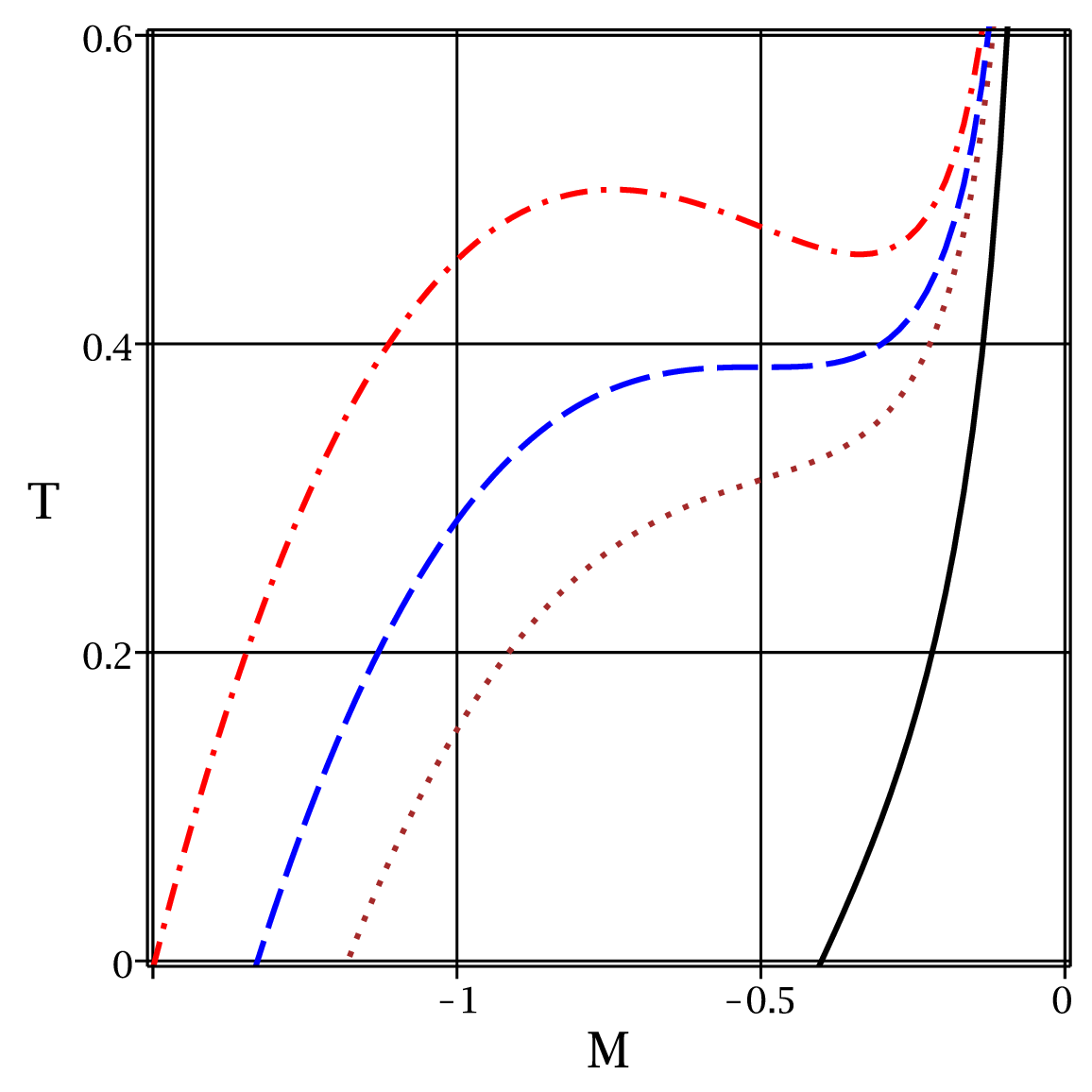} & \epsfxsize=5.5cm %
\epsffile{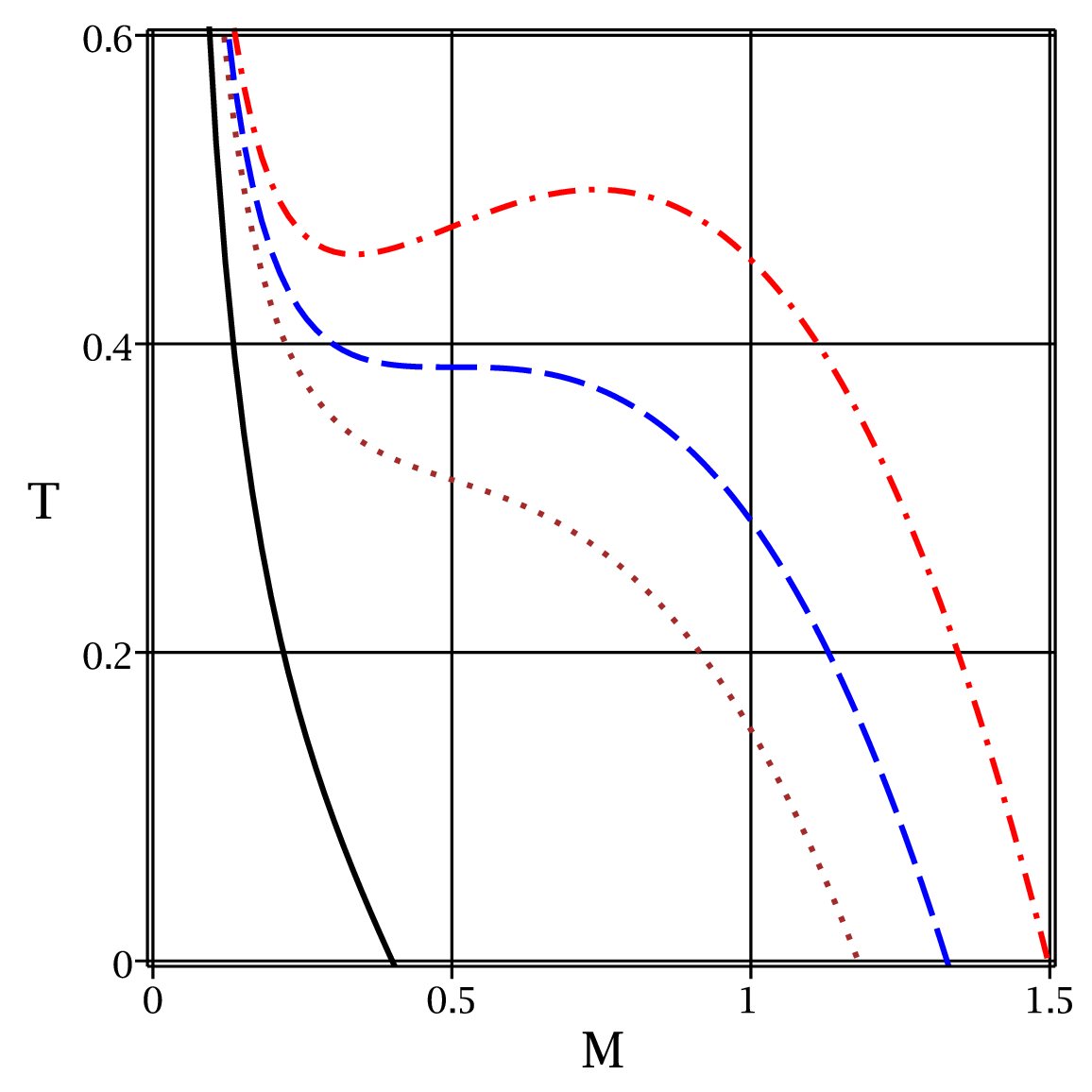} \\
\epsfxsize=5.5cm \epsffile{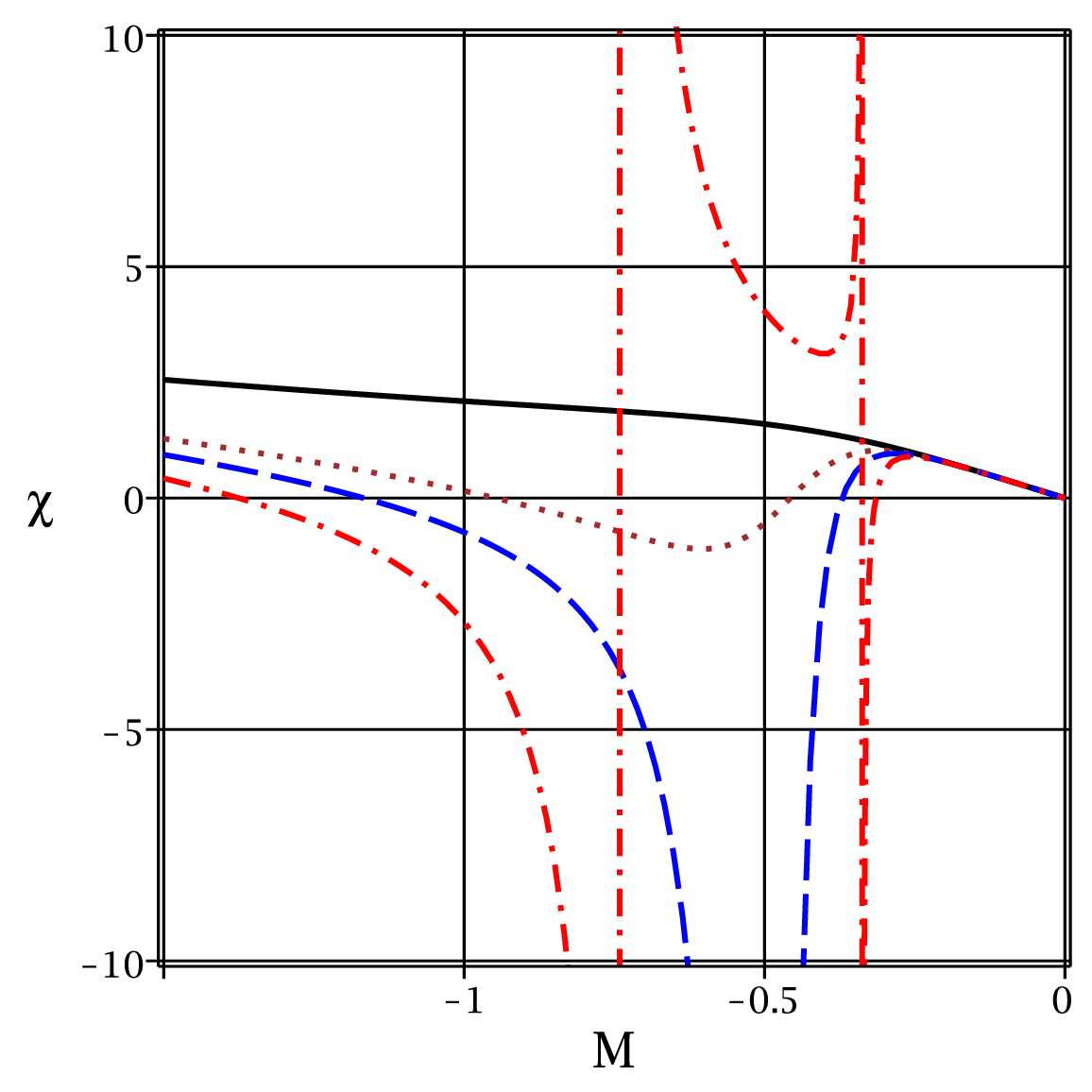} & \epsfxsize=5.5cm %
\epsffile{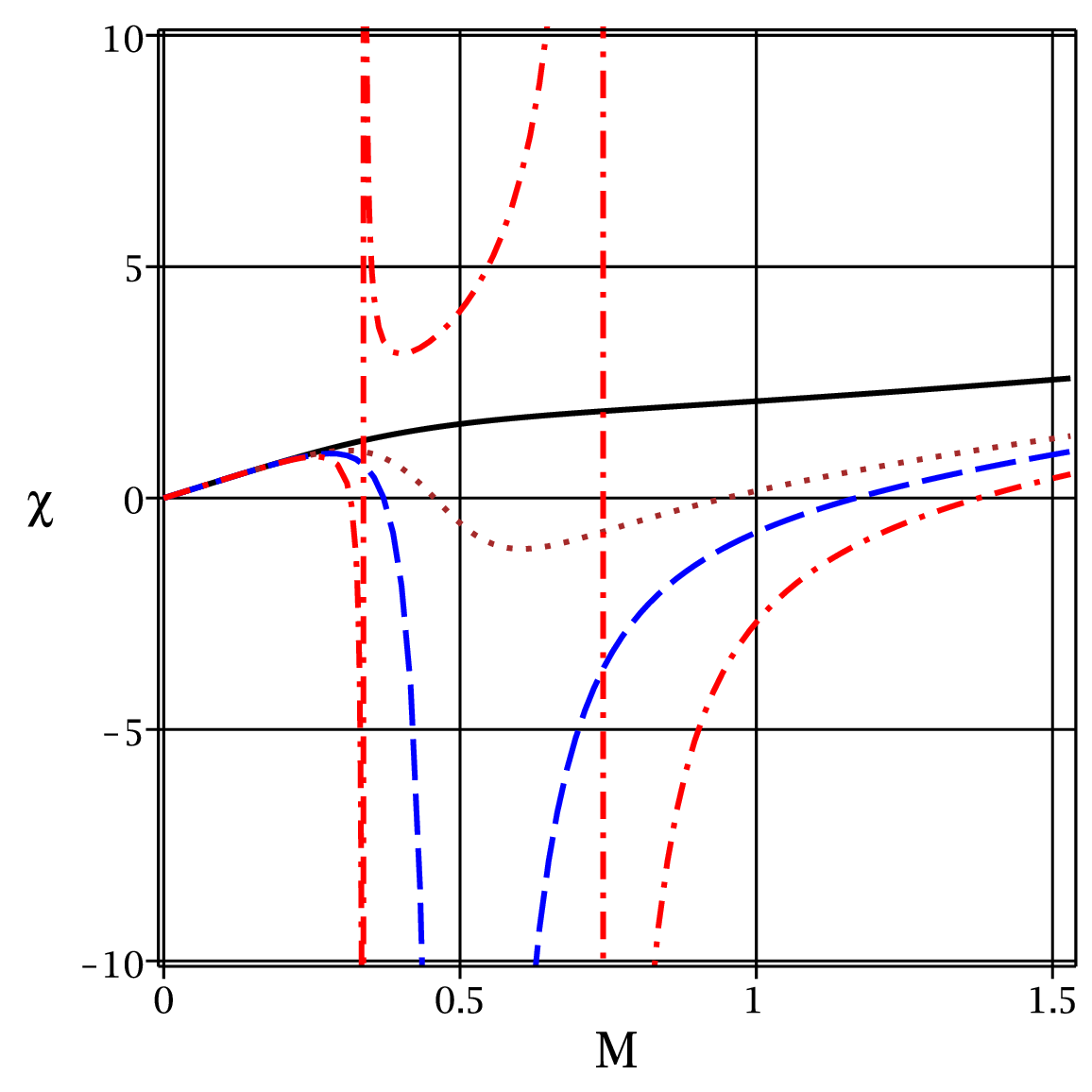}%
\end{array}
$%
\caption{$T$ (up panels) and $\protect\chi$ (down panels) versus
$M$ diagrams for $c=c_{1}=2$, $c_{2}=3$, $\Phi _{E}=0.1$, $b=1$
and $k=-1$; $m=0$ (continuous line), $m=0.42$ (dotted line),
$m=0.4573474245$ (dashed line) and $m=0.5$ (dashed-dotted line).
\newline \textbf{Left panels:} $B=0.25$; \textbf{Right panels:}
$B=-0.25$.} \label{Fig11}
\end{figure}

\begin{figure}[tbp]
$%
\begin{array}{cc}
\epsfxsize=5.5cm \epsffile{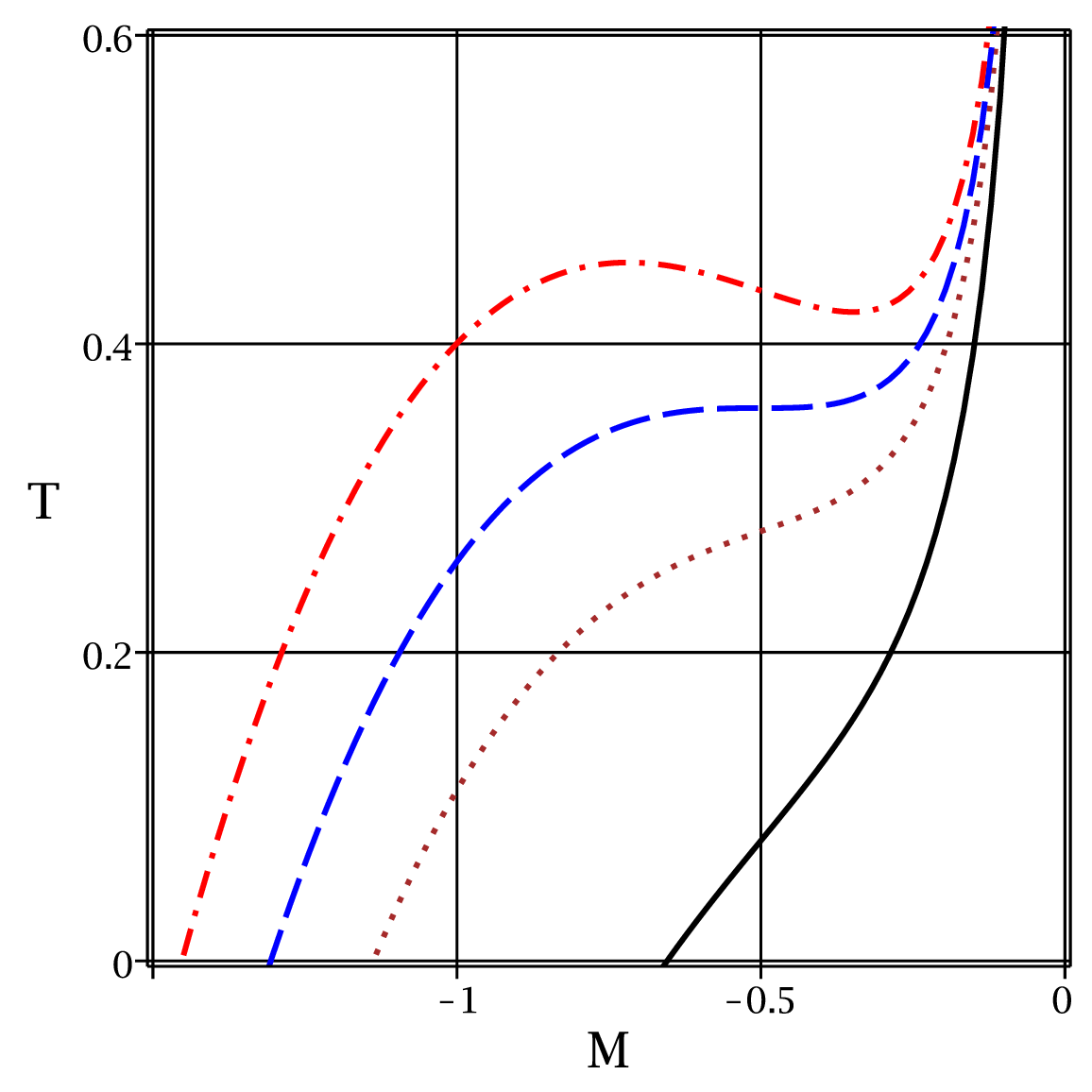} & \epsfxsize=5.5cm %
\epsffile{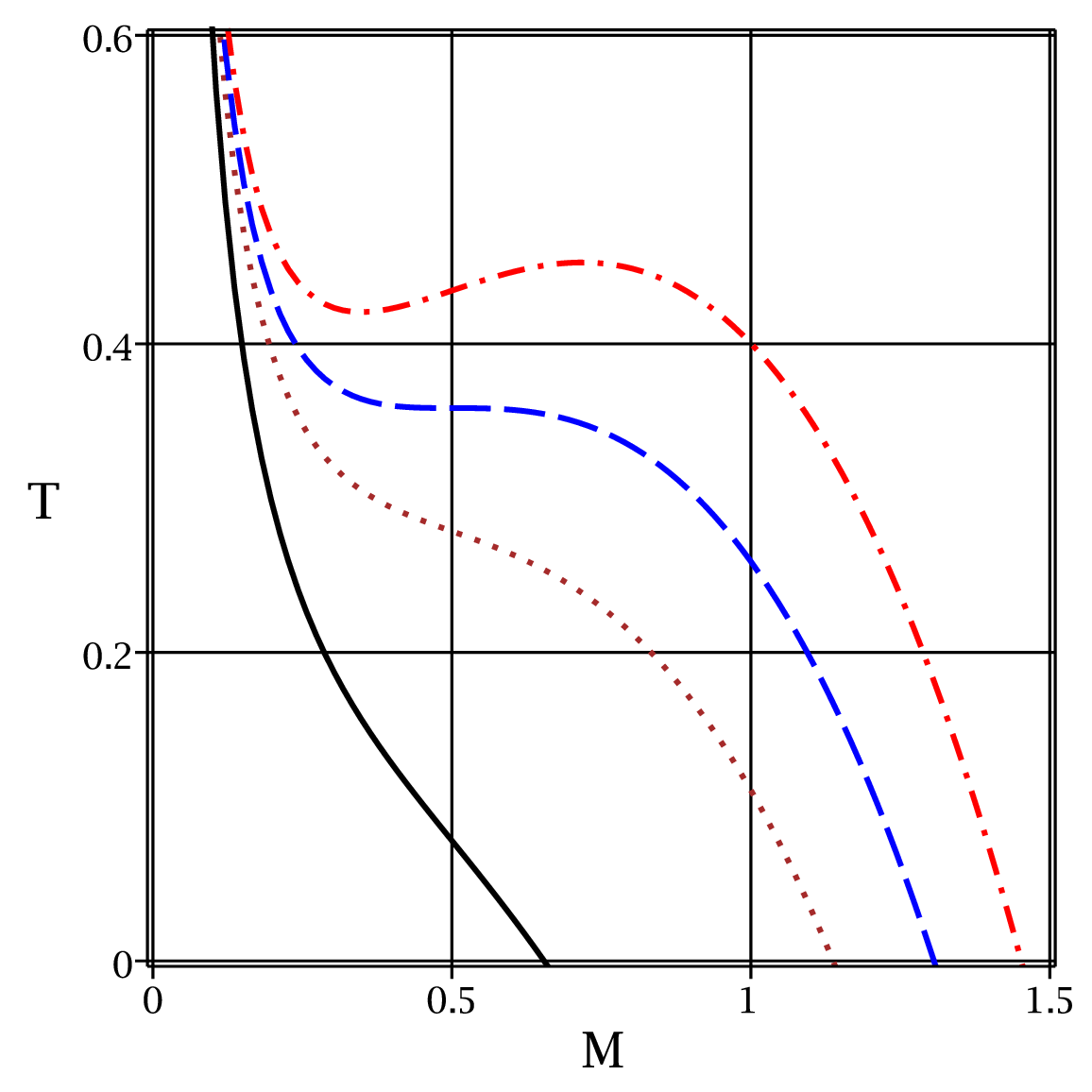} \\
\epsfxsize=5.5cm \epsffile{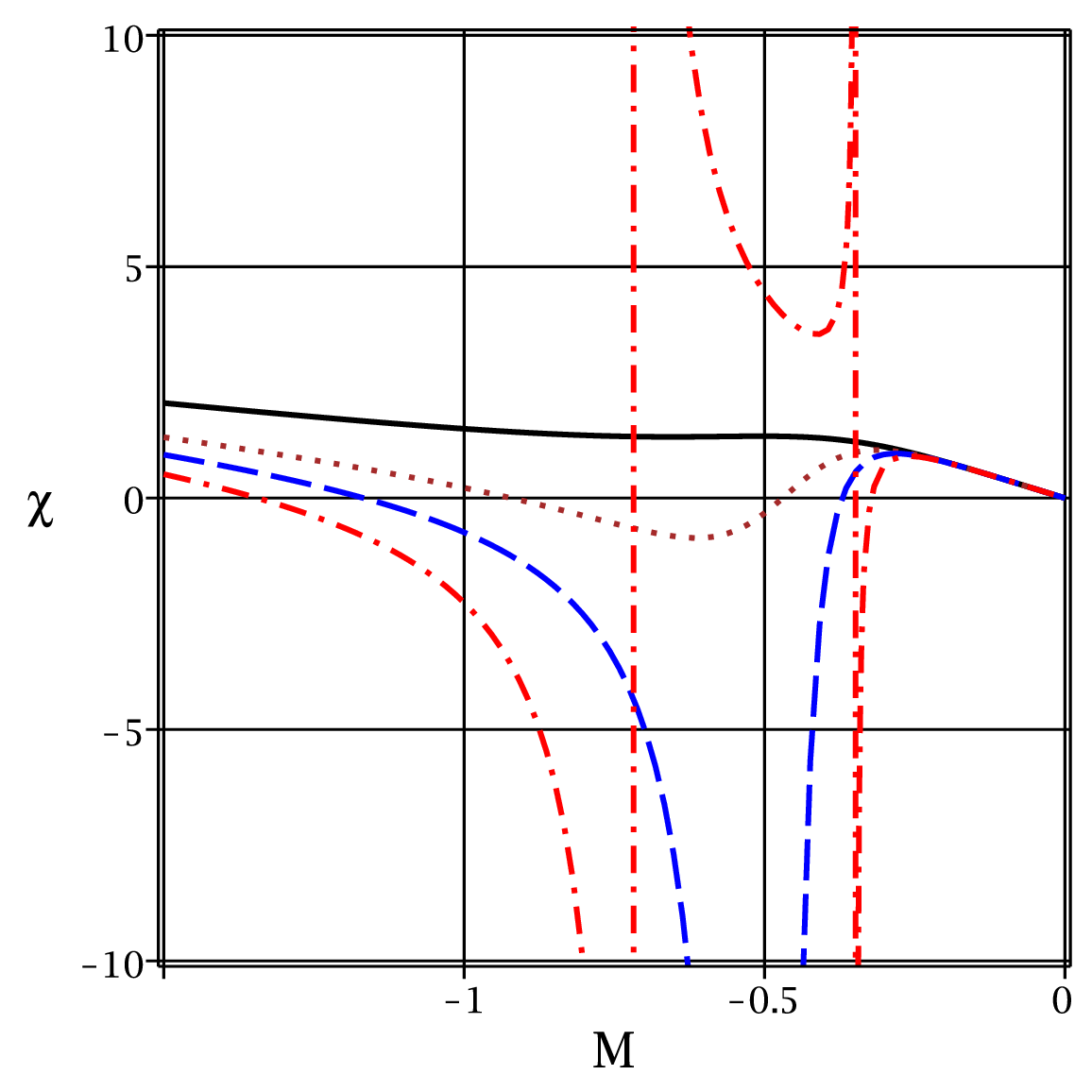} & \epsfxsize=5.5cm %
\epsffile{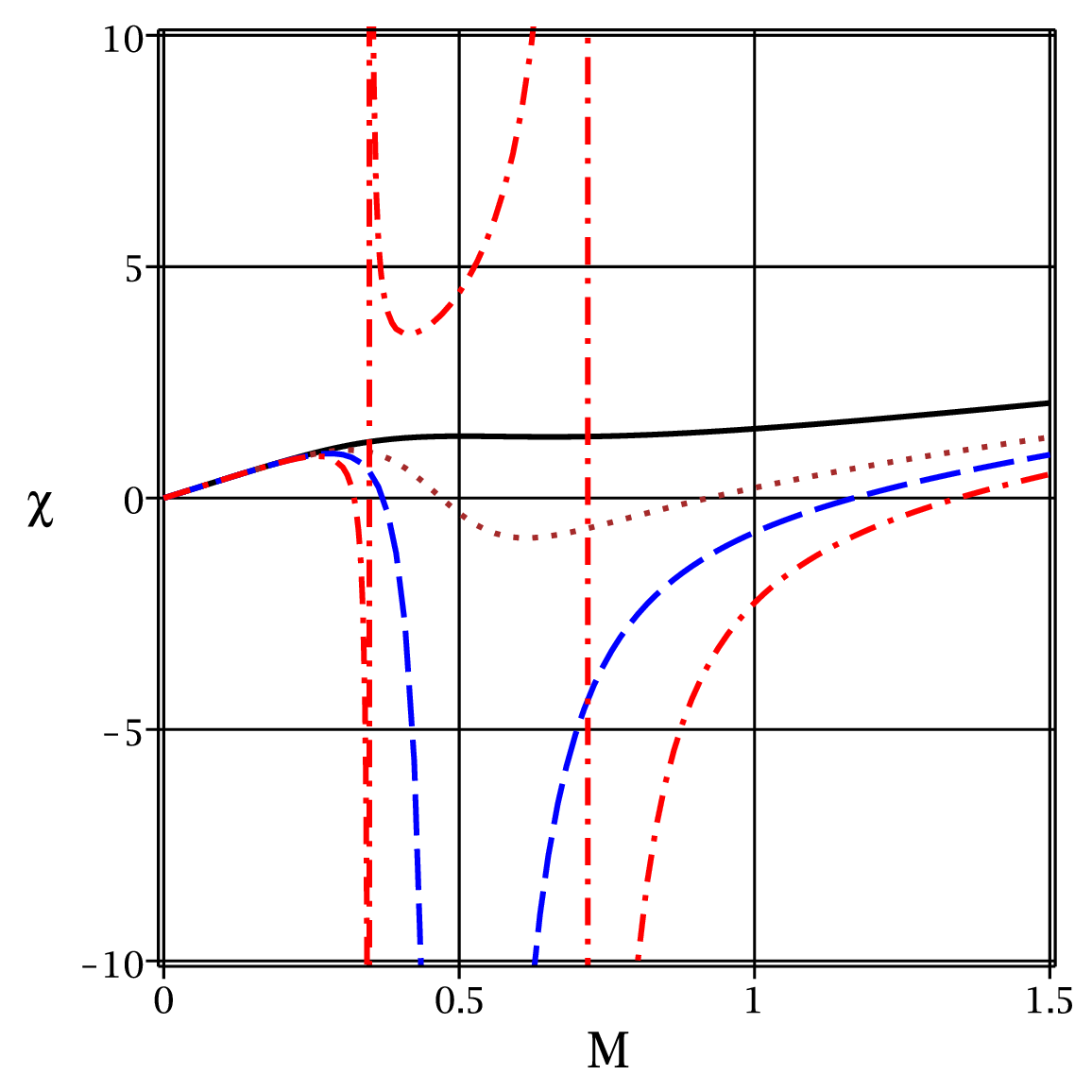}%
\end{array}
$%
\caption{$T$ (up panels) and $\protect\chi$ (down panels) versus
$M$ diagrams for $c=c_{1}=2$, $c_{2}=3$, $\Phi _{E}=0.1$, $b=1$
and $k=0$; $m=0$ (continuous line), $m=0.3$ (dotted line),
$m=0.3547299442$ (dashed line) and $m=0.4$ (dashed-dotted line).
\newline \textbf{Left panels:} $B=0.25$; \textbf{Right panels:}
$B=-0.25$.} \label{Fig12}
\end{figure}

\begin{figure}[tbp]
$%
\begin{array}{cc}
\epsfxsize=5.5cm \epsffile{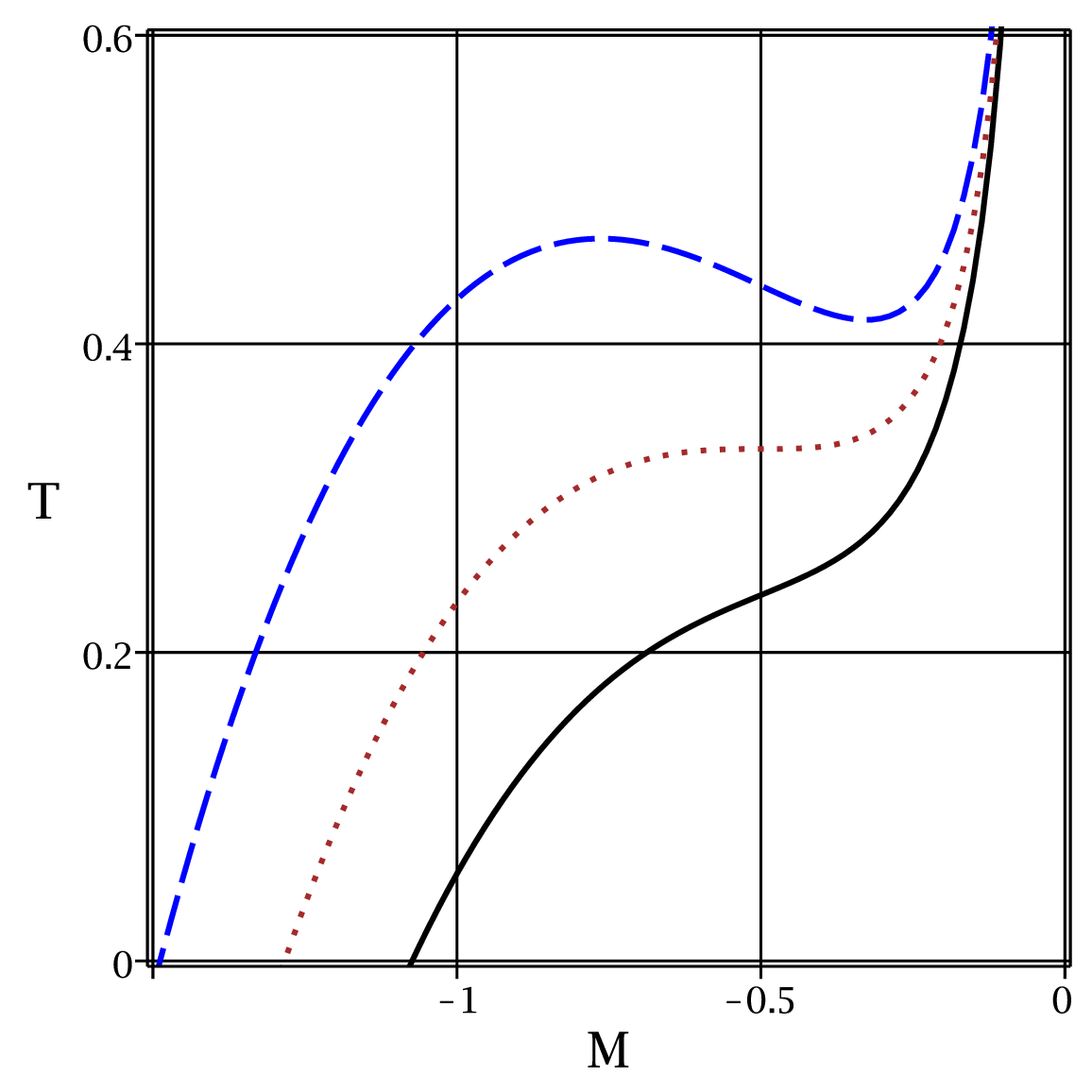} & \epsfxsize=5.5cm %
\epsffile{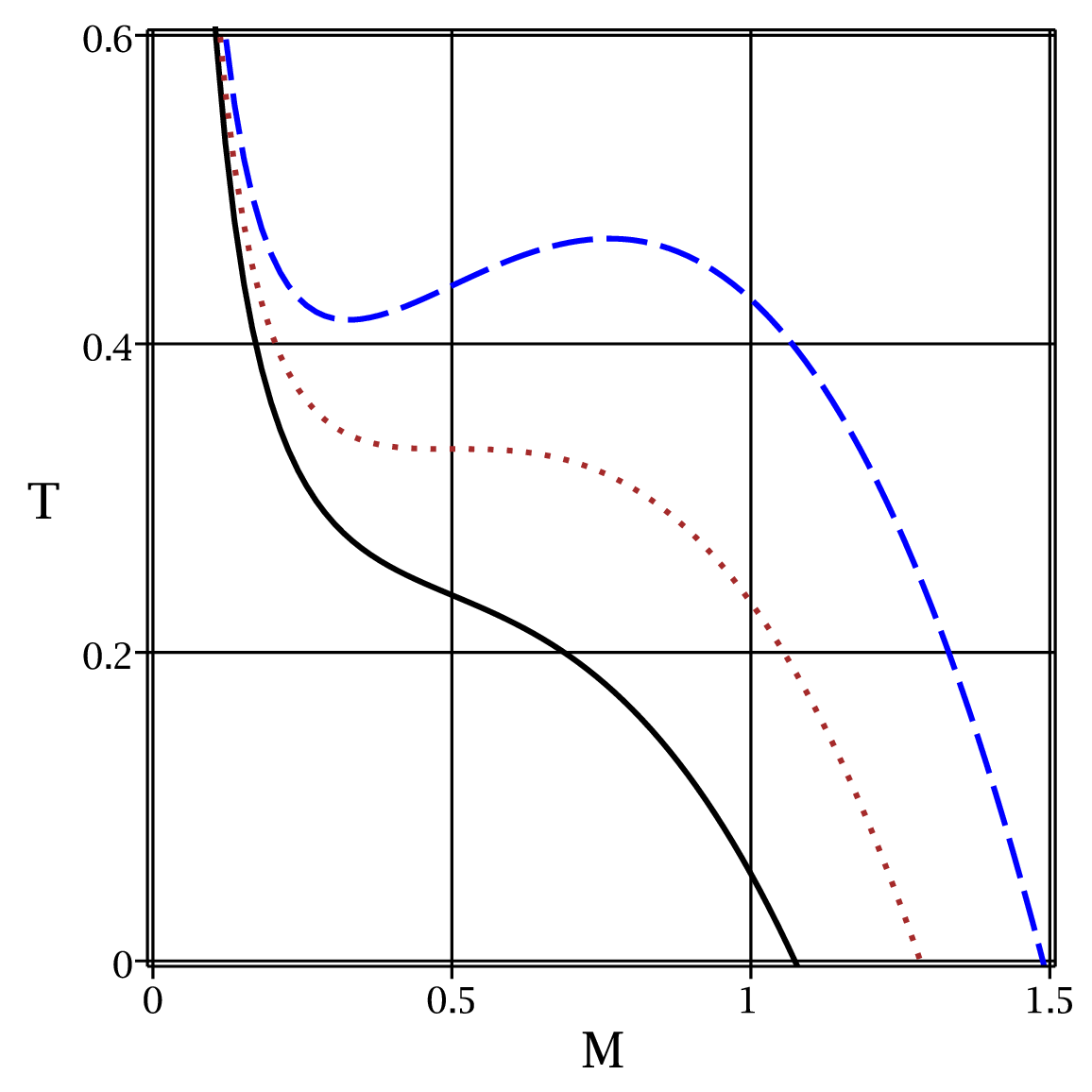} \\
\epsfxsize=5.5cm \epsffile{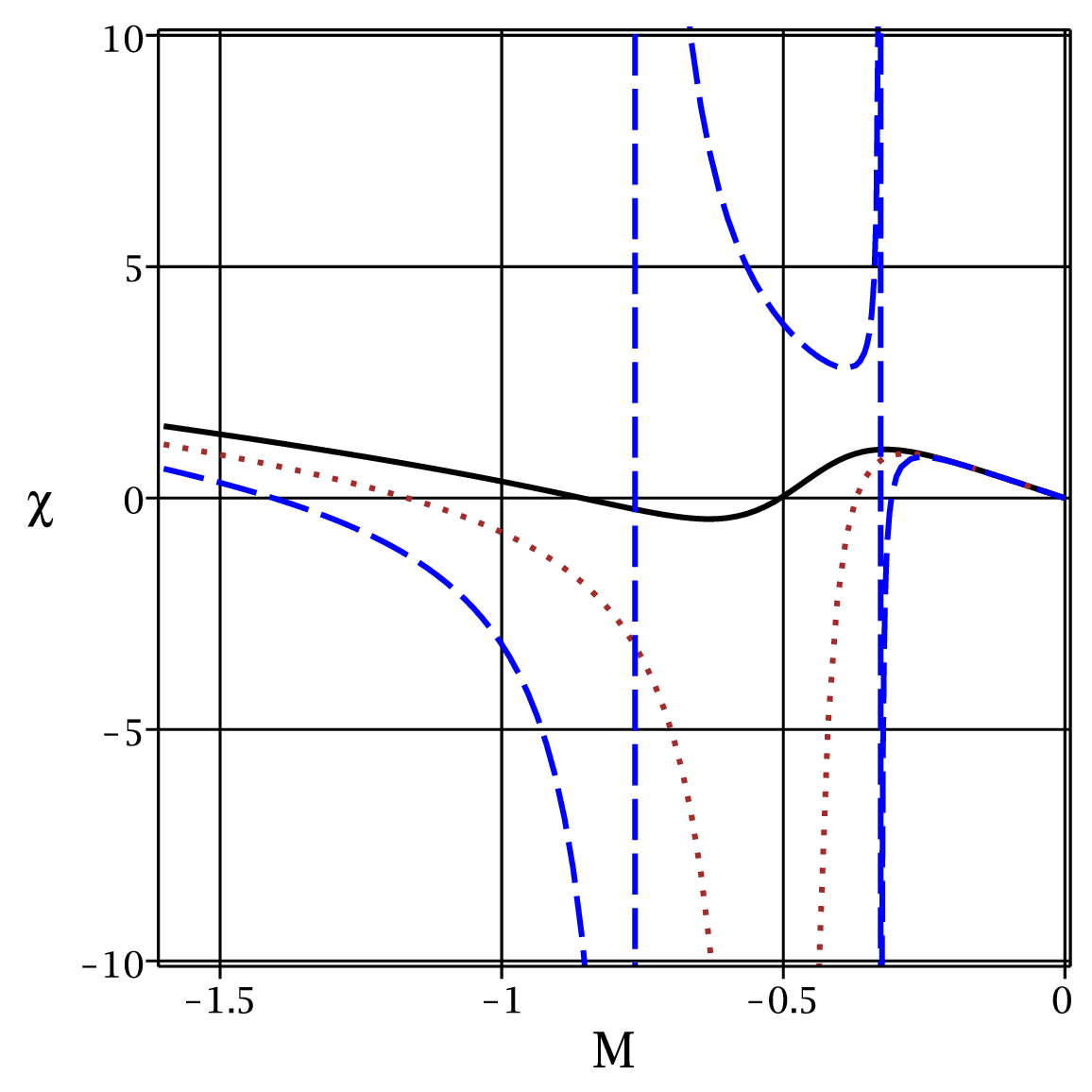} & \epsfxsize=5.5cm %
\epsffile{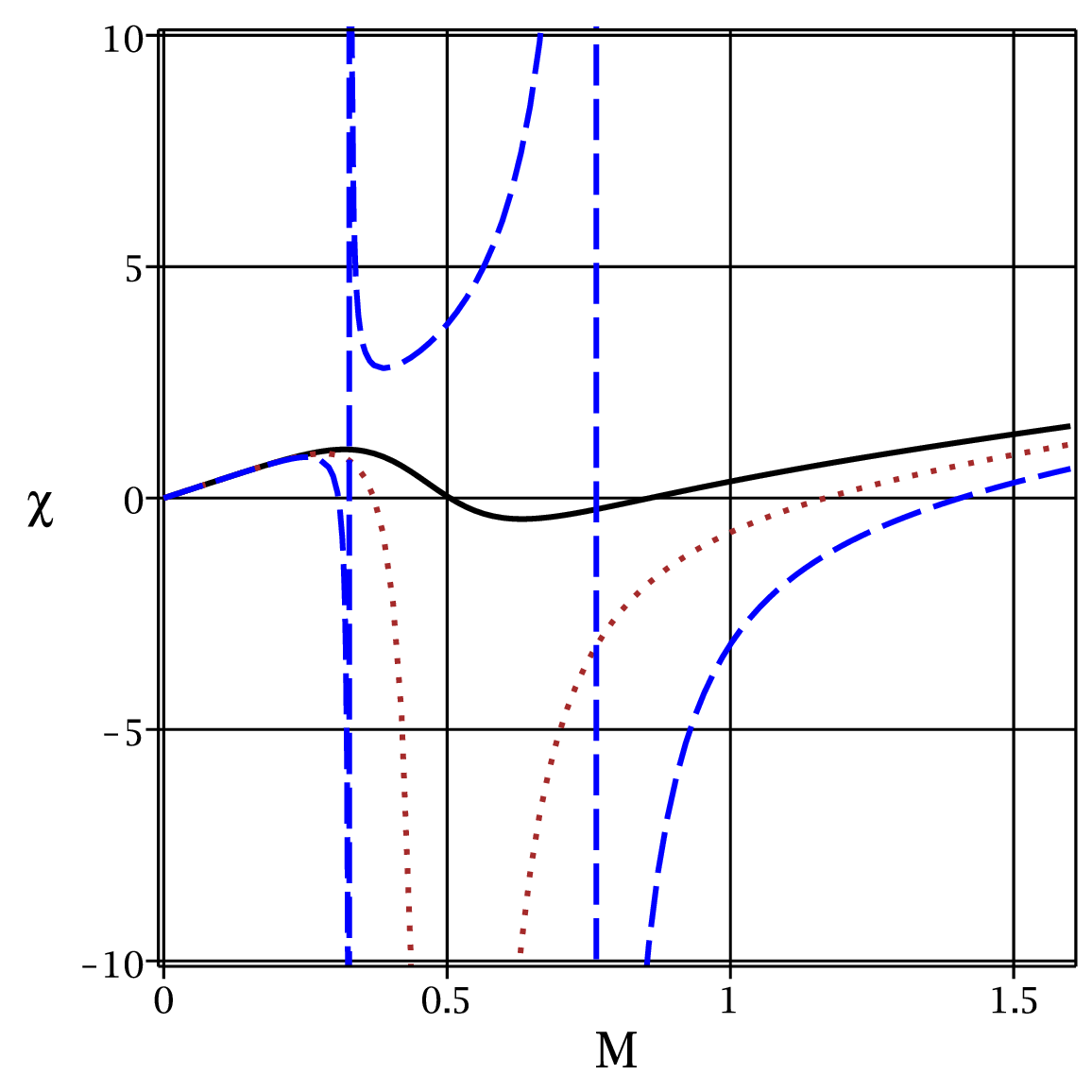}%
\end{array}
$%
\caption{$T$ (up panels) and $\protect\chi$ (down panels) versus
$M$ diagrams for $c=c_{1}=2$, $c_{2}=3$, $\Phi _{E}=0.1$, $b=1$
and $k=1$; $m=0$ (continuous line), $m=0.2061552812$ (dashed line)
and $m=0.3$ (dashed-dotted line). \newline \textbf{Left panels:}
$B=0.25$; \textbf{Right panels:} $B=-0.25$.} \label{Fig13}
\end{figure}

In this section, we will employ the obtained relations in the
previous
section to plot $T-M$ and $\chi -M$ diagrams (Figs. \ref{Fig10}--\ref{Fig13}%
) for studying the diamagnetic and paramagnetic behavior of the
solutions.


Fig. \ref{Fig10} is provided as an example of the massless case.
The massless case was studied in an interesting paper (we refer
the reader to Ref. \cite{dyonicmassless}). We just point out that
for large (small) values of magnetic field the phase transition
points are eliminated and only paramagnetic phase exists for the
boundary theory. Here, we only focus on the effects of massive
gravity and topological structure of the solutions.

The first important effect of massive gravity is the existence of
phase transitions and roots for non-spherical boundary solutions.
Taking a closer look at the obtained roots (\ref{rootchi}) and
divergencies (\ref{Mcchi}) for the susceptibility, one can confirm
that in the absence of massive gravity, no root and divergency
exists for the susceptibility in non-spherical cases. This
indicates that the sign of susceptibility for non-spherical cases
remains fixed and only one phase of diamagnetic/paramegnetic
exists. But generalization to massive gravity provides the
possibility of existence of root and divergency for the
susceptibility in non-spherical cases. This leads to the presence
of different diamagnetic and paramagnetic behaviors for spherical
and non-spherical boundary solutions (see dashed and dashed-dotted
lines in Figs. \ref{Fig11}-\ref{Fig13} in which dashed line is for
the existence of only one phase transition while dashed-dotted
lines is related to the cases of existence of two phase transition
points).

Since $B$, could be positive or negative, we have plotted two sets
of panels in Figs. \ref{Fig11}-\ref{Fig13} and investigated the
effects of the topological factor on the magnetic behavior of the
solutions for both branches. Studying dashed lines for Figs.
\ref{Fig11}-\ref{Fig13} shows that divergency is an increasing
(decreasing) function of the topological factor. In other words,
for the hyperbolic case, obtained divergency is located at larger
(smaller) magnetization comparing to horizon flat and spherical
cases. One can conclude that phase transition takes place at
smaller (larger) magnetization in spherical solutions while such
transition happens at larger (smaller) magnetization. This
emphasizes the role of the topological structure of the boundary
theory.

Generally, the effects of the graviton mass should be separated
into four groups which are marked with $m_{*}$ and $m_{c}$. For
gravitons lighter than $m_{*}$, $0 \leq m < m_{*}$, for hyperbolic
and flat cases (Figs. \ref{Fig11} and \ref{Fig12}, respectively),
the susceptibility is positive valued and
only paramagnetic behavior exists. Increasing mass of the graviton to reach $%
m_{*} < m < m_{c}$ leads to the existence of two crossovers
between paramagnetic and diamagnetic phases for non-spherical
cases. Evidently, the small and large boundary cases have
paramagnetic behavior while the medium solutions are diamagnetic.
Therefore, here, we have three phases where small and large ones
have paramagnetic behavior whereas the medium phase has
diamagnetic properties. The smaller phase of the boundary theory
corresponds to the large black holes while vice versa could be
said about the larger phase of the boundary theory. For the
spherical case of considered parameters without massive term, the
crossovers are observed and single phase is not seen.

For the case $m = m_{c}$ two crossovers between paramagnetic and
diamagnetic are observed with a phase transition located at the
diamagnetic region. The crossovers are marked as roots for the
susceptibility while the phase transition could be recognized by
its divergency. Here two (before smaller root and after larger
roots of the susceptibility) solutions have paramagnetic behavior
while the intermediate diamagnetic solutions are located between
these two roots. Interestingly, the phase transition takes place
at the diamagnetic region and it is between two diamagnetic
phases. The phase transition here is matched with an extremum in
temperature diagrams. For this case, we have two phases of
paramagnetism and two phases of diamagnetism.

Finally, for $m_{c} < m$, the susceptibility acquires two
divergencies and
two roots. Therefore, here, we have four points which are smaller root ($%
M_{smallr-root}$), smaller divergency ( $M_{smallr-\infty}$),
larger divergency ($M_{larger-\infty}$) and larger root
($M_{larger-root}$). These points follow the inequality
$M_{smallr-root} < M_{smallr-\infty} < M_{larger-\infty} <
M_{larger-root}$. As it is clear, these four points provide five
different phases.

For $M < M_{smallr-root}$ and $M_{larger-root} < M$, the
susceptibility is positive valued which indicates that
paramagnetic phases exist in these two regions. On the other hand,
in the cases of $M_{smallr-root} < M < M_{smallr-\infty}$ and
$M_{larger-\infty} < M < M_{larger-root}$, the susceptibility is
negative which shows that the solutions in these cases have
diamagnetic behavior. Interestingly, between two divergencies, we
have positive susceptibility which presents a paramagnetic phase.
It is worthwhile to mention that both divergencies in the
susceptibility are matched with two extrema in the temperature.
Remembering the results of previous sections regarding the regions
of thermal stability, one can conclude that the region of
paramagnetic solutions which is between two divergencies is
thermally unstable. Here, when the boundary theory reaches smaller
divergency, it jumps to the larger divergency. Therefore, the
system never experiences this specific region of the
paramagnetism. This phase transition between two different values
of magnetization also indicates that specific values of
magnetization are not experienced by the boundary theory.

We therefore see that the variation of massive parameter
introduces new phase transitions and different phases of
diamagnetic/paramagnetic into the phase space of the boundary
theory. This shows that depending on the graviton mass, the phase
structure of the boundary theory might be modified and new
phenomena could happen.

Another interesting result appearing from plotted diagrams is that
phase transitions of the susceptibility match with extrema of the
temperature. In previous sections, we showed that extrema of the
temperature, divergencies of the heat capacity, extrema of the
free energy and the critical points of van der Waals diagrams
coincide with each other. Combining the results of these two
sections with each other, one can conclude that the van der Waals
like phase transitions of the black holes, divergencies of the
heat capacity of black holes and the phase transitions of the
boundary theory coincide with each other. Therefore, studying heat
capacity of black holes or their van der Waals like phase
transitions enables one to make statements regarding phase
transition of the boundary theory and these phase transitions are
connected.

\subsection{High temperature limit}

Our final study for this section concerns the high temperature
limit. For the high temperature limit, we can regard the dominant
terms of $T $ as
\begin{equation}
T\approx \frac{m^{2}cc_{1}}{4\pi }-\frac{3Bl^{2}}{4\pi M}.
\label{highT}
\end{equation}

The obtained relation contains massive term which is the
correction and the effect of generalization to massive gravity.
Here, the massive term is not coupled with magnetic field and
magnetization, which indicates that it plays the role of a
constant. As for the susceptibility, we see that the high energy
limit leads to
\begin{equation}
\chi \approx -\frac{M}{B},  \label{highchi1}
\end{equation}%
in which by using Eq. (\ref{highT}), one can find

\begin{equation}
\chi \approx \frac{3l^{2}}{4\pi T-m^{2}cc_{1}} \approx \frac{3l^{2}}{4\pi T}+%
\frac{3l^{2}m^{2}cc_{1}}{16\pi ^{2}T^{2}}+O(T^{-3}).
\end{equation}

Here, we see the Curie constant which is given by $\left( \frac{4\pi }{3b^{2}%
}\right) ^{-1}$ and a correction a correction of Curie law which
is due to generalization to massive gravity for high temperature
limit of the susceptibility.

\section{Closing Remarks}

This paper included a thermodynamical study of charged dyonic
black holes in the presence of massive gravity and holographical
aspects of its dual boundary $CFT$. Thermodynamical quantities
were extracted and it was shown that existence of the massive
gravity term modifies conserved and thermodynamical quantities of
the black holes. These modifications introduce new phenomena with the most
 important ones being: existence of remnant for the
temperature in evaporation of black holes, van der Waals behavior
and phase transition for non-spherical black holes.

Thermal stability and conditions for having physical solutions
were investigated. In addition, the free energy and its properties
were discussed. The concept of extended phase space was also
employed to extract phase transition points and plot the van der
Waals like diagrams. The limiting cases were also studied and it
was shown that existence of critical behavior is magnetic charge
dependent. It was also pointed out that generalization to massive
gravity leads to the presence of van der Waals like behavior for
non-spherical black holes. This essentially resulted into the
existence of second order phase transitions in the boundary theory
with non-spherical horizons (hyperbolic and planar) as well.
Furthermore, an alternative method was employed to extract
critical points which presented themselves as maxima in the
phase diagrams. More importantly, it was shown that the critical
points that were extracted through different methods were
consistent. In addition, a comparative study regarding thermal
stability and van der Waals like phase transition was done and
thermodynamical behavior of the black holes in unstable regions
was discussed.

Next, the properties of the boundary $CFT$ dual to the bulk theory
were investigated. It was shown that depending on the choices of
different allowed parameters (see appendix), the boundary theory
may have; only one paramagnetic phase, two paramagnetic and one
diamagnetic phases, two paramagnetic and two diamagnetic phases
with a phase transition between the two diamagnetic phases and
finally, two paramagnetic and two diamagnetic phases with two
phase transitions and one thermally unstable paramagnetic phase.

Finally, it was pointed out that the high temperature limit of
holographical aspect is free of massive term up to first dominant
term. However, the graviton mass effects will appear as the second
dominant term of high temperature limit. The correction terms were
shown to be necessary in the expressions for temperature, the
susceptibility of boundary theory and the Curie law.

In this paper, we also investigated linear electrodynamics in the
context of massive gravity. It will be interesting to generalize
our results to the cases of nonlinear electrodynamics and higher
derivative gravity. In addition, it is worthwhile to study new
massive gravity in this regards. The massive gravity considered in
this paper enjoys the Lorentz violating property. The existence of
Lorentz violating property provides the possibility of
constructing other consistent massive gravity theories which could
be expressed in Stueckelberg form. Several of these theories were
introduced and employed in the context of holography
\cite{NN1,NN2,NN3,NN4}. The thermodynamical quantities and
behavior that were reported for obtained black holes in this paper
will be modified if other generic theories of the massive gravity
are employed. It would be interesting to see how the behaviors
reported in this paper could be modified if other theories of the
massive gravity are employed. In addition, we have investigated
thermodynamic stability of the solutions in this work. It is
interesting to
study dynamical stability and anti-evaporation of such solutions \cite%
{Evapor1,Evapor2,Evapor3} in a separate work.

\begin{figure}[tbp]
$%
\begin{array}{c}
\epsfxsize=5.5cm \epsffile{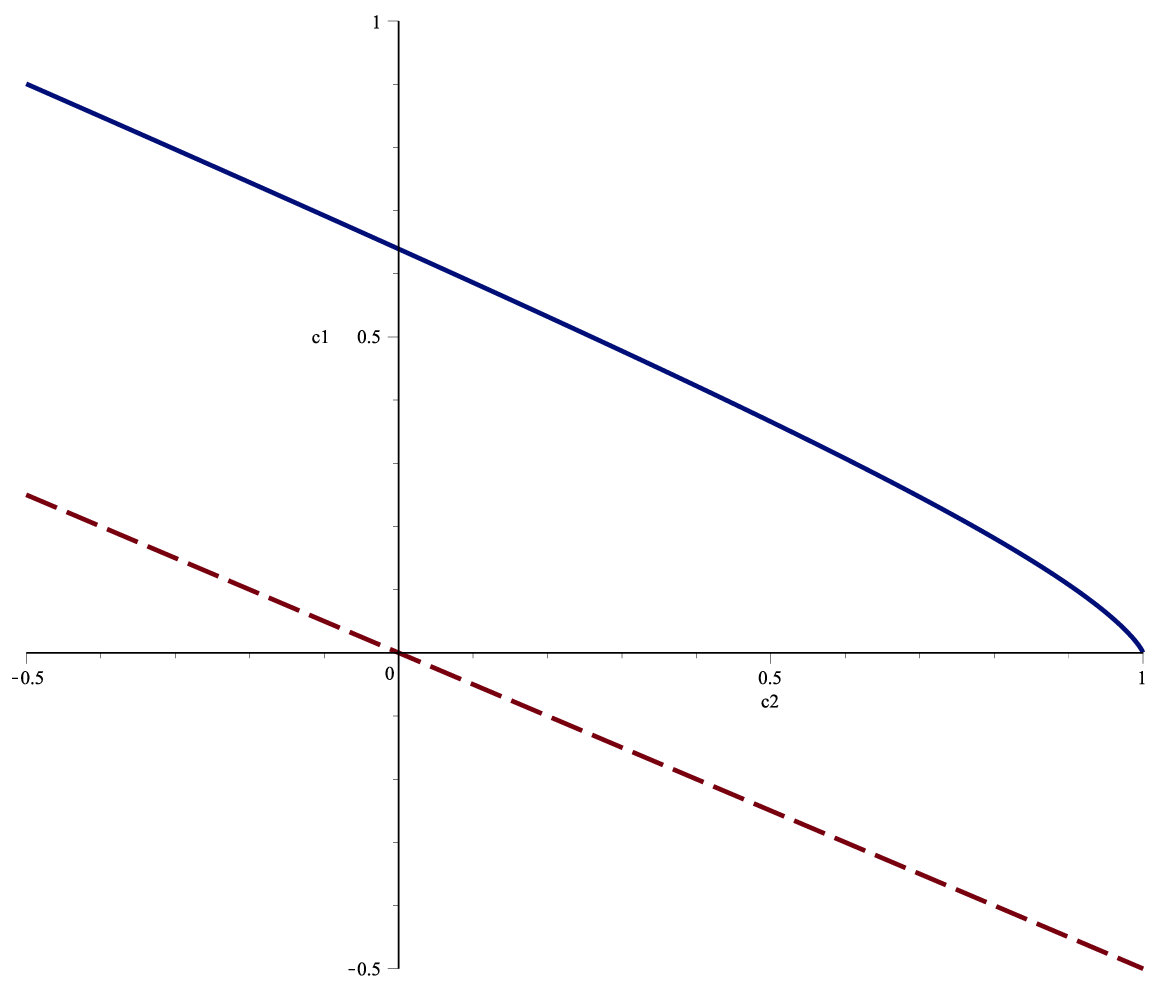}
\end{array}
$%
\caption{$c_{1}$ versus $c_{2}$ for $r_{h}=2$, $c=m=\Phi_{E}=1$
and $q _{M}=P=k=0$.\newline \textbf{solid diagram:}
$c_{1}=\frac{1}{2}(c_{2}-1)\left(-1-LambertW(\frac{e^{-1}}{1-c_{2}})
\right)$ and \textbf{dashed diagram:} $c_{1}=-\frac{c_{2}}{2} $.}
\label{CC12}
\end{figure}

\begin{center}
\textbf{Appendix}
\end{center}

Here, we are going to study the regime where the graviton mass is
positive \cite{Vegh,app2}. Following the method of Ref.
\cite{Vegh}, one finds two relations for the massive parameters,
which could be employed to find the stability (positive mass of
gravitons) condition in the parameter space. One of the mentioned
relations comes from identification of entropy density calculated
from the finite Euclidean action and what we obtain from the area
law. Such identification leads to
\begin{equation}
\sum_{R=R_{0}}\frac{R^{3}\ln (r_{h}-R)}{%
3m^{2}cc_{1}R^{2}+2m^{2}c^{2}c_{2}R+32\pi PR^{3}-2\Phi
_{E}^{2}R+2kR}=0, \label{S-LIMITED}
\end{equation}%
where $R_{0}$ is the root of the following equation%
\begin{equation*}
8\pi PR_{0}^{4}+m^{2}cc_{1}R_{0}^{4}+(m^{2}c^{2}c_{2}-\Phi
_{E}^{2}+k)R_{0}^{4}-q_{M}^{2}=0.
\end{equation*}

The other relation can be obtained from maximizing the chemical
potential by setting $q_{E}=q_{M}=0$ in the vanishing temperature.
After straightforward calculations, we find
\begin{equation}
c_{2}=-\frac{r_{h}}{c}c_{1}-\frac{8\pi Pr_{h}^{2}}{m^{2}c^{2}}-\frac{k}{%
m^{2}c^{2}}.  \label{T-LIMITED}
\end{equation}

Because of the existence of cosmological constant (pressure), in
the first relation (\ref{S-LIMITED}), it is not possible to obtain
the required relation analytically, therefore we have employed
numerical method. For more clarity, we plot Fig. \ref{CC12} as an
example of parameter space. It is worth mentioning that the solid
line is related to
Eq. (\ref{S-LIMITED}), while dashed line is plotted based on Eq. (\ref%
{T-LIMITED}). As it mentioned in Ref. \cite{Vegh}, the region
between these lines is allowed parameter space, while other zones
are forbidden.

\begin{acknowledgements}
We thank the anonymous reviewers for their careful reading of our
manuscript and their many insightful comments and suggestions. We
also thank both Shiraz University and Shahid Beheshti University
Research Councils. This work has been supported financially by the
Research Institute for Astronomy and Astrophysics of Maragha,
Iran.
\end{acknowledgements}

\end{document}